\documentclass[twocolumn,prd,floatfix,preprintnumbers,nofootinbib,superscriptaddress, aps]{revtex4-1}
\usepackage{graphicx}  
\usepackage{amsmath}   
\usepackage{amssymb}   
\usepackage{bm} 
\usepackage{dcolumn}
\usepackage{color}
\usepackage{mathrsfs}
\usepackage{amsfonts}
\usepackage{varioref}
\usepackage{textcomp}
\usepackage{mathtools}
\usepackage[normalem]{ulem}
\usepackage{nccmath}
\usepackage{url}
\RequirePackage[colorlinks,citecolor=blue,urlcolor=magenta,linkcolor=blue]{hyperref}
\allowdisplaybreaks

\def\hl{\hat{\ell}}
\def\tl{\tilde{\ell}}
\def\nz{near-zone }
\def\fz{far-zone }

\newcommand{\mce}{\mathcal{E}}
\newcommand{\mcb}{\mathcal{B}}
\newcommand{\mc}[1]{\mathcal{#1}}
\newcommand{\hp}{\hat{\partial}}
\labelformat{section}{Section #1} 
\labelformat{subsection}{Section \thesection #1} 
\labelformat{subsubsection}{Section \thesection \thesubsection #1}
\labelformat{subsubsubsection}{Section #1}
\labelformat{equation}{Eq.~(#1)} 
\labelformat{figure}{Fig.~#1} 
\labelformat{subfigure}{Fig.~\thefigure#1} 
\labelformat{table}{Tab.~#1} 
\labelformat{appendix}{Appendix #1}
\begin{document}

\title{Dynamical tidal Love numbers of black holes under generic perturbations: \\
Connecting black hole perturbation theory with effective field theory}

\author{Sumanta Chakraborty}
\email{tpsc@iacs.res.in} 
\affiliation{School of Physical Sciences, Indian Association for the Cultivation of Science, Kolkata-700032, India}

\author{M.V.S Saketh}
\email{msaketh@aei.mpg.de}
\affiliation{Max-Planck-Institute for Gravitational Physics (Albert-Einstein-Institute), \\
Am m\"{u}hlenberg 1, 14476 Potsdam-Golm, Germany, European Union}
\affiliation{International Centre for Theoretical Sciences, Tata Institute of Fundamental Research, Bangalore 560089, India}

\author{Tanja Hinderer}
\email{t.p.hinderer@uu.nl }
\affiliation{Institute for Theoretical Physics, Utrecht University,\\
Princetonplein 5, 3584 CC Utrecht, Netherlands, European Union}

\author{Jan Steinhoff}
\email{jan.steinhoff@aei.mpg.de}
\affiliation{Max-Planck-Institute for Gravitational Physics (Albert-Einstein-Institute), \\
Am m\"{u}hlenberg 1, 14476 Potsdam-Golm, Germany, European Union}

\begin{abstract}
The foundation for modeling the coupling of the internal structure of compact objects in binary systems to their dynamics and emitted gravitational waves is a systematic effective field theory (EFT) framework, where each compact object is replaced by a worldline endowed with a set of internal degrees of freedom. These degrees of freedom encode finite-size effects and thereby distinguish between different classes of compact objects. Among finite-size effects, tidal interactions play a central role, as they are associated to various kinds of deformations of a body under the influence of external tidal fields. 
In this work, we analyze the dynamical tidal response of Kerr black holes to generic-spin perturbations, focusing primarily on the scalar and gravitational cases, and working to linear order in frequency. We establish an EFT description of the perturbed black hole that accounts for the couplings between the spin, gravitoelectric and -magnetic tidal fields. We match this to wave-like solutions to the full black hole perturbation equations in order to determine the tidal response coefficients. In particular, we obtain the dynamical Love number, which appears at linear order in frequency for spinning black holes, and derive an approximate expression for the dynamical tidal response, including both dissipative and conservative pieces. We also examine several technical subtleties that arise in the matching procedure, with special emphasis on the mixing of multipolar modes induced by the spin of the compact object, which proves to be essential for a consistent treatment.
\end{abstract}
\maketitle
\tableofcontents
\section{Introduction}
Relativistic tidal effects in binary systems of compact objects play a key role for probing the nature, interior structure, and strong internal gravity of the objects with gravitational waves (GWs). During a binary inspiral, tidal deformations lead to small but clean and cumulative GW signatures that involve internal-structure-dependent parameters such as tidal Love numbers~\cite{Flanagan:2007ix}. Love numbers characterize the ratio of the induced deformation of an object's exterior spacetime to the strength of the perturbing tidal field in the adiabatic limit, where the tidal forcing frequency is far from resonance with a normal mode of the object~\cite{Hinderer:2007mb, Damour:2009vw,Binnington:2009bb}. They were first measured in the binary neutron star event GW170817~\cite{LIGOScientific:2018cki}, which yielded valuable information about the equation of state of matter at supranuclear density. For black holes (BHs) in vacuum in General Relativity and four spacetime dimensions, the tidal Love numbers vanish~\cite{Kol:2011vg,Ivanov:2022qqt, LeTiec:2020bos, Chia:2020yla, Poisson:2021yau}, which is a consequence of special symmetries~\cite{Berens:2025okm, Sharma:2025xii, Parra-Martinez:2025bcu, Hui:2021vcv, Sharma:2024hlz, Barbosa:2025uau, BenAchour:2022uqo, Hui:2022vbh, Charalambous:2022rre, Katagiri:2022vyz} (however, see \cite{Ghosh:2026vig}). 
Relaxing any of these restrictions generally leads to non-vanishing Love numbers for BHs. For a review, see \cite{Chakraborty:2026qru, Rodriguez:2026iot}.

Tidal Love numbers are only a limiting case of the richer physics encoded in an object's frequency-dependent response to a tidal field. During an inspiral, 
the tidal forcing frequency, related to the orbital motion, sweeps up over many decades in magnitude. When it passes through an interval where it approximately matches the internal-structure-dependent mode frequency of one of the objects, the tidal deformation is resonantly enhanced~\cite{Bildsten:1992my}. Even far from a resonance, the frequency-dependent tidal response can be significantly larger than the strictly static case~\cite{Hinderer:2016eia}. Such \emph{dynamic} tides lead to subdominant effects in GWs that involve additional parameters, including mode frequencies. This opens the possibility of spectroscopic studies of compact objects with isolated modes during the inspiral, which is of interest for a wide range of science goals with GWs~\cite{ET:2025xjr, Evans:2021gyd, LISA:2022kgy}. As GW detectors are becoming more sensitive and delivering more precise and larger, more diverse datasets, and with the advent of novel detectors such as Einstein Telescope~\cite{EinsteinTelescope}, Cosmic Explorer~\cite{Reitze:2019iox} and LISA~\cite{LISA}, such effects are expected to become important and must be accurately modeled. This is because the GW data analysis for coalescing binaries relies on cross-correlating template waveforms with the datastream.
Although the statistical errors for GW events detected thus far have been relatively large, the field is already at a turning point. Systematic uncertainties due to shortcomings in the models are becoming noticeable~\cite{LIGOScientific:2025rsn}, and they may become the dominant limitation for precision GW physics in the future.

Including effects of dynamical tides in waveform models has been an active area of research, with a large influx of new methods and approaches. Obtaining the imprints on the GW signal involves a description of the binary system at large separation as being effectively point masses but with extra degrees of freedom associated to each worldline that account for finite-size effects. This is known as skeletonization or in more modern formulations as wordline effective theory~\cite{Goldberger:2004jt,Goldberger:2005cd}, which we hereafter refer to as 'EFT'. The EFT involves coupling coefficients that must be matched to results from fully relativistic calculations of a compact object's response to external tidal perturbations. This matching is a key step determining the information flow from the strong-field region of the compact object to the much larger scales of the binary system and GWs. For static tides, the matching identifies the Love numbers extracted from calculations of relativistic tidal perturbations to compact objects with Wilson coefficients that enter the coarse-grained EFT description. However, more generally, the matching is subtle and requires care, leading to a nontrivial link between parameters of a perturbed object and the coefficients appearing it its EFT, see, e.g., Refs.~\cite{Gupta:2020lnv,Caron-Huot:2025tlq,Katagiri:2024wbg,Diedrichs:2025vhv}. The EFT underpins further calculations of the effects on binary dynamics and GWs, using e.g. post-Newtonian theory for two-body interactions and radiation.       

Incorporating tidal effects beyond the static limit in such EFT descriptions is of particular interest for BHs, both from a theoretical perspective to understand their peculiarities and because they are currently the most frequently detected GW signals~\cite{LIGOScientific:2025slb}. For instance, tidal effects can probe physics on horizon scales~\cite{Coviello:2025pla, Maselli:2018fay, nair2023dynamical-a19, chakraborty2024dynamical-799}, the theory of gravity~\cite{Cano:2025zyk, Cardoso:2018ptl, Singha:2025xah}, the number of dimensions~\cite{Kol:2011vg, Cardoso:2019vof, Chakravarti:2018vlt, Rodriguez:2023xjd}, and distinguish Kerr BHs from other objects~\cite{Silvestrini:2025lbe, Gurlebeck:2015xpa, Cardoso:2017cfl, Maggio:2021ans, Sennett:2017etc, Mendes:2016vdr, Crescimbeni:2024cwh, chakraborty2024dynamical-799}. An understanding of tidal phenomena for BHs is also needed as the baseline for identifying matter effects, for instance, in neutron stars~\cite{Chatziioannou:2020pqz} or the BH's environment~\cite{Chakraborty:2024gcr, DeLuca:2021ite, Cannizzaro:2024fpz}. The frequency-dependent, dynamical tidal response function (DTRF) for a given mode (characterized by $\ell$,$m$) of BHs is generally non-zero and complex. The imaginary contributions relate to dissipative effects, while the real part, hereafter referred to as the conservative tidal response function (CTRF), encapsulates frequency-dependent corrections to the Love numbers~\cite{Goldberger:2005cd, Saketh:2022xjb, creci2021tidal-42e, chakrabarti2013new-fa4, chia2020tidal-b97}. An explicit decomposition of the DTRF into dissipative and conservative parts is needed for calculating the effects in GWs, as phenomena in the two sectors enter in different ways. There has been much recent progress on computing the DTRFs of BHs, such as for non-rotating BHs in the scalar~\cite{creci2021tidal-42e} and gravitational cases~\cite{Combaluzier--Szteinsznaider:2025eoc,Kobayashi:2025vgl}, slowly rotating BHs~\cite{Chakraborty:2025wvs, Bhatt:2024rpx}, and rotating BHs for scalar~\cite{Caron-Huot:2025tlq} and gravitational cases~\cite{Saketh:2023bul, Perry:2023wmm, Perry:2024vwz, chakrabarti2013new-fa4, Bhatt:2024rpx}; see, e.g., Refs.~\cite{Andersson:2025iyd, HegadeKR:2024agt} for related work in the context of neutron stars. These works used various approaches and scenarios, and altogether revealed several important insights: (i) the matching of BH perturbations to an EFT is essential to extract the relevant coefficients in the DTRF in a gauge-invariant way, (ii) the connection between the near-horizon behavior and the EFT becomes more subtle in the frequency-dependent case, and (iii) basing the calculations on wave scattering rather than considering stationary perturbations bypasses several difficulties, e.g. due to the matching being carried out near null infinity. These conclusions motivate addressing the problem in a unified framework for the scalar, electromagnetic, and gravitational perturbations to rotating BHs that is based on scattering amplitudes and matching to an EFT. This paper centers around this aim, and also fills in remaining gaps in the literature along the way, as we explain.  

Specifically, we use scattering amplitudes for BH perturbations matched to an EFT to determine the DTRF of rotating BHs with arbitrarily large spins up to linear order in frequency, or more precisely to linear order in the dimensionless quantity $(GM\omega/c^3)$, where $M$ is the mass of the black hole, $G$ and $c$ are the gravitational constant and speed of light respectively, and $\omega$ is a Fourier frequency associated to the perturbations. This regime applies to situations far from a mode resonance, which is a good approximation for the early inspiral as the quasi-normal mode frequencies are typically of order $f\sim 1/M$, much higher than tidal forcing frequencies of order $f_{\rm orb}\sim \sqrt{(M+M_2)/d^3}$ at large separation $d$. Here $M_{2}$ is the mass of the secondary BH, with $M+M_{2}$ being the total mass of the binary. We compute the full DTRF in this linear approximation for generic-spin perturbations to BHs, which encompass the scalar, electromagnetic, and gravitational cases, and decompose it to extract the CTRF. We further show explicitly that while  intermediate calculational steps in our framework appear to become singular in the case of extremal BH spins, our final results for the DTRF in fact apply for BHs with arbitrarily large rotation, including extremal spins. 

This paper is organized as follows. In \ref{sec:EFT} we discuss the connection between the tidal coefficient in the EFT action and the response function in BH perturbation theory. Subsequently, in~\ref{scalar_scattering}, we consider the scattering amplitudes for scalar tidal perturbations to a rotating BH. These have no closed-form solution and we obtain approximate solutions in the near-horizon regime and far from the BH. We connect the information of these approximations in~\ref{subsec:scalar_matching_intermediate}, which determines how the horizon boundary conditions get imprinted in properties of the far-zone solution. Then we match the EFT coefficients from \ref{sec:EFT} with those from the far-zone solution, thereby extracting the dynamical response function and verify agreement with previous results in limiting cases. ~\ref{genspin} generalizes the methodology to generic spin fields. We follow a similar workflow as for the scalar case and explain several subtleties that arise along the way. We discuss the near-horizon behavior of generic-spin BH perturbations in~\ref{subsec:NZgeneric}, the far-zone asymptotics in~\ref{subsec:FZgeneric}, the matching of this information in~\ref{subsec:BHPTmatching}, and the resulting response function in~\ref{subsec:tidalresponsegeneric}. We also verify that our results for zero-spin fields agree with those derived in~\ref{scalar_scattering}. We discuss how the results extend to extremal BH spins in~\ref{sec:extremal}. We summarize the key results, discuss physical implications and compare with other works in~\ref{sec:discussion}. 

{\it Notation and conventions}: We use units $G=c=1$ throughout. Overbars denote complex conjugation. Greek letters denote covariant spacetime indices, Latin letters $i,j,k,\ldots $ indicate three-dimensional spatial indices. Capital Latin letters indicate multi-index strings, e.g. $T^L=T^{ i_1 i_2 \ldots i_\ell}$. Angular brackets around indices denote the symmetric-trace-free (STF) projection, see~\cite{RevModPhys.52.299} for details. For quantities that are always STF, we use the special multi-index notation $T^{I_\ell}=T^{\langle  L\rangle}$.

\begin{figure*}
\includegraphics[scale=0.4]{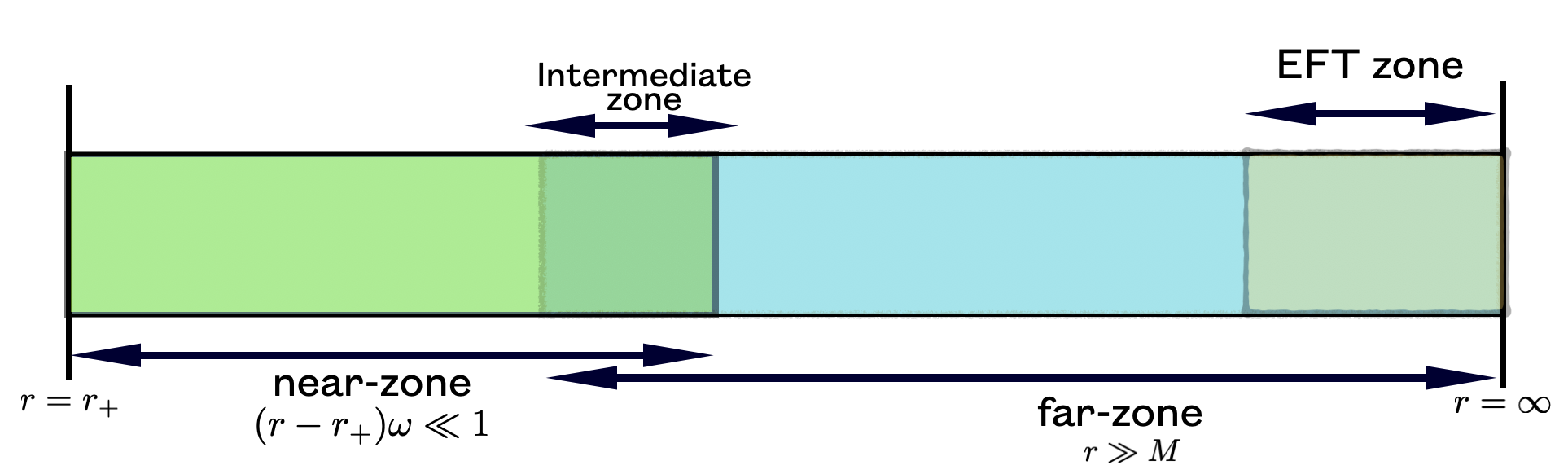}
\caption{Relevant zones in the problem of matching of BH perturbation theory with EFT. As illustrated, BH perturbation theory covers two zones, the far- and near-zones, which overlap in the intermediate zone. The asymptotic limit of the far-zone is matched with the EFT in order to obtain gauge invariant DTRFs. See text for further discussion.}
\label{fig:zoneEFT}
\end{figure*}

\section{The tidal response in the effective worldline theory}\label{sec:EFT}

A coarse-grained effective description of a tidally perturbed compact object such as a BH can be obtained in an (worldline) effective field theory (EFT), where the object is modelled as a point-entity described by a worldline augmented with certain additional multipolar degrees of freedom to account for finite size effects. For linear tidal effects, the multipolar degrees of freedom are chosen to be induced by external tidal fields via a linear functional, which in turn is characterized by tidal response functions. These functions are a priori undetermined, but can be fixed by matching the solutions to the perturbation equations in EFT to the solutions of the perturbation theory in the far-zone ~(see \ref{fig:zoneEFT}).

In this section, we briefly outline such an EFT description, and the solutions to the perturbation equations in the EFT for scalar and gravitational fields which are later used to fix the tidal response in the low-frequency, large-wavelength regime via matching. Some crucial but navigable subtleties arise since the equations for BH perturbation theory (BHPT) are most simply solved when the angular basis is spanned by spheroidal harmonics, while the tidal degrees of freedom in the EFT given by the multipole moments, and the fields they couple to are more naturally described using STF tensors which are closely related to spherical harmonics as both are representations of the rotation group in 3-dimensions (SO(3)).

We will work with perturbations of different spin weights corresponding to scalar, vector, and tensor fields, coupled to the worldline. Although the detailed structure of the action varies for different fields, the resulting actions can all be written schematically as the sum of three components as
\begin{equation}
\mathcal{A}_{\rm EFT} = \mathcal{A}_{\rm p.p} + \mathcal{A}_{\rm field} + \mathcal{A}_{\rm tidal}.    
\end{equation}
Here, $\mathcal{A}_{\rm p.p}$ contains the terms describing the dynamics of a ``point particle''. More precisely, it includes only the couplings between the spin and four-velocity of the representative worldline $z^\mu(\tau)$ and the external field, along with any terms describing the internal dynamics of the compact object without involving fields~\cite{steinhoff2016dynamical-df7, Vines:2018gqi, Vines:2016unv}. The term $\mathcal{A}_{\rm field}$ contains the action of the external field itself, and is therefore not localized on the worldline. Finally, $\mathcal{A}_{\rm tidal}$ encodes the interaction between the induced tidal moments of the compact object and the external field.

\subsection{Scalar case}    
\label{subsec:scalarEFT}
\subsubsection{Tidal fields and induced multipoles}
For scalar perturbations of a Schwarzschild black hole, tidal effects  can be described by setting~\cite{creci2021tidal-42e}, 
\begin{align}
\mathcal{A}_{\rm tidal}=-K_{\rm tide}\int d\tau \sqrt{-u_{\mu}u^{\mu}}\sum_{\ell=0}^{\infty}\frac{1}{\ell!}Q^{I_\ell}\nabla_{I_\ell}\phi~.
\label{eq:scalartides}
\end{align}
Here, $u_{\mu}$ is the tangent to the worldline and $\tau$ the proper time along it. Further $Q^{I_\ell}$, is the $\ell$th multipole moment, and $K_{\rm tide}$ is an overall constant. 

We will work with scattering amplitudes near null infinity, where we can neglect the effect of nonlinear gravitational interactions between external tidal perturbations and the stationary metric sourced by the particle. Thus, we treat the tidal field as metric perturbations over flat spacetime, and work in the particle's rest frame. This implies that $\nabla_{\mu}\to \partial_{\mu}$, and $u^{\mu}=(1,0,0,0)$.

We choose the action $\mc{A}_{\rm field}$ and $\mc{A}_{\rm p.p}$ to reproduce the scalar field Klein-Gordon equation~\cite{creci2021tidal-42e} and the dynamics of a generic spinning object, i.e., the Matthison Pappapetrou Dixon equations, respectively. We further assume that there is no non-tidal or non-gravitational coupling to the scalar field. Then, the variation of the total action in the flat spacetime limit leads to the following equation of motion for the scalar field~\cite{creci2021tidal-42e}
\begin{align}
\label{eq:eomphiEFT1}
\partial_{\mu}\partial^{\mu}\Phi=\frac{K_{\rm tide}}{\sqrt{2\pi}}\sum_{\ell=0}^{\infty}\frac{(-1)^{\ell}}{\ell!}Q^{I_\ell}\partial_{I_\ell}\delta(x^{i})~,
\end{align}
where $K_{\rm tide}$ is a normalization constant.
Thus, away from the worldline, which we take to be at the origin $x^i=0$, and assuming there are no other sources, the scalar field satisfies the flat spacetime Klein-Gordon equation. A general monochromatic solution can be written as
\begin{equation}
\begin{split}
&\phi=\phi_{\rm reg}+\phi_{\rm irreg}\,, 
\\
&\phi_{\rm reg/irreg} = e^{-i\omega t}\sum_{\ell=0}^{\infty}\frac{1}{\omega^{\ell}}C_{\rm reg/irreg}^{I_\ell}\partial_{I_\ell}\psi_{\rm reg/irreg}\,, 
\\
&\psi_{\rm reg} = \frac{\sin(\omega r)}{\omega r},~\psi_{\rm irreg}=\frac{\cos(\omega r)}{\omega r}\,,
\end{split}
\label{eq:regireggsplit}
\end{equation}
where $C^{I_\ell}_{\rm reg,irreg}$ are dimensionless STF tensors. Here, $\phi_{\rm reg}$ is the part of the field that is regular at origin, i.e., satisfies the Klein-Gordon equation everywhere including on the worldline. This may be interpreted as the externally applied field that induces the multipole moments. The response field on the other hand, is the irregular part of the field $\phi_{\rm irreg}$, and it is supported by the induced multipole moments in accordance with \ref{eq:eomphiEFT1}. 

The multipole moments are induced by the tidal fields, $E_{I_\ell}$, which are derived from $\phi_{\rm reg}$ at the worldline as
\begin{align}
\begin{split}
E_{ I_\ell } &= \partial_{ L }\phi_{\rm reg}|_{\vec{x}=0}
\\
& = \mc{N}_{\ell}\omega^{\ell} C_{\rm reg}^{I_\ell}\,,
\label{eq:regfieldcoeff}
\end{split}
\end{align}
where 
\begin{align}
\mc{N}_\ell = \frac{2^\ell (-1)^\ell \ell!\ell!}{(2\ell+1)!}\,,
\end{align}
as shown in \ref{app:scalartidescomp}. We can write down an ansatz relating the scalar tidal fields to the induced multipole moments, consistent with the symmetries of the system   as\footnote{Alternatively, the dynamics of the induced multipole moments may be fixed via additional terms in $\mc{A}_{\rm p.p}$, such as a harmonic-oscillator-like action for the multipole moments. However, it is difficult to incorporate dissipative tidal effects in this manner without the use of the in-in formalism.}
\begin{align}
\nonumber
Q^{I_{\ell}}(\omega)&= M^{2\ell+1} F^{I_\ell}{}_{I'_\ell}E^{I'_{\ell}}(\omega) 
\nonumber
\\
& +\chi^2 \Big[M^{2\ell-1} F^{(-2)I_\ell}{}_{I'_\ell} E^{\langle I_{\ell-2}}(\omega) \hat{s}^{i'_{\ell-1}}\hat{s}^{i'_{\ell}\rangle}
\nonumber
\\
&+M^{2\ell+3} F^{(2)I_\ell}{}_{ I'_\ell} E^{I'_{\ell+2}}(\omega)\hat{s}_{i'_{\ell+1}}\hat{s}_{i'_{\ell+2}}\Big]  
\nonumber 
\\
& + \chi^4\left[\cdots\right]+\cdots\,,
\label{eq:genexpand}
\end{align}
where $\chi$ is the spin parameter of the Kerr BH, and $\hat{s}^i$ is the unit-vector oriented along the spatial spin direction.

Note that $Q^{I_\ell}$ can be induced by tidal fields at multipolar orders $\ell,\ell\pm 2,\dots$. Parity invariance forbids induction by terms at relative-odd multipolar orders ($\ell\pm1,\dots$). The first term, containing $F^{I_\ell}_{I'_\ell}$, parametrizes the diagonal tidal response and is the only term which persists in the absence of spin. We refer to the remaining tidal response tensors as mixing tidal-response tensors. It is important to note that \ref{eq:genexpand} is {\it not} a Taylor expansion in $\chi$, and each tidal response tensor $F^{(n),I_\ell}{}_{I'_\ell}$ ($F^{(0),I_\ell}{}_{I'_\ell}=F^{I_\ell}{}_{I'_\ell}$), can contain arbitrary dependence on spin-parameter $\chi$ and perturbation frequency $\omega$.

\subsubsection{Structure of the response functions}
To proceed, we need to further specify the structure of each $F^{(n)I_\ell}{}_{I'_\ell}$. This can be done by writing all possible STF combinations involving the spin vector, as done for gravitational perturbations at $\ell=2$ in Refs.~\cite{Goldberger:2005cd, Saketh:2022xjb}. However, it is simpler to switch to the basis of scalar spherical harmonics instead, where the axisymmetry about the spin vector is manifest. Following~\cite{RevModPhys.52.299}, for any STF tensor $M^{I_\ell}$ (say $M=Q,C$), we can write
\begin{align}
M^{\ell m} &= \int d\Omega M^{I_\ell} N_{I_\ell} \bar{Y}^{\ell m} \,,
\label{eq:ellemmdef}
\end{align}
where bar denotes complex conjugation,  $N_{I_\ell} = \hat{n}_{\langle i_1}\hat{n}_{i_2}\dots\hat{n}_{i_\ell\rangle}$ is a multilinear of unit vectors, and $(\theta,\phi)$ represent the angles on a unit sphere pointed to by $\hat{n}$. We choose the $z$-axis to be parallel to the spin vector $\hat{s}^i$. Additionally, defining a suitable basis set of STF tensors through the relation
\begin{align}
Y_{\ell m}(\theta,\phi) = \mc{Y}^{I_{\ell}}_{\ell m}N_{I_{\ell}}\,,
\end{align}
we can rewrite \ref{eq:ellemmdef}, and its inverse as
\begin{align}
\begin{split}
M^{\ell m} &= 4 \pi \frac{l!}{(2l+1)!!}M^{I_{\ell}}\bar{\mc{Y}}^{\ell m}_{I_\ell},\\ M^{I_\ell} &= \sum_{m=-\ell}^\ell M^{\ell m}\mc{Y}_{\ell m}^{I_\ell}\,.
\end{split}
\end{align}
The up/down position of the $\ell m$ labels has no physical meaning and no summation convention applies to them.

To obtain the tidal response in the $\ell m$ basis, consider projecting out the action of a tidal response tensor on another STF tensor ($F^{(n)I_\ell}{}_{I_\ell'}$) as
\begin{align}
\begin{split}
& 4\pi \frac{\ell !}{(2\ell+1)!!}F^{(n)I_{\ell}}{}_{I'_{\ell}}E^{I'_{\ell}} \bar{\mc{Y}}_{I_\ell}^{\ell m} \\
&= 4\pi \frac{\ell !}{(2\ell+1)!!}F^{(n)I_{\ell}}{}_{I'_{\ell}} \sum_{m'=-\ell}^\ell E^{\ell m'}\mc{Y}^{I'_\ell}_{\ell m'} \bar{\mc{Y}}_{I_\ell}^{\ell m} 
\\
&= \sum_{m'=-\ell}^\ell  F^{(n)\ell m}{}_{\ell m'}E^{\ell m'}\,,
\end{split}
\label{eq:tensorspher}
\end{align}
where 
\begin{align}
F^{(n)\ell m}{}_{\ell m'}=4\pi \frac{\ell !}{(2\ell+1)!!}F^{(n)I_{\ell}}{}_{I'_{\ell}}\mc{Y}^{I'_{\ell}}_{\ell m'}\bar{\mc{Y}}_{I_\ell}^{\ell m}\,.
\end{align}
The tidal response tensors $F^{(n)I_\ell}{}_{I'_\ell}$ linearly map STF tensors (of rank $\ell$) to STF tensors of the same rank. Thus, their action upon the tidal fields is akin to a rotation (and a scaling), where different $m-$ modes may be mixed. However, axisymmetry prevents the mixing of different $m$ modes. Thus,
\begin{align}
F^{(n)\ell m}{}_{\ell m'} = F^{(n)}_{\ell m}\delta^m_{m'}\,.
\end{align}
Then, the last line of \ref{eq:tensorspher} is simply $F_{\ell m}^{(n)}E^{\ell m}$. The coefficients $F_{\ell m}^{(n)}\equiv F_{\ell m}^{(n)}(\epsilon,\chi)$
may be regarded as tidal response in the $\ell m$ basis, with 
\begin{equation}
\label{eq:epsilondef}\epsilon=M\omega.
\end{equation}

Finally, we need to treat STF tensors involving direct products or contractions with the spin-vector which also feature in \ref{eq:genexpand}. Consider for instance the projection onto $\ell,m$ basis of the following tensor product involving unit spin vectors: 
\begin{align}
\label{eq:productsSTFtolm}
\begin{split}
& 4\pi \frac{\ell !}{(2\ell+1)!!}C^{\langle I_{\ell-k}}\hat{s}^{i_{\ell-k+1}}\dots\hat{s}^{i_{\ell}\rangle} \bar{\mc{Y}}^{\ell m}_{I_\ell} 
\\
&=  C^{ I_{\ell-k}}\hat{s}^{i_{\ell-k+1}}\dots\hat{s}^{i_{\ell}} \bar{\mc{Y}}^{\ell m}_{I_\ell} 
\\
& =   4\pi \frac{\ell !}{(2\ell+1)!!}\sum_{m'} C^{\ell-k, m'} \mc{Y}^{I_{\ell-k}}_{\ell-k, m'}\hat{s}^{i_{\ell-k+1}}\dots\hat{s}^{i_{\ell}} \bar{\mc{Y}}^{\ell m}_{I_\ell}\,.
\end{split}
\end{align}
Since we have oriented the spin-axis to be the $\hat{z}$ direction, the spin-vector has $\ell=1,~m=0$, and we can thus write
\begin{align}
\label{eq:Zdef}
Z_{\ell}^{k} \delta^m_{m'}= 4\pi \frac{\ell !}{(2\ell+1)!!}\mc{Y}^{I_{\ell-k}}_{\ell-k,m'}\hat{s}^{\langle i_{\ell-k+1}}\dots\hat{s}^{i_{\ell}\rangle} \bar{\mc{Y}}^{\ell m}_{I_\ell}\,.
\end{align}
Using this, we can write~\ref{eq:productsSTFtolm} as 
\begin{align}
\begin{split}
& 4\pi \frac{\ell !}{(2\ell+1)!!}C^{\langle I_{\ell-k}}\hat{s}^{i_{\ell-k+1}}\dots\hat{s}^{i_{\ell}\rangle} \bar{\mc{Y}}^{\ell m}_{I_\ell} =  C^{\ell-k, m}Z_{\ell}^k\,.
\end{split}
\end{align}
Now, consider the contraction
\begin{align}
\begin{split}
& \frac{ 4\pi\ell !}{(2\ell+1)!!}C^{I_{\ell+k}}\hat{s}_{i_{\ell+k}}\dots\hat{s}_{i_{\ell+1}} \bar{\mc{Y}}^{\ell m}_{I_\ell} 
\\
& =   \frac{ 4\pi\ell !}{(2\ell+1)!!}\sum_{m'=-\ell-k}^{\ell+k}
C^{\ell+k,m}\mc{Y}^{I_{\ell+k}}_{\ell+k,m'}\hat{s}_{i_{\ell+k}}\dots\hat{s}_{i_{\ell+1}}\bar{\mc{Y}}^{\ell m}_{I_{\ell}}\,.
\end{split}
\end{align}
Once again, the accompanying product must vanish when $m\neq m'$, and we can thus write
\begin{align}
\begin{split}      
\frac{ 4\pi\ell !}{(2\ell+1)!!}C^{I_{\ell+k}}\hat{s}_{i_{\ell+k}}\dots\hat{s}_{i_{\ell+1}} \bar{\mc{Y}}^{\ell m}_{I_\ell}  = C^{\ell+k,m}\bar{Z}^k_{\ell+k}\,.
\end{split}
\label{eq:Zbdef}
\end{align}
Now, we can finally rewrite \ref{eq:genexpand} in $\ell,m$ basis as
\begin{align}
\begin{split}
M^{-\ell} Q^{\ell m}&=M\epsilon^\ell \mc{N}_{\ell} F_{\ell m} C_{\rm reg}^{\ell m} + M \chi^2 \Big[Z^2_{\ell} \mc{N}_{\ell-2}\epsilon^{\ell-2} F^{(-2)}_{\ell m} C_{\rm reg}^{\ell-2, m} 
\\
&+\bar{Z}^2_{\ell+2} \mc{N}_{\ell+2}\epsilon^{\ell+2}  F^{(2)}_{\ell m} C_{\rm reg}^{\ell+2, m}\Big]+M \chi^4\left[\cdots\right]+\cdots\,, 
\\ &=M \sum_{\ell'=0}^\infty \mc{Z}^{(\ell-\ell')}_{\ell'}\mc{N}_{\ell'}\chi^{|\ell-\ell'|}\epsilon^{\ell'}F^{(\ell'-\ell)}_{\ell m}C^{\ell' m}_{\rm reg}\,,
\end{split}
\label{eq:Qexpandsc}
\end{align}
where we have defined 
\begin{equation}\label{eq:calZdef}\mc{Z}^{0}_{\ell'}=1, \; \; \; \mc{Z}^{k>0}_{\ell'}= Z^k_{\ell'+k}, \;\; \; \mc{Z}^{k<0}_{\ell'} = \bar{Z}^{|k|}_{\ell'},
\end{equation}
where $Z,~\bar{Z}$ were defined in~\ref{eq:Zdef} and \ref{eq:Zbdef} respectively.

\subsubsection{Relation to amplitudes of scattering states}
Now, as mentioned, the multipole moments source $\phi_{\rm irreg}$ in accordance with \ref{eq:eomphiEFT1}. Substituting the expression for the irregular field from \ref{eq:regireggsplit} into \ref{eq:eomphiEFT1}, we obtain 
\begin{align}
\frac{4 \pi}{\omega^{\ell+1}}C^{I_\ell}_{\rm irreg}\partial_{I_\ell} \delta^{(3)}(x^i)= \frac{K_{\rm tide}}{\sqrt{2\pi}} \frac{(-1)^\ell}{\ell!}Q^{I_\ell}\partial_{I_\ell} \delta^{(3)}(x^i)\,,
\end{align}
implying $C^{I_\ell}_{\rm irreg} = 2\mc{M}_{\ell} \omega^{\ell+1} Q^{I_\ell}$,
where we have defined 
\begin{equation}
\mc{M}_{\ell} = (-1)^{\ell+1}(K_{\rm tide})/[4(2\pi)^{3/2}\ell!].
\end{equation}
Thus, we can rewrite \ref{eq:Qexpandsc} as
\begin{align}
\begin{split}
C_{\rm irreg}^{\ell m} &=  \mc{M}_{\ell} \sum_{\ell'=0}^\infty \mc{Z}_{\ell'}^{\ell-\ell'} \mc{N}_{\ell'}\chi^{|\ell-\ell'|}\epsilon^{\ell+\ell'+1}F^{(\ell'-\ell)}_{\ell m}C^{\ell' m}_{\rm reg}\,.
\end{split}
\end{align}
Now, the right hand side only has applied tidal field coefficients, which we can freely tune in principle. Let us now manually choose to set $C^{\ell m}_{\rm reg}\neq 0$ iff $\ell=\ell_0$ for some desired $\ell_0$. Then, we get 
\begin{align}
C_{\rm irreg}^{\ell m} =\mc{M}_{\ell} \mc{Z}_{\ell_0}^{\ell-\ell_0} \mc{N}_{\ell_0}\chi^{|\ell-\ell_0|}\epsilon^{\ell+\ell_0+1}F^{(\ell_0-\ell)}_{\ell m}C^{\ell_0 m}_{\rm reg}\,.
\label{eq:onemodetoallprv}
\end{align}
Furthermore, it is convenient to define a rescaled response as
\begin{equation}
    f_{\ell_0 m}^{(\ell-\ell_0)}=\mc{M}_{\ell} \mc{Z}_{\ell_0}^{\ell-\ell_0} \mc{N}_{\ell_0}F^{(\ell_0-\ell)}_{\ell m}.
    \label{eq:rescaledsc}
\end{equation}
    This can be considered a convenient choice of normalization in the tidal response. We thus have 
\begin{align}
C_{\rm irreg}^{\ell m}& =\chi^{|\ell-\ell_0|} \epsilon^{\ell+\ell_0+1}f_{\ell_0 m}^{(\ell-\ell_0)}C^{\ell_0 m}_{\rm reg}\,.
\label{eq:onemodetoall}
\end{align}
In this work, we will be primarily interested in computing the diagonal tidal response, i.e., $f_{\ell_0 m}=f^{(0)}_{\ell_0 m}$. In \ref{eq:onemodetoall}, setting $\ell_0=\ell$ yields
\begin{alignat}{3}
C^{\ell m}_{\rm irreg} = \epsilon^{2\ell+1} f_{\ell m} C^{\ell m}_{\rm reg}\,.
\end{alignat}
However, this relation is not as straightforward to fix via perturbation theory as it seems.  This is because the perturbation equations in the full theory (linear perturbation theory of BHs) are separable in the spheroidal harmonic basis. The full theory, which will be discussed in~\ref{scalar_scattering} a priori yields a relation of the form
\begin{align}
D^{\ell m}_{\rm irreg} = \epsilon^{2l+1}\lambda_{\ell m} D^{\ell m}_{\rm reg}\,,
\label{eq:pert_response}
\end{align}
where $D$s are coefficients in the spheroidal basis, and $\ell,m$ label the spheroidal modes. The computation of the $\lambda_{\ell m}$ coefficients will be discussed in~\ref{scalar_scattering}. We can relate the two sets of coefficients as follows. The the scalar spheroidal and spherical harmonics are related by 
\begin{equation}
\label{eq:scalarspheroidaltosphericalharmonics}
\begin{split}
Y_{\ell m}   &= S_{\ell m} + \sum_{k=1}^{[\ell/2]} (\chi \epsilon)^{2k} \tilde{r}^{\ell m}_{-2k}S_{\ell-2k,m}
\\
&+\sum_{k=1}^{\infty} (\chi \epsilon)^{2k} \tilde{r}^{\ell m}_{2k}S_{\ell+2k,m}\,,
\end{split}
\end{equation}
where $\tilde{r}^{\ell m}_{2k}$ are the associated Clebsch-Gordon like coefficients. This perturbation expansion can be generated, for instance, using the BHPT toolkit~\cite{BHPToolkit} in Mathematica. We can use this, along with the asymptotic radial solution, to relate the sets of coefficients as
\begin{align}
\begin{split}
D^{\ell m}  &= C^{\ell m} + \sum_{k=1}^{[\ell/2]} (\chi \epsilon)^{2k} r^{\ell m}_{-2k}C^{\ell-2k,m}
\\
& +\sum_{k=1}^{\infty} (\chi \epsilon)^{2k} r^{\ell m}_{2k}C^{\ell+2k,m}\,.
\end{split}
\label{eq:spherexps0}
\end{align}
The relationship between $r_{2k}^{\ell m}$ and $\tilde{r}_{2k}^{\ell m}$ can be determined by imposing that the full solution be identical in spherical and spheroidal basis. We show this explicitly in \ref{ref:asymp_EFT}.

\ref{eq:spherexps0} holds for both regular and irregular sets of coefficients. As mentioned before, there is freedom to choose the regular fields in spherical basis. In particular, we set that only $C^{\ell_0 m}_{\rm reg}$ is non-zero. Then, we have 
\begin{align}
\begin{split}
D_{\rm reg}^{\ell m } = (\chi \epsilon)^{|l-l_0|} r^{\ell m}_{l_0-l}C_{\rm reg}^{\ell_0m}\,,
\end{split}
\label{eq:sphersphreg}
\end{align}
where we have defined $r^{\ell m}_0=1$.

Now substituting \ref{eq:spherexps0} and \ref{eq:sphersphreg}  into \ref{eq:pert_response}, using the spherical-basis induction equation(s) in \ref{eq:onemodetoall}, and imposing $\ell_{0}\geq \ell$ we get constraint equations as
\begin{align}
\begin{split}
&\left[f_{\ell_0 m}^{(l-l_0)} + \sum_{k=1}^{[l/2]}\chi^{4k}f^{(\ell-2k-\ell_0)}_{\ell_0,m}r^{\ell m}_{-2k} \right] 
\\
&+ \mathcal{O}(\epsilon^{4}) = \lambda_{\ell m}r^{\ell m}_{\ell_0-\ell}, \quad \ell_0\geq \ell\,.
\end{split}
\label{eq:recursor}
\end{align}
These constraint equations can be solved iteratively. To start, we set $\ell=\ell_0=0$ or $\ell=\ell_0=1$, where we simply get
\begin{align}
f_{\ell m} = \lambda_{\ell m}\,,~\ell=0,1\,.
\end{align}
For $\ell\geq 2$, we describe the relation between BHPT reponse and the EFT response below.

\subsubsection{Iterative fixing of EFT response coefficients}
    
For say $\ell_0=2$ and $\ell=0$, we have the relation
\begin{align}
f_{2m}^{(-2)} = \lambda_{0m} r^{0m}_{2}\,. 
\end{align}
Now, keeping $\ell=\ell_0=2$, we get
\begin{align}
\begin{split}
& f_{2m}+\chi^4 r^{2m}_{-2}f^{(-2)}_{2m} = \lambda_{2m}\,.
\end{split}
\end{align}
Using the result that $f^{(-2)}_{2m}=\lambda_{0m}r^{0m}_{2}$, we finally obtain,
\begin{align}
\begin{split}
f_{2m} = \lambda_{2m} -\chi^4 r_{-2}^{2m} r_2^{0m} \lambda_{0m}\,.
\end{split}
\label{eq:l2resscalar}
\end{align}
Thus, we could fix the EFT response coefficient $f_{2m}$. We can similarly fix $f_{3m}$ as
\begin{align}
f_{3m} = \lambda_{3m} - \chi^4 r_{-2}^{3m}\lambda_{1m}r_2^{1m}\,.\label{eq:l3resscalar}
\end{align}
Now, consider $\ell_0=4$ and $\ell=2,0$, for which we have the relations
\begin{equation}
\begin{split}
\chi^2 f_{4 m}^{(-2)} + \chi^6 f^{(-4)}_{4m}r_{-2}^{2m} &= \lambda_{2m}\chi^{2}r^{2m}_{2},\text{ at}~\ell=2\,,
\\
\chi^4 f^{(-4)}_{4m} &= \lambda_{0m}\chi^{4}r_{4}^{0m},\text{ at}~\ell=0\,.
\end{split}
\end{equation}
We can solve them to get
\begin{align}
&f_{4m}^{(-4)}=\lambda_{0m}r_4^{0m}\,,
\\
&f_{4m}^{(-2)}=\lambda_{2m}r_2^{2m} - \chi^4 \lambda_{0m} r_4^{0m}r_{-2}^{2m}\,.
\end{align}
Now, for $\ell_0=\ell=4$, we obtain,
\begin{align}
\begin{split}
f_{4m} + \chi^4 r^{4m}_{-2}f^{(-2)}_{4m}+\chi^8r^{4m}_{-4}f^{(-4)}_{4m} = \lambda_{4m}\,,
\end{split}
\end{align}
which using the previous results connecting $f_{4m}$ with $\lambda_{0m},\lambda_{2m}$, yields,
\begin{align}
\begin{split}
f_{4m}&=\lambda_{4m} - \chi^4 r_{-2}^{4m}\lambda_{2m}r_2^{2m} + \chi^8 r_{-2}^{4m} \lambda_{0m}r_{4}^{0m}r_{-2}^{2m} 
\\
& -\chi^8 \lambda_{0m}r_{-4}^{4m}r_4^{0m}\,.
\end{split}
\end{align}
We can similarly proceed to obtain relation between EFT response $f_{\ell m}$ and BHPT response $\lambda_{\ell m}$ to all orders in spin and the angular number $\ell$.

Once the coeffcients $f_{\ell m}^{(0)}$ are fixed, we can obtain the EFT coefficients in $\ell m$ basis trivially using the definition in \ref{eq:rescaledsc} as
\begin{equation}
    F_{\ell m}^{(0)} = \frac{f_{\ell m}^{(0)}}{M_\ell N_{\ell}} = -\frac{2^{\frac{9}{2}+\ell}\pi \Gamma(\frac{3}{2}+\ell)f_{\ell m}}{K_{\rm tide}}
\end{equation}
\subsubsection{Asymptotic form of the EFT scalar field}
\label{ref:asymp_EFT}

Before turning to Gravitational case, it is useful to use the relations between tensor and spherical harmonic basis to rewrite the asymptotic field, obtained by taking the limit $\omega r\gg 1$ of \ref{eq:regireggsplit} as,
\begin{equation}
\begin{split}
\phi &= \frac{e^{-i\omega t}}{\omega r}\sum_{\ell=0}^{\infty}\frac{1}{\omega^{\ell}}\left[C_{\rm reg}^{I_\ell}\partial_{I_\ell}\sin(\omega r)+C_{\rm irreg}^{I_\ell}\partial_{I_\ell}\cos(\omega r)\right]
\\
&= \frac{e^{-i\omega t}}{\omega r}\sum_{\ell=0}^{\infty}(-1)^{\ell}\Big[N_{ I_{\ell}}\Big(C_{\rm reg}^{I_\ell}\sin\left(\omega r-\ell \frac{\pi}{2}\right)
\\
&+C_{\rm irreg}^{I_\ell}\cos\left(\omega r-\ell \frac{\pi}{2}\right)\Big)\Big]+\mc{O}\left(\frac{1}{r^2}\right)\,.
\end{split}
\end{equation}
Now, we use
\begin{equation}
C^{I_\ell}_{\rm reg/irreg}N_{I_\ell } = \sum_{m=-\ell}^\ell C^{\ell m}_{\rm reg/irreg} Y_{\ell m}(\theta,\phi)\,,
\end{equation}
to rewrite the asymptotic form of the scalar field as
\begin{equation}
\begin{split}
\phi &=  \frac{e^{-i\omega t}}{\omega r}\sum_{\ell=0}^{\infty}\sum_{m=-\ell}^\ell Y_{\ell m}(\theta,\phi)(-1)^{\ell}\Big\{C_{\rm reg}^{\ell m}\sin\left(\omega r-\ell \frac{\pi}{2}\right)
\\
&+C_{\rm irreg}^{\ell m}\cos\left(\omega r-\ell \frac{\pi}{2}\right)\Big\}+\mc{O}\left(\frac{1}{\omega^2 r^2}\right)\,.
\\& = \frac{e^{-i\omega t}}{\omega r}\sum_{\ell=0}^\infty \sum_{m=-\ell}^\ell R_{\ell m}^{\rm EFT} Y_{\ell m}(\theta,\phi) +\mc{O}\left(\frac{1}{\omega^2 r^2}\right)\,.
\end{split}
\label{asymptotic-EFT}
\end{equation}
This is the asymptotic results for the scalar field obtained from EFT. Finally, using \ref{eq:scalarspheroidaltosphericalharmonics}, we can rewrite this in spheroidal basis as 
\begin{equation}
\begin{split}
          \phi & \approx \frac{e^{-i\omega t}}{\omega r} \sum_{\ell,m} S_{\ell m}(\theta,\phi)\Big \{ R^{\rm EFT}_{\ell m}
+\sum_{k=1}^{[l/2]}(\chi \epsilon)^{2k} \tilde{r}^{\ell-2k, m}_{2k} R^{\rm EFT}_{\ell-2k,m}
\\& + \sum_{k=1}^\infty (\chi \epsilon)^{2k} \tilde{r}^{\ell+2k,m}_{-2k}R^{\rm EFT}_{\ell+2k,m}\Big\}\,.
\\& =  \frac{e^{-i\omega t}}{\omega r}\sum_{\ell=0}^{\infty}\sum_{m=-\ell}^\ell S_{\ell m}(\theta,\phi)(-1)^{\ell}\Big\{D_{\rm reg}^{\ell m}\sin\left(\omega r-\ell \frac{\pi}{2}\right)
\\&+C_{\rm irreg}^{\ell m}\cos\left(\omega r-\ell \frac{\pi}{2}\right)\Big\}+\mc{O}\left(\frac{1}{\omega^2 r^2}\right)\,.
\end{split}
\label{rvsplit}
\end{equation}
This yields 
\begin{equation}
    \begin{split}
        D^{\ell m} &= C^{\ell m}_{\rm reg} +\sum_{k=1}^{[l/2]}(i \chi \epsilon)^{2k} \tilde{r}^{\ell-2k, m}_{2k} C^{\ell-2k,m}\\& + \sum_{k=1}^\infty (i \chi \epsilon)^{2k} \tilde{r}^{\ell+2k,m}_{-2k}C^{\ell+2k,m}\,.
    \end{split}
    \label{eq:asymptotic_EFT_sph}
\end{equation}
Comparing with \ref{eq:spherexps0}, we obtain the relations $\tilde{r}^{\ell-2k, m}_{2 k} = (-1)^{k}r^{\ell m}_{-2k}$, $\tilde{r}^{\ell+2k, m}_{-2 k} = (-1)^{k}r^{\ell m}_{2k}$. We can then explicitly write the relationship between the tidal response coefficients in spheroidal and spherical basis. For instance, we can rewrite ~\ref{eq:l2resscalar} as 
\begin{equation}
\begin{split}
        f_{2m} &= \lambda_{2m} -\chi^4 r_{-2}^{2m} r_2^{0m} \lambda_{0m}\,,
        \\& =  \lambda_{2m} -\chi^4 \tilde{r}_{2}^{0m} \tilde{r}_{-2}^{2m} \lambda_{0m}\,,
        \\& = \lambda_{2m} -\chi^4 \frac{(1-m^2)(4-m^2)}{1620} \lambda_{0m}\,,
\end{split}
\end{equation}
Similarly, we can fix \ref{eq:l3resscalar} as
\begin{equation}
\begin{split}
        f_{3m} &= \lambda_{3m} -\chi^4 r_{-2}^{3m} r_2^{1m} \lambda_{1m}\,,
        \\& =  \lambda_{3m} -\chi^4 \tilde{r}_{2}^{1m} \tilde{r}_{-2}^{3m} \lambda_{1m}\,,
        \\& = \lambda_{3m} -\chi^4 \frac{\left(4-m^2\right) \left(9-m^2\right)}{52500} \lambda_{1m}\,,
\end{split}
\end{equation}
and similarly proceed further for higher values of $\ell$.

The above shows how to obtain the EFT tidal response coefficients given the ratio of $D^{\ell m}_{\rm irreg}/D^{\ell m}_{\rm reg}$. To compute the ratio, we need expressions for $D^{\ell m}_{\rm reg/irreg}$. These are obtained by matching the asymptotic EFT solution~\ref{eq:asymptotic_EFT_sph} with the corresponding far-zone solution in perturbation theory. We work this out in detail in~\ref{subsec:EFTscalar}.

\subsection{Gravitational case}
\label{sec:eft_grav_disc}
In the previous subsection, we showed how scalar tidal responses can be matched between EFT and perturbation theory. We now review how that procedure extends to generic ($s\neq 0$) perturbations. The extension is relatively straightforward, up to two main differences. Firstly, there are now couplings between neighbouring modes as well since there are two parities (electric and magnetic) for a single $\ell m$ mode. Additionally, the tidal responses of electric and magnetic parts of the perturbation need to be related in some way. In particular, we will need to set them equal to each other. This is because there is only a single second order perturbation equation governing a complex curvature scalar in the perturbation theory, with no explicit parity dependence. Below, we discuss the procedure in detail for $s=-2$, i.e., gravitational perturbations. The same steps apply for $s=-1$ but we do not give them explicitly.

\subsubsection{EFT tidal fields and induced multipole moments}
The tidal/finite size action in \ref{eq:scalartides} can be generalized for the case of gravitational perturbation as
\begin{align}
\begin{split}
\mathcal{A}_{\rm tidal}&=-\int d\tau \sqrt{-u_{\mu}u^{\mu}}\sum_{\ell=0}^{\infty}\frac{1}{\ell!}Q_E^{I_\ell}E_{I_\ell}\\&-\int d\tau \sqrt{-u_{\mu}u^{\mu}}\sum_{\ell=0}^{\infty}\frac{1}{\ell!}Q_B^{I_\ell}B_{I_\ell}\,.    
\end{split}
\label{eq:scalartides}
\end{align}
where the gravito-electric and -magnetic tidal fields are related to the Riemann curvature tensor by
\begin{align}
\begin{split}
E_{\mu_L} &= \nabla_{\langle \mu_{L-2}}R_{\mu_{\ell-1}|\alpha|\mu_{\ell}\rangle \beta}u^\alpha u^\beta\,,
\\ 
B_{\mu_L}&=\frac{1}{2}\nabla_{\langle\mu_{L-2}}\epsilon_{|\gamma|\mu_{\ell-1}}{}^{\alpha\beta}R_{|\alpha\beta|\mu_{\ell}\rangle \delta}u^\gamma u^\delta,~\ell\geq 2\,, 
\end{split}
\label{eq:EBtiddef}
\end{align}
are the electric and magnetic tidal fields. As before, the tidal fields induce the multipole moments, and the induction relation for the scalar case, see \ref{eq:genexpand}, is modified for the gravitational case as,
\begin{align}
\begin{split}
Q_E^{I_{\ell}}(\omega)&= M^{2\ell+1} F^{I_\ell}{}_{I'_\ell}E^{I'_{\ell}}(\omega) 
\\
&+ \chi\Big[ M^{2\ell} F^{(-1)I_\ell}{}_{I'_\ell} B^{\langle I_{\ell-1}}(\omega) \hat{s}^{i'_{\ell}\rangle}
\\
&+ M^{2\ell+2} F^{(1)I_\ell}{}_{ I'_\ell} B^{I'_{\ell+1}}(\omega)\hat{s}_{i'_{\ell+1}}\Big]  
\\
& + \chi^2\left[\cdots\right]+\cdots\,,
\\
Q_B^{I_{\ell}}(\omega)&= M^{2\ell+1} F^{I_\ell}{}_{I'_\ell}B^{I'_{\ell}}(\omega) 
\\
& - \chi \Big[ M^{2\ell} F^{(-1)I_\ell}{}_{I'_\ell} E^{\langle I_{\ell-1}}(\omega) \hat{s}^{i'_{\ell}\rangle}
\\
&+ M^{2\ell+2} F^{(1)I_\ell}{}_{ I'_\ell} E^{I'_{\ell+1}}(\omega)\hat{s}_{i'_{\ell+1}}\Big]  \\
&+\chi^2\left[\cdots\right]+\cdots\,.
\end{split}
\label{eq:genexpandgens}
\end{align}
We note that coupling between multipolar orders starts at linear order in spin, between neighbouring modes. This is allowed since the gravito-electric (-magnetic) fields transform as $(-1)^\ell$ ($(-1)^{\ell+1}$) under parity. It is also worth noting that the tidal response tensors are not labelled as $E$ or $B$. This is because the linear tidal response is expected to be identical (with proper normalizations and signs) for both parities for Kerr BHs~\cite{Goldberger:2005cd, Saketh:2022xjb, saketh2023dynamical-b30, Bhatt:2024rpx}. This can be understood as a symmetry under $E^{I_\ell}\rightarrow i B^{I_\ell} $, $B^{I_\ell}\rightarrow -i E^{I_\ell}$. 

\subsubsection{Relating response functions to asymptotic wave amplitudes}
Now, the metric perturbations asymptotically far away from the object are given (for monochromatic perturbations) in terms of $h^{\mu\nu}\equiv\sqrt{-g}g^{\mu\nu}-\eta^{\mu\nu}$, where
\begin{align}
\begin{split}
h_{ij} &= \sum_{\ell=2}^\infty \frac{1}{\omega^{\ell-2}}\Bigg(C^{K_\ell}_{\mce,\rm reg}\Pi^{k_{\ell-1}k_{\ell}}_{ij}\partial_{K_{\ell-2}}\\&+\omega^{-1}C_{\mcb}^{K_{\ell}}\Pi^{k_\ell m}_{ij}\epsilon_{k_{\ell-1}mn}\partial_{K_{\ell-2}n} \Bigg)\psi_{\rm reg}+(\rm reg\leftrightarrow \rm irreg)\,,
\end{split}
\label{eq:metric}
\end{align}
where, the regular part is $\psi_{\rm reg}=e^{-i\omega t}\sin(\omega r)/\omega r$, and the irregular part is $\psi_{\rm irreg}=e^{-i\omega t}\cos(\omega r)/\omega r$. The projector is given by $\Pi^{ij}_{kl}=(1/2)(P^i{}_kP^j{}_l+P^j{}_kP^i{}_l-P^{ij}P_{kl})$, with $P^{ij}=\delta^{ij}+\partial^i\partial^j/\omega^2$.

The tidal fields at the position of the BH, which we choose to be the origin ($x_{i}=0$), are obtained by evaluating \ref{eq:EBtiddef} at the origin. This yields,
\begin{align}
E^{I_\ell}&=(-1)^{\ell+1} \mc{N}_{\ell} \omega^\ell C_{\mce,\rm reg}^{I_\ell}e^{-i\omega t}\,, 
\\ 
B^{I_\ell} &= i(-1)^{\ell} \mc{N}_{\ell}\omega^{\ell} C_{\mcb,\rm reg}^{I_\ell}e^{-i\omega t}\,,
\end{align}
where $\mc{N}_\ell=2^{-\ell-3}(1+\ell)(2+\ell)\sqrt{\pi}/\Gamma(\frac{3}{2}+\ell)$, as shown in~\ref{app:gravity}.

The tidal fields induce multipole moments via \ref{eq:genexpandgens}, which in turn connects to the irregular coefficients $C^{L}_{\rm \mce/\mcb,\rm irreg}$ via the equations of motion similar to the scalar case. Working with linearized gravity, the equations of motion are given by~\cite{Saketh:2022xjb}
\begin{align}
\partial_\mu \partial^\mu h_{ij} = 16 \pi G |g| \Pi^{kl}_{ij}T_{kl}.
\end{align}
The stress energy tensor can be obtained from the action by varying the metric perturbations as shown in Ref.~\cite{Saketh:2022xjb}. This yields relations between the multipole moments and the coefficients parametrizing the irregular part of the metric perturbation as
\begin{align}
e^{-i\omega t}C_{\mce,\rm irreg}^{L}&=4 (\ell!)^{-1}(-1)^{\ell}\omega^{\ell+1}Q^L_E\,, 
\\ 
e^{-i\omega t}C_{\mcb, \rm irreg}^{L} &= i 4 (\ell!)^{-1}(-1)^{\ell}\omega^{\ell+1}Q^L_B\,.
\end{align}
Thus, the relation \ref{eq:genexpandgens} can be rewritten to relate regular and irregular components as 
\begin{align}
\begin{split}
C_{\mce,\rm irreg}^{I_{\ell}}&=
\frac{-4}{l!}\mc{N}_\ell  \epsilon^{2\ell+1} F^{I_\ell}{}_{I'_\ell}C^{I'_{\ell}}_{\rm \mce, reg} 
\\
& 
-\frac{4}{l!}i
\chi \Big[ \mc{N}_{\ell-1}\epsilon^{2\ell} F^{(-1)I_\ell}{}_{I'_\ell} C^{\langle I'_{\ell-1}}_{\mcb, \rm reg}\hat{s}^{i'_{\ell}\rangle}
\\
&+  \mc{N}_{\ell+1} \epsilon^{2\ell+2} F^{(1)I_\ell}{}_{ I'_\ell} C^{I'_{\ell+1}}_{\mcb, \rm reg}\hat{s}_{i'_{\ell+1}}\Big]+\mathcal{O}(\chi^{2})\,,
\end{split}
\label{eq:genexpandgenscoeff}
\end{align}
and an identical relation for $\mce\leftrightarrow\mcb$.
Now, it would be convenient to combine the above relations into a single one, suitable for comparison against the relevant quantity governed by the Teukolsky equation. The Teukolsky equation governs the Teukolsky scalar $\psi_4$, which admits a similar decomposition into regular and irregular parts. The corresponding coefficients $C_{\rm reg,irreg}$ that enters $\psi_4$, in the spherical basis, are related to the coefficients in the metric perturbation as $C_{\rm reg,irreg}=C_{\mce,\rm reg/irreg}+C_{\mcb, \rm reg/irreg}$ for all $(\ell m)$ modes~\cite{Saketh:2022xjb}. Thus, we add the electric and magnetic parts of \ref{eq:genexpandgens} to get
\begin{align}
\begin{split}
C_{\rm irreg}^{I_{\ell}}&=  \frac{-4}{l!}\mc{N}_\ell  \epsilon^{2\ell+1} F^{I_\ell}{}_{I'_\ell}C^{I'_{\ell}}_{\rm  reg} 
\\
& - \frac{4}{l!}i\chi \Big[ \mc{N}_{\ell-1}\epsilon^{2\ell} F^{(-1)I_\ell}{}_{I'_\ell} C^{\langle I'_{\ell-1}}_{ \rm reg}\hat{s}^{i'_{\ell}\rangle}
\\
&+  \mc{N}_{\ell+1} \epsilon^{2\ell+2} F^{(1)I_\ell}{}_{ I'_\ell} C^{I'_{\ell+1}}_{ \rm reg}\hat{s}_{i'_{\ell+1}}\Big] + \mathcal{O}(\chi^2)\,,
\end{split}
\label{eq:genexpandgenspsi4}
\end{align}
Now, we can just repeat the steps as in the scalar case, with the only difference being that we should switch to the $\ell,m$ basis corresponding to spin-weight (-2) spherical/spheroidal harmonics. In this manner, all the tidal response coefficients can be fixed iteratively. Explicit steps in the gravitational case for $\ell=2$ and the mixing coefficients between $\ell=2$ and $\ell=3$ modes may be found in Ref.~\cite{Saketh:2022xjb, saketh2023dynamical-b30}.

\subsubsection{Structure of the response functions}
To proceed further, we define
\begin{equation}
C^{\ell m} = \int d\Omega C^{I_\ell}\hat{N}_{\langle I_{\ell-2}}\bar{m}_{i_{\ell-1}}\bar{m}_{i_\ell\rangle} {}_{-2}\bar{Y}_{\ell m}\,,
\label{eq:analog1}
\end{equation}
where $m_i=(1/\sqrt{2})(\hat{\theta}+i\hat{\phi})$, where $\hat{\theta}.\hat{\theta}=\hat{\phi}.\hat{\phi}=m.\bar{m}=1$. Meanwhile, $m.m=\bar{m}.\bar{m}=0$. Now, we can define, in a similar manner as before, a tensor ${}_{-2}\mc{Y}^{I_{\ell}}_{\ell m}$, via the relation
\begin{equation}
{}_{-2}Y_{\ell m}(\theta,\phi) = {}_{-2}\mc{Y}^{I_{\ell}}_{\ell m}\hat{N}_{\langle I_{\ell-2}}\bar{m}_{i_{\ell-1}}\bar{m}_{i_\ell\rangle}\,.
\label{eq:analog2}
\end{equation}
Note that we have the relation (derived in \ref{normalization})
\begin{equation}
\begin{split}
&\int d\Omega \hat{N}_{\langle I_{\ell-2}}m_{i_{\ell-1}}m_{i_\ell\rangle}  \hat{N}^{\langle J_{\ell'-2}}\bar{m}^{j_{\ell'-1}}\bar{m}^{j_\ell'\rangle } 
\\ 
& = \delta_{\ell \ell'} \delta^{J_{\ell}}_{I_\ell} \frac{8\pi(\ell-2)!}{\ell(\ell-1)(2\ell+1)(2\ell-3)!!}\,.
\end{split}
\label{eq:legend}
\end{equation}
Substituting \ref{eq:analog2} into \ref{eq:analog1}, and using \ref{eq:legend}, we get
\begin{equation}
C^{\ell m} = C^{I_{\ell}}{}_{-2}\bar{\mc{Y}}_{I_\ell}^{\ell m} \frac{8\pi(\ell-2)!}{\ell(\ell-1)(2\ell+1)(2\ell-3)!!}\,,
\end{equation}
and the inverse as 
\begin{equation}
C^{I_\ell} = \sum_{m=-\ell}^\ell C^{\ell m}\,_{-2}\mc{Y}_{\ell m}^{I_\ell}\,.
\end{equation}
Going forward, for convenience, we define 
\begin{equation}
\zeta_\ell =\frac{8\pi(\ell-2)!}{\ell(\ell-1)(2\ell+1)(2\ell-3)!!} .
\end{equation}
Now, similar to the scalar case, we can define the following quantities:
\begin{equation}
\begin{split}
& \zeta_\ell~ F^{(n)I_{\ell}}{}_{I'_{\ell}}E^{I'_{\ell}} {}_{-2}\bar{\mc{Y}}_{I_\ell}^{\ell m} 
\\
&=  \zeta_\ell ~F^{(n)I_{\ell}}{}_{I'_{\ell}} \sum_{m'=-\ell}^\ell E^{\ell m'}{}_{-2}\mc{Y}^{I'_\ell}_{\ell m'} {}_{-2}\bar{\mc{Y}}_{I_\ell}^{\ell m} 
\\
&= \sum_{m'=-\ell}^\ell  F^{(n)\ell m}{}_{\ell m'}E^{\ell m'}\,,
\end{split}
\label{eq:tensorspherGR}
\end{equation}
where 
\begin{align}
F^{(n)\ell m}{}_{\ell m'}= \zeta_\ell~F^{(n)I_{\ell}}{}_{I'_{\ell}}{}_{-2}\mc{Y}^{I'_{\ell}}_{\ell m'}{}_{-2}\bar{\mc{Y}}_{I_\ell}^{\ell m}\,.
\end{align}
Just as in the scalar case, axisymmetry ensures that different $m$ modes cannot mix, and thus 
\begin{align}
F^{(n)\ell m}{}_{\ell m'} = F^{(n)}_{\ell m}\delta^m_{m'}\,.
\end{align}
As before, we need to treat STF tensors involving direct products or contractions with the spin-vector which also feature in \ref{eq:genexpand}. Consider for instance the projection onto $\ell,m$ basis of the following tensor product involving unit spin vectors: 
\begin{align}
\label{eq:productsSTFtolm}
\begin{split}
& \zeta_\ell C^{\langle I_{\ell-k}}\hat{s}^{i_{\ell-k+1}}\dots\hat{s}^{i_{\ell}\rangle} {}_{-2}\bar{\mc{Y}}^{\ell m}_{I_\ell} 
\\
&=  \zeta_\ell C^{ I_{\ell-k}}\hat{s}^{i_{\ell-k+1}}\dots\hat{s}^{i_{\ell}} {}_{-2}\bar{\mc{Y}}^{\ell m}_{I_\ell} 
\\
& =   \zeta_\ell \sum_{m'} C^{\ell-k, m'} {}_{-2}\mc{Y}^{I_{\ell-k}}_{\ell-k, m'}\hat{s}^{i_{\ell-k+1}}\dots\hat{s}^{i_{\ell}} {}_{-2}\bar{\mc{Y}}^{\ell m}_{I_\ell}\,.
\end{split}
\end{align}
Since we have oriented the spin-axis to be the $\hat{z}$ direction, the spin-vector has $\ell=1,~m=0$, and we can thus write
\begin{align}
\label{eq:ZdefGR}
{}_{-2}Z_{\ell}^{k} \delta^m_{m'}= \zeta_\ell \,_{-2}\mc{Y}^{I_{\ell-k}}_{\ell-k,m'}\hat{s}^{i_{\ell-k+1}}\dots\hat{s}^{i_{\ell}} \,_{-2}\bar{\mc{Y}}^{\ell m}_{I_\ell}\,.
\end{align}
Using this, we can write~\ref{eq:productsSTFtolm} as
\begin{align}
\begin{split}
& \zeta_\ell C^{\langle I_{\ell-k}}\hat{s}^{i_{\ell-k+1}}\dots\hat{s}^{i_{\ell}\rangle } {}_{-2}\bar{\mc{Y}}^{\ell m}_{I_\ell} =  C^{\ell-k, m}{}_{-2}Z_{\ell}^k\,.
\end{split}
\end{align}
Now, consider the contraction
\begin{align}
\begin{split}
&\zeta_\ell C^{I_{\ell+k}}\hat{s}_{i_{\ell+k}}\dots\hat{s}_{i_{\ell+1}} ~{}_{-2}\bar{\mc{Y}}^{\ell m}_{I_\ell} 
\\
& =   \zeta_\ell\sum_{m'=-\ell-k}^{\ell+k}
C^{\ell+k,m}\,_{-2}\mc{Y}^{I_{\ell+k}}_{\ell+k,m'}\hat{s}_{i_{\ell+k}}\dots\hat{s}_{i_{\ell+1}}~{}_{-2}\bar{\mc{Y}}^{\ell m}_{I_{\ell}}\,.
\end{split}
\end{align}
Once again, the accompanying product must vanish when $m\neq m'$, and we can thus write
\begin{align}
\begin{split}      
\zeta_\ell C^{I_{\ell+k}}\hat{s}_{i_{\ell+k}}\dots\hat{s}_{i_{\ell+1}} \,_{-2}\bar{\mc{Y}}^{\ell m}_{I_\ell}  = C^{\ell+k,m}~{}_{-2}\bar{Z}^k_{\ell+k}\,.
\end{split}
\end{align}
Now, we can finally rewrite \ref{eq:genexpandgenspsi4} in $\ell,m$ basis as
\begin{align}
\begin{split}
C^{\ell m}_{\rm irreg}&= \frac{-4}{\ell!}\mc{N}_\ell  \epsilon^{2\ell+1} F_{\ell m}C^{\ell m}_{\rm reg} 
\\
& - \frac{4}{\ell!} (i\chi) \Big[ \mc{N}_{\ell-1}\epsilon^{2\ell} F^{(-1)}_{\ell m} {}_{-2}\mc{Z}_{\ell-1}^1 C^{\ell-1,m}_{\rm reg}         
\\
&+  \mc{N}_{\ell+1} \epsilon^{2\ell+2} F^{(1)}_{\ell m} {}_{-2}\mc{Z}_{\ell+1}^{-1} C^{\ell+1,m}_{\rm reg}\Big]  
\\
& + \chi^2\left[\cdots\right]+\cdots\,.
\end{split}
\label{eq:Qexpandgw}
\end{align}
Here we have defined 
\begin{equation}\label{eq:calZdef}
{}_{-2}\mc{Z}^{0}_{\ell}=1, \; \; \; {}_{-2}\mc{Z}^{k>0}_\ell = {}_{-2}Z^k_{\ell+k}, \;\; \; {}_{-2}\mc{Z}^{k<0}_\ell = {}_{-2}\bar{Z}^{|k|}_{\ell}\,,
\end{equation}
where ${}_{-2}Z^{k}_{\ell}$ has been defined in~\ref{eq:ZdefGR}. We can write the relation between the irregular and the regular part in a more generic manner as,
\begin{equation}
C^{\ell m}_{\rm irreg} =\frac{-4}{\ell!} \sum_{\ell'=2}^{\infty} i^{\beta} \chi^{|\ell-\ell'|}{}_{-2}\mc{Z}^{\ell-\ell'}_{\ell'}
\mc{N}_{\ell'} \epsilon^{\ell+\ell'+1}F^{(\ell'-\ell)}_{\ell m} C^{\ell' m}_{\rm reg}\,,
\end{equation}
where $\beta$ is 1 when $|\ell-\ell'|$ is odd, and 0 otherwise.

As before, switching on just one mode of regular field at a time, at say $\ell'=\ell_0$, we obtain
\begin{equation}
\begin{split}
C^{\ell m}_{\rm irreg} =\frac{-4}{\ell!} i^{\beta} \chi^{|\ell-\ell_0|} {}_{-2}\mc{Z}^{\ell-\ell_0}_{\ell_0}\mc{N}_{\ell_0} \epsilon^{\ell+\ell_0+1}F^{(\ell_0-\ell)}_{\ell m} C^{\ell_0 m}_{\rm reg}\,,
\end{split}
\end{equation}
and then defining a rescaled response as previously done for the scalar case, 
\begin{equation}
f_{\ell_0 m}^{(\ell-\ell_0)}=\frac{-4}{\ell!} i^{\beta-|\ell-\ell_0|} {}_{-2}\mc{Z}_{\ell_0}^{\ell-\ell_0} \mc{N}_{\ell_0}F^{(\ell_0-\ell)}_{\ell m}\,.
\label{eq:rescaledgw}
\end{equation}
This can be considered a convenient choice for the normalization of the tidal response for the EFT. Thus, we have the relation between irregular and regular part using the new response function as,
\begin{align}
C_{\rm irreg}^{\ell m}& =(i\chi)^{|\ell-\ell_0|} \epsilon^{\ell+\ell_0+1}f_{\ell_0 m}^{(\ell-\ell_0)}C^{\ell_0 m}_{\rm reg}\,.
\label{eq:onemodetoallGR}
\end{align}
This is similar to the scalar case, except that in the gravitational case $\ell-\ell_0$ can be odd as well.

In this work, we will be primarily interested in computing the diagonal tidal response, i.e., $f_{\ell_0 m}=f^{(0)}_{\ell_0 m}$. In \ref{eq:onemodetoallGR}, setting $\ell_0=\ell$ yields
\begin{alignat}{3}
C^{\ell m}_{\rm irreg} = \epsilon^{2\ell+1} f_{\ell m} C^{\ell m}_{\rm reg}\,.
\end{alignat}
Once again, in perturbation theory, we obtain a relationship in spheroidal harmonic basis of spin weight `-2' as
\begin{align}
D^{\ell m}_{\rm irreg} = \epsilon^{2\ell+1}\,_{-2}\lambda_{\ell m} D^{\ell m}_{\rm reg}\,,
\label{eq:pert_response}
\end{align}
where $D^{\ell m}_{\rm reg/irreg}$s are the coefficients from BH perturbation theory in the spheroidal basis, and $\ell,m$ label the spheroidal modes. The computation of the $\lambda_{\ell m}$ coefficients will be discussed in~\ref{genspin}. To connect BH perturbation theory response $\lambda_{\ell m}$ with the EFT response $f_{\ell m}$, the first thing is to connect the spin weighted spheroidal and spherical harmonics of spin weight `-2',
\begin{equation}
\label{eq:scalarspheroidaltosphericalharmonicssm2}
\begin{split}
{}_{-2}Y_{\ell m}   &= {}_{-2}S_{\ell m} + \sum_{k=1}^{\ell-2
} (\chi \epsilon)^{k} {}_{-2}\tilde{r}^{\ell m}_{-k}~{}_{-2}S_{\ell-k,m}
\\
&+\sum_{k=1}^{\infty} (\chi \epsilon)^{k} {}_{-2}\tilde{r}^{\ell m}_{k}~{}_{-2}S_{\ell+k,m}\,,
\end{split}
\end{equation}
where ${}_{-2}\tilde{r}^{\ell m}_{k}$ are the associated Clebsch-Gordon-like coefficients. As in the scalar case, we can use \ref{eq:scalarspheroidaltosphericalharmonicssm2} to obtain a relationship between the coefficients of spheroidal and spherical basis as
\begin{align}
\begin{split}
D^{\ell m}  &= C^{\ell m} + \sum_{k=1}^{\ell-2} (i \chi \epsilon)^{k} {}_{-2}r^{\ell m}_{-k}C^{\ell-k,m}
\\
& +\sum_{k=1}^{\infty} (i \chi \epsilon)^{k} {}_{-2}r^{\ell m}_{k}C^{\ell+k,m}\,,
\end{split}
\label{eq:spherexpsm2}
\end{align}
where the relationship between $r^{\ell m}_k$ and $\tilde{r}^{\ell m}_{k}$ will be fixed in the following subsection.

The relation in \ref{eq:spherexpsm2} holds for both regular and irregular field coefficients. Once again, there is freedom to choose the regular fields in the spherical basis. In particular, we set that only $C^{\ell_0 m}_{\rm reg}$ is non-zero. Then, we have 
\begin{align}
\begin{split}
D_{\rm reg}^{\ell m } = (i\chi \epsilon)^{|l-l_0|}\,_{-2}r^{\ell m}_{l_0-l}\,C_{\rm reg}^{\ell_0m}\,,
\end{split}
\label{eq:sphersphregGR}
\end{align}
where we have defined $r^{\ell m}_0=1$. Now substituting \ref{eq:spherexpsm2} and \ref{eq:sphersphregGR}  into \ref{eq:pert_response}, using the spherical-basis induction equation(s) in \ref{eq:onemodetoallGR}, and imposing $\ell_{0}\geq \ell$ we get constraint equations just as before
\begin{align}
\begin{split}
&\left[f_{\ell_0 m}^{(l-l_0)} + \sum_{k=1}^{\ell}(i\chi)^{2k}f^{(\ell-k-\ell_0)}_{\ell_0,m}\,_{-2}r^{\ell m}_{-k} \right] 
\\
&+ \mathcal{O}(\epsilon^{2}) =\,_{-2}\lambda_{\ell m}\,_{-2}r^{\ell m}_{\ell_0-\ell}, \quad \ell_0\geq \ell\,.
\end{split}
\label{eq:recursor}
\end{align}
This provides the connection between EFT tidal response $f_{\ell m}$ with the BH perturbation theory tidal response $\lambda_{\ell m}$. In what follows we will briefly illustrate the above connection for the $\ell=2$ and the $\ell=3$ modes, respectively.

\subsubsection{Iterative fixing of the tidal coefficients}

Here we will solve the connection formula, \ref{eq:Qexpandgw}, iteratively. To start with, we set $\ell=\ell_0=2$ (since for gravitational perturbation the lowest mode is $\ell=2$), where we simply get
\begin{align}
\label{eq:lambdatofquad}
f_{\ell m} = \,_{-2}\lambda_{\ell m}\,,~\ell=2\,.
\end{align}
We can solve it similar to the scalar case for $\ell>2$. For instance, we can set $\ell=2,\ell_0=3$, to obtain
\begin{equation}
f_{3m}^{(-1)}  = \,_{-2}\lambda_{2m}\,_{-2}r^{2m}_{1}\,.
\end{equation}
Subsequently, setting $\ell=3$, $\ell_0=3$, we get
\begin{equation}
\begin{split}
& f_{3m}  -\chi^2 f_{3m}^{(-1)} \,_{-2}r^{3m}_{-1}= \,_{-2}\lambda_{3m}\,.
\\
&  f_{3m}  = \chi^2 \,_{-2}\lambda_{2m} \,_{-2}r^{2m}_{1} \,_{-2}r^{3m}_{-1} +  \,_{-2}\lambda_{3m}\,.
\end{split}
\label{eq:l3resgrav}
\end{equation}
The above explicitly depicts the connection between EFT and BH perturbation theory responses for the two lowest lying modes, $\ell=2,3$, and can be generalized to any higher modes by using the connection formula repeatedly.

Once the coeffcients $f_{\ell m}^{(0)}$ are fixed, we can obtain the EFT coefficients in $\ell m$ basis trivially using the definition in \ref{eq:rescaledgw} as
\begin{equation}\label{grav_eft_scale}
F_{\ell m}^{(0)} = \frac{-\ell! f_{\ell m}^{(0)}}{4 N_\ell} = -\frac{2^{-\ell}\Gamma(2+2\ell) f_{\ell m}}{(2+3\ell+\ell^2)}\,.
\end{equation}

\subsubsection{Asymptotic form of gravitational field in EFT}

We showed earlier how the asymptotic structure of the scalar field can be used to match between EFT coefficients $C^{\ell m}_{\rm reg/irreg}$ and perturbation theory coefficients $D^{\ell m}_{\rm reg/irreg}$ in the far-zone. We show the equivalent matching in the gravitational case below. For this purpose one needs to expand $h_{ij}$ in the $\omega r\gg 1$ region using \ref{eq:metric}. However, for the sake of gauge invariance it is useful to perform the computation using the Weyl scalar $\psi_4$, rather than metric perturbation $h_{ij}$, in the large $r$ limit. This has already been done in \cite{Saketh:2022xjb}, and we obtain
\begin{equation}
\begin{split}
\psi_4 &= \omega^2 \frac{e^{-i\omega(t-r)}}{\omega r} \sum_{\ell=2}^\infty i^{\ell-2}(C^{K_{\ell-2}ij}_{\mce,\rm out}+C^{K_{\ell-2}ij}_{\mcb,\rm out})N_{K_{\ell-2}}\bar{m}_i\bar{m}_j
\\
&+\mathcal{O}(r^{-2})\,,
\end{split}
\end{equation}
where $C_{\mce/\mcb,\rm out}^{K_{\ell-2}ij} =(1/2)(C_{\mce/\mcb,\rm irr}^{K_{\ell-2}ij}-i C_{\mce/\mcb,\rm reg}^{K_{\ell-2}ij})$. We can decompose the above expression in spherical basis by using the following result,
\begin{equation}
\begin{split}
&\int d\Omega~ C^{K_{\ell-2}ij}\bar{m}_i \bar{m}_j N_{K_{\ell-2}}~{}_{-2}\bar{Y}^{\ell m}= C^{\ell m}\,,
\end{split}
\end{equation}
Moreover, using the normalization of ${}_{-2}Y_{\ell m}(\theta,\phi)$, and the completion relations
\begin{equation}
\begin{split}
& \int {}_{-2}Y_{\ell m}(\theta,\phi) {}_{-2}\bar{Y}_{\ell m}(\theta,\phi) d\Omega =1\,,
\\
&\sum_{\ell,m}{}_{-2}Y_{\ell m}(\theta,\phi) {}_{-2}\bar{Y}_{\ell m}(\theta',\phi') = \frac{1}{\sin^2\theta}\delta(\theta-\theta')\delta(\phi-\phi')\,,
\end{split}
\end{equation}
and defining $C^{K_{\ell-2},ij}_{\rm reg/irreg}=C^{K_{\ell-2},ij}_{\mce,\rm reg/irreg}+C^{K_{\ell-2},ij}_{\mcb,\rm reg/irreg}$, as before, we obtain
\begin{equation}
\begin{split}
\psi_4 &= \omega^2 \frac{e^{-i\omega(t-r)}}{\omega r} \sum_{\ell=2}^\infty i^{\ell-2}C^{\ell m}_{\rm out} {}_{-2}Y_{\ell m}(\theta,\phi)\,,
\\
& = \omega^2 \frac{e^{-i\omega(t-r)}}{\omega r} \sum_{\ell=2}^\infty i^{\ell-2}D^{\ell m}_{\rm out} {}_{-2}S_{\ell m}(\theta,\phi)\,,
\end{split}
\label{eq:psi4}
\end{equation}
where in the second line we have used the result from BH perturbation theory, and in each of the lines we have ignored terms $\mathcal{O}(r^{-2})$. Using \ref{eq:scalarspheroidaltosphericalharmonicssm2} on the first line of the above expression, we can write it in terms of spheroidal harmonics as,
\begin{equation}
\begin{split}
\psi_4 = \omega^2 &\frac{e^{-i\omega(t-r)}}{\omega r} \sum_{\ell=2}^\infty i^{\ell-2}[C^{\ell m}_{\rm out}
\\
& +\sum_{k=1}^{\ell-2}(-i \chi \epsilon)^{k} \,_{-2}\tilde{r}^{\ell-k, m}_{k} C^{\ell-k,m}_{\rm out}
\\
& + \sum_{k=1}^\infty (i \chi \epsilon)^{k} \,_{-2}\tilde{r}^{\ell+k,m}_{-k}C^{\ell+k,m}_{\rm out}] {}_{-2}S_{\ell m}(\theta,\phi)\,.
\end{split}
\end{equation}
Comparing the above expression with the last equality in \ref{eq:psi4}, we obtain the relation between EFT and BH perturbation theory coefficients, this yields,
\begin{equation}
\begin{split}
D^{\ell m} &= C^{\ell m}+\sum_{k=1}^{\ell-2}(-i \chi \epsilon)^{k} \,_{-2}\tilde{r}^{\ell-k, m}_{k} C^{\ell-k,m}
\\
& + \sum_{k=1}^\infty (i \chi \epsilon)^{k} \,_{-2}\tilde{r}^{\ell+k,m}_{-k}C^{\ell+k,m}\,.
\end{split}
\end{equation}

Comparing with \ref{eq:spherexpsm2}, this yields $(-1)^k\,_{-2}\tilde{r}^{\ell-k, m}_{ k} = \,_{-2}r^{\ell m}_{-k}$, $\,_{-2}\tilde{r}^{\ell+k, m}_{- k} = \,_{-2}r^{\ell m}_{k}$, where $k>0$. We can now explicitly write down the relationship between the tidal response in spheroidal and spherical basis. For instance, \ref{eq:l3resgrav} can be rewritten as 
\begin{equation}\label{grav_eft_bh}
\begin{split}
f_{3m}&= \,_{-2}\lambda_{3m}+ \chi^2 \,_{-2}\lambda_{2m} \,_{-2}r^{2m}_{1} \,_{-2}r^{3m}_{-1}
\\    
&=\,_{-2}\lambda_{3m}- \chi^2 \,_{-2}\lambda_{2m} \,_{-2}\tilde{r}^{3m}_{-1} \,_{-2}\tilde{r}^{2m}_{1}
\\ 
& =\,_{-2}\lambda_{3m}- \chi^2 \,_{-2}\lambda_{2m} \frac{4(9-m^2)}{567}\,.
\end{split}
\end{equation}
Thus, we have shown how the ratio $D^{\ell m}_{\rm irreg}/D^{\ell m}_{\rm reg}$ can be used to fix the tidal-response coefficients in EFT. But as in the scalar case, we need to obtain expressions for $D^{\ell m}_{\rm irreg/reg}$ from perturbation theory. We obtain this by matching \ref{eq:psi4} with the far-zone solution obtained from perturbation theory in \ref{subsec:EFTgenerics}.

\section{Scalar field on a Kerr background: Amplitude of scattering states}\label{scalar_scattering}

In this section, we focus on external scalar perturbations, playing the role of an external tidal field, on the Kerr background. We know that such a perturbation, in the static limit, does not lead to  conservative deformations of the Kerr multipole moments, which translates to vanishing tidal Love numbers~\cite{Damour:2009va,Kol:2011vg,Ivanov:2022qqt, LeTiec:2020bos, Chia:2020yla, Poisson:2021yau, Berens:2025okm, Sharma:2025xii, Parra-Martinez:2025bcu, Hui:2021vcv, Sharma:2024hlz, Barbosa:2025uau, BenAchour:2022uqo, Hui:2022vbh, Charalambous:2022rre, Katagiri:2022vyz}. Here we focus on dynamical scalar perturbations, in the frequency domain, and investigate the dynamical (frequency-dependent) tidal response. Typically, the computations performed using BH perturbation theory are not gauge invariant. This makes it essential to include the additional step of matching the BH perturbation theory with the EFT discussed in the previous section, which provides a coarse-grained description of the perturbed BH and is the essential baseline for computing dynamics and GWs in a binary system. We closely follow the methods of~\cite{creci2021tidal-42e}, and perform the computation of the DTRF in the following steps (also see \ref{fig:zoneEFT}) --- (a) solving the scalar perturbation in the \nz as well as in the \fz regime, (b) imposing the ingoing boundary conditions at the horizon, (c) matching the \nz and the \fz solutions in the intermediate-zone, and finally (d) taking the asymptotic limit of the \fz solution for the scalar field and matching with the EFT computation. 

Following the above prescription, we first present the behaviour of the scalar field in the \nz regime, obtained with the approximation $\omega(r-r_{+})\ll 1$, where $r_{+}$ is the horizon radius of the BH. Subsequently, we solve the perturbation equations in the \fz are using the approximation $(M/r)\ll 1$, where $M$ is the mass of the BH. The matching of the near-zone with the asymptotic solution determines how information on the near-horizon behavior is imprinted in far-field amplitudes, which we in turn match with the EFT. Throughout this computation, we use a small frequency approximation, $M\omega \ll 1$ and work only up to linear order in this quantity. 

\subsection{The basic equations}

We start with BH perturbation theory for scalar perturbations. Given the stationary and axi-symmetric nature of the background Kerr geometry, we decompose the scalar field $\Phi$ as, 
\begin{align}
\Phi=\sum_{\ell,m}\int d\omega e^{-i\omega t}R_{\ell m}(r)S_{\ell m}(\theta)e^{im\phi}\,,
\label{eq:phiexp}
\end{align}
where $R_{\ell m}(r)$ is the radial function and $S_{\ell m}(\theta)$ the angular function. The BH perturbation equations in this case are equivalent to the Klein-Gordon equation for a massless scalar field, $\square \Phi=0$. This leads to the following decoupled radial and angular equations, 
\begin{widetext}
\begin{align}
&\frac{1}{\sin \theta}\dfrac{d}{d\theta}\left(\sin \theta\dfrac{dS_{\ell m}}{d\theta} \right)-\left(a^{2}\omega^{2}\sin^{2}\theta+\frac{m^{2}}{\sin^{2}\theta}-\widetilde{A}_{\ell m}\right)S_{\ell m}=0~,
\label{angeq_scalar}
\\
&\dfrac{d}{dr}\left(\Delta\dfrac{dR_{\ell m}}{dr}\right)+\left[\frac{\omega^{2}(r^{2}+a^{2})^{2}-4arM\omega m+a^{2}m^{2}}{\Delta}-\widetilde{A}_{\ell m}\right]R_{\ell m}=0~.
\label{radeq_scalar}
\end{align}
\end{widetext}
Here, $\Delta=r^{2}-2Mr+a^{2}$, and $a=(J/M)$ is the rotation parameter, where $J$ is the angular momentum and $M$ is the mass of the BH. The quantity $\widetilde{A}_{\ell m}$ is the separation constant between radial and angular equations. The above angular equation can be converted to the one obtained by Teukolsky~\cite{1973ApJ...185..635T} with the substitution $\widetilde{A}_{\ell m}=E_{\ell m}+a^{2}\omega^{2}$, and the radial equation also converts to the corresponding Teukolsky equation with $\lambda=\widetilde{A}_{\ell m}-2am\omega=E_{\ell m}-2am\omega+a^{2}\omega^{2}$, where $E_{\ell m}$ is the separation constant between the radial and the angular sectors in Teukolsky's formalism. The angular equation determines that $S_{\ell m}(\theta)e^{im\phi}$ are the spheroidal harmonics. In what follows, we will use the parametrization used by Teukolsky~\cite{1973ApJ...185..635T} for the radial equation.

\subsection{Near-zone solution for the radial equation}

In this section, we will determine the \nz solution to~\ref{radeq_scalar}. We introduce the dimensionless radial coordinate 
\begin{equation}
\label{eq:zdef}
z=\frac{(r-r_{+})}{(r_{+}-r_{-})}\,,
\end{equation}
where $r_{\pm}=M\pm\sqrt{M^{2}-a^{2}}$ are the event and the Cauchy horizons. Thus it follows that $\Delta=(r-r_{+})(r-r_{-})=(r_{+}-r_{-})^{2}z(1+z)$. Therefore we obtain the following structure for the radial equation describing scalar perturbation of the Kerr background:
\begin{align}
\dfrac{d}{dz}\left[z(1+z)\dfrac{dR_{\ell m}}{dz}\right]+\frac{\left[P_{+}^{2}-\lambda z(1+z)\right]}{z(1+z)}R_{\ell m}=0~,
\end{align}
where we have assumed $\omega(r-r_{+})\ll 1$ and have defined 
\begin{equation}
\label{eq:Pplusdef}
P_{+}\equiv(2Mr_{+}\omega-am)/(r_{+}-r_{-})=2Mr_{+}\bar{\omega}/(r_{+}-r_{-}), 
\end{equation}with $\bar{\omega}\equiv \omega-m\Omega_{+}$, and $\Omega_{+}\equiv (a/2Mr_{+})$ being the angular velocity of the event horizon, located at $r=r_{+}$.

The general solution of the above second order differential equation, describing the radial part of perturbations of the Kerr background due to an external scalar field in the \nz regime, becomes, 
\begin{align}\label{scalar_gen_near}
&R^{\rm (near)}_{\ell m}=(1+z)^{iP_{+}}\Big[Az^{-iP_{+}}\,_{2}F_{1}\left(-\hat{\ell},\hat{\ell}+1;1-2iP_{+};-z\right)
\nonumber
\\
&+Bz^{iP_{+}}\,_{2}F_{1}\left(-\hat{\ell}+2iP_{+},\hat{\ell}+1+2iP_{+};1+2iP_{+};-z\right)\Big]~,
\end{align}
where we have expressed the separation constant as $\lambda=\hat{\ell}(\hat{\ell}+1)$, such that the quantity $\hl$ has the following expansion in powers of $M\omega$, 
\begin{align}
&\hat{\ell}(\hat{\ell}+1)=E_{\ell m}-2am\omega+a^{2}\omega^{2}
\nonumber
\\
&=\ell(\ell+1)-2am\omega+a^{2}\omega^{2}
\nonumber
\\
&\qquad +2a^{2}\omega^{2}\left[\frac{m^{2}+\ell(\ell+1)-1}{(2\ell-1)(2\ell+3)} \right]+\mathcal{O}(M^{3}\omega^{3})\,. 
\end{align}
This can be solved to yield, 
\begin{equation}
\hat{\ell}=\ell-\{2am\omega/(2\ell+1)\}+\mathcal{O}(M^{2}\omega^{2}).
\label{eq:lhat}
\end{equation}
The fact that $\hat{l}-l=\mathcal{O}(M\omega)$, will be useful in our later analysis. The above \nz radial solution for scalar tidal perturbations, presented in \ref{scalar_gen_near}, has two unknown constants $A$ and $B$, one of which is fixed by the relevant boundary condition condition at the BH horizon, namely purely ingoing modes. 

\subsubsection{Ingoing boundary condition at the horizon}

To fix one of the two arbitrary constants appearing in \ref{scalar_gen_near}, we consider the limit to the BH horizon ($z\to 0$). Using the fact that the Hypergeometric function becomes unity as their argument vanishes, the near-zone radial perturbation function becomes, 
\begin{align}
R^{\rm (near)}_{\ell m}(r\to r_{+})&\simeq Az^{-iP_{+}}+Bz^{iP_{+}}
\nonumber
\\
&=Ae^{-iP_{+}r_{*}}+Be^{iP_{+}r_{*}}~.
\end{align}
Here, $r_{*}$ is the Tortoise coordinate, defined through the differential equation: $(dr/dr_{*})=(\Delta/r^{2})$, which in the near-zone region can be approximated as $r_{*}\sim r_{+}\ln z$. As the term $e^{iP_{+}r_{*}}$ in the decomposition of~\ref{eq:phiexp} belongs to outgoing waves at the horizon, it follows that the coefficient $B$ must be set to zero, since for BHs we expect only ingoing modes to be present. Hence the radial part of the scalar perturbation associated with the Kerr BH takes the form,
\begin{align}\label{radial_scalar_near}
R^{\rm (near)}_{\ell m}&=A_{\rm in}^{\rm H}(1+z)^{iP_{+}}z^{-iP_{+}}
\nonumber
\\
&\times\,_{2}F_{1}\left(-\hat{\ell},\hat{\ell}+1;1-2iP_{+};-z\right)~.
\end{align}
This provides the \nz solution for the radial perturbation function due to a scalar field on the Kerr background, with $A_{\rm in}^{\rm H}$ being an arbitrary constant. To verify the above result, consider Ref.~\cite{Chia:2020yla}, which yields an identical expression for the radial function with $\hl\to \ell$. This is because Ref.~\cite{Chia:2020yla} ignores quite a few terms of $\mathcal{O}(M\omega)$. In this work, however, we have kept all such terms but ignored terms of $\mathcal{O}(M\omega z)$. This leads to \ref{radial_scalar_near}, with $\hl=\ell+\mathcal{O}(M\omega)$. In contrast, Ref.~\cite{sasaki2003analytic-827} retains all possible terms and expresses the solution as an infinite series of hypergeometric functions, which does not match with the above result. We plan to revisit this infinite series expansion of the hypergeometric function to determine the TLNs in future work.

\subsubsection{Limit to the intermediate zone}

The same solution in the intermediate region, i.e., in the region satisfying $(r_{+}/r)\ll 1$ can be determined by computing the above radial function in the large $z$ regime. Given the asymptotic expansion of the hypergeometric function, the \nz radial function in \ref{radial_scalar_near}, in the corresponding limit becomes
\begin{widetext}
\begin{align}\label{radial_near_far}
R_{\ell m\,\textrm{(int)}}^{\rm (near)}&=A_{\rm in}^{\rm H}\left[\frac{\Gamma(1-2iP_{+})\Gamma(1+2\hat{\ell})}{\Gamma(1+\hat{\ell})\Gamma(1+\hat{\ell}-2iP_{+})}\frac{r^{\hl}}{(r_{+}-r_{-})^{\hl}}\left(1+\cdots \right)+\frac{\Gamma(1-2iP_{+})\Gamma(-1-2\hat{\ell})}{\Gamma(-\hat{\ell})\Gamma(-\hat{\ell}-2iP_{+})}\frac{r^{-\hat{\ell}-1}}{(r_{+}-r_{-})^{-\hl-1}}\left(1+\cdots \right)\right]\,.
\end{align}
\end{widetext}
Thus, asymptotically the radial function consists of two independent power law branches, namely $r^{\hat{\ell}}$ and $r^{-1-\hat{\ell}}$. We have only kept the leading order pieces in each branch. The ``$\cdots$" denote sub-leading terms, decaying as a power law in the radial coordinate $r$, denoting fractional corrections to the leading order contribution. Since $M\omega\ll 1$, it follows that $\hat{\ell}>0$, and hence the $r^{\hat{\ell}}$ branch describes a growing mode related to the tidal field, while the $r^{-1-\hat{\ell}}$ branch describes a decaying mode and yields the response of the body. For static perturbations, we have $\hat{\ell}=\ell$ and hence the usual static tidal and response fields of the body follow. But note that the tidal field has sub-leading terms, which might raise concerns about ambiguities in the definition of the response function \cite{gralla2017ambiguity-aad, Kol:2011vg, ivanov2022revisiting-a24}. For rotating BHs, $\hat{\ell}$ is given by~\ref{eq:lhat}, and it involves $\ell$ corrected by a frequency-dependent quantity. As the frequency can take arbitrary values, it follows that $\hat{\ell}$ is analytically continued and hence those sub-leading terms cannot mimic the response function, thus avoiding ambiguities. For the static case and for the case $a=0$, where $\hl=\ell$, the ambiguity problem needs to be avoided by an explicit analytic continuation of the angular number $\ell$.

It is worth mentioning that the large $z$ limit of the \nz solution seems a bit odd, but effectively, we need to satisfy the following condition: $\omega r\ll 1$. Thus in the dynamical scenario, the growing and decaying behaviour of the perturbing field holds true within a sphere of radius $(\omega)^{-1}$. Thus we have three characteristic scales in the problem: (a) the horizon at $r_{+}$, (b) the frequency scale $\omega^{-1}$ and (c) the asymptotic region. The \nz solution is valid in a region $r_{+}<r\ll \omega^{-1}$, while the intermediate zone is centered about $\omega^{-1}$, connecting the \nz and the far-zone, whereas the \fz corresponds to $r\gg M$ (see \ref{fig:zoneEFT}). Therefore, \ref{radial_near_far} describes the extension of the \nz solution to the intermediate zone.

In the next section, we will determine the \fz solution and shall match its small distance behaviour with the \nz solution in the intermediate region, as obtained above.

\subsection{Far-zone Solution}

The \fz solution is obtained by taking the asymptotic limit, i.e., $(r/M)\gg 1$, of the radial equation~\ref{radeq_scalar}. In the asymptotic limit, we keep the leading order terms in front of the derivatives of $R_{\ell m}$, while in the coefficient of $R_{\ell m}$ we keep the sub-leading terms as well. This is necessary because the leading order term $\omega^{2}r^{2}$ is not that large in the small frequency approximation. Thus, in the far-zone, the radial part of the scalar perturbation in the Kerr background satisfies the following differential equation,
\begin{align}
\dfrac{d^{2}R^{\rm (far)}_{\ell m}}{dr^{2}}+\frac{2}{r}\dfrac{dR^{\rm (far)}_{\ell m}}{dr}&+\left[\omega^{2}-\frac{\lambda+2ma\omega}{r^{2}}\right]R^{\rm (far)}_{\ell m}
\nonumber
\\
&=\mathcal{O}(M^{2}\omega^{2}r^{-2},r^{-3})~.
\end{align}
Here, we assume $(r/M)\gg 1$. Since we are keeping terms linear in $M\omega$, it follows that $\lambda+2am\omega=E_{\ell m}+\mathcal{O}(M^{2}\omega^{2})=\ell(\ell+1)+\mathcal{O}(M^{2}\omega^{2})$. Hence, the above differential equation has the following general solution, 
\begin{align}\label{scalar_far}
R_{\ell m}^{\rm (far)}=\frac{A^{\infty}_{\ell(\rm reg)}}{\sqrt{r}}J_{\ell+\frac{1}{2}}(\omega r)+\frac{A^{\infty}_{\ell(\rm irreg)}}{\sqrt{r}}Y_{\ell+\frac{1}{2}}(\omega r)~.
\end{align}
Note that the \fz solution depends on $\ell$, rather than $\hl$. This will lead to additional Logarithmic terms in the response function due to improper matching at the intermediate zone. We will discuss this issue in detail later. Thus the \fz solution is given by the Bessel functions with two arbitrary constants $A^{\infty}_{\ell(\rm reg)}$ (since the Bessel function $J_{\alpha}(x)$ is regular as $x\to 0$) and $A^{\infty}_{\ell(\rm irreg)}$ (as the associated Bessel function $Y_{\alpha}(x)$ is irregular as $x\to 0$). We will now consider two limits, $\omega r\ll 1$, which defines the intermediate zone, and $r\to \infty$, describing the asymptotic limit. We will first discuss the limit to the intermediate zone and then shall move on to discuss the asymptotic limit, which will be used to match with the EFT.  

\subsubsection{Limit to the intermediate zone}

The behaviour of the \fz solution in the intermediate zone is obtained by taking the $\omega r\ll 1$ limit of~\ref{scalar_far}. As is well known, Bessel functions for small arguments scale as power law for both first and second kind. Thus in the intermediate region, the radial solution of the \fz behaves as, 
\begin{widetext}
\begin{align}\label{radial_far_ints}
R_{\ell m}^{\rm (far)}(\omega r\ll1)&=\frac{A^{\infty}_{\ell(\rm reg)}}{\sqrt{r}\Gamma(\ell+\frac{3}{2})}\left(\frac{\omega r}{2}\right)^{\ell+\frac{1}{2}}-\frac{A^{\infty}_{\ell(\rm irreg)}}{\sqrt{r}}\left(\frac{\omega r}{2}\right)^{-\ell-\frac{1}{2}}\frac{\Gamma(\ell+\frac{1}{2})}{\pi}
\nonumber
\\
&=\frac{A^{\infty}_{\ell(\rm reg)}}{\Gamma(\ell+\frac{3}{2})}\left(\frac{\omega r}{2}\right)^{\ell}\left[1-\frac{A^{\infty}_{\ell(\rm irreg)}}{A^{\infty}_{\ell(\rm reg)}}\frac{1}{\pi}\left(\frac{2}{\omega}\right)^{2\ell+1}\Gamma\left(\ell+\frac{3}{2}\right)\Gamma\left(\ell+\frac{1}{2}\right)r^{-2\ell-1}\right]~.
\end{align}
\end{widetext}
As evident, the \fz solution in the intermediate regime also has a growing mode as well as a decaying mode, identical to the behaviour observed in the asymptotic limit of the \nz solution, as presented in \ref{radial_near_far}. There is one important difference though, the power of the growing and the decaying modes differ by $\hl-\ell$. We will discuss the implications of this in the subsequent section. This facilitates matching between the \nz and the \fz solution, which we will discuss in~\ref{subsec:scalar_matching_intermediate} below. Before discussing this matching, we also consider the asymptotic limit of~\ref{radial_far_ints} needed to connect with the EFT. 

\subsubsection{Limit to the asymptotic region}

Consider now the asymptotic limit of the \fz solution~\ref{scalar_far}, which is obtained by taking $\omega r\gg 1$. In this limit, the Bessel functions can be expressed in terms of sinusoidal functions, such that, 
\begin{align}\label{asympradial}
R_{\ell m}^{\rm (far)}(\omega r\gg 1)&\simeq \sqrt{\frac{2}{\pi \omega}} \Big[\frac{A_{\ell(\textrm{reg})}^{\infty}}{r}\sin\left(\omega r-\frac{\ell \pi}{2}\right)
\nonumber
\\
&\qquad \qquad -\frac{A_{\ell(\textrm{irreg})}^{\infty}}{r}\cos\left(\omega r-\frac{\ell \pi}{2}\right)\Big]~.
\end{align}
Note that the above represents only the radial part of the perturbation, which will be multiplied by the angular part $S_{\ell m}(\theta,\phi)e^{im\phi}$, along with $e^{-i\omega t}$, and then summed over $\ell$ and $m$ in order to determine the total scalar perturbation $\Phi$. This result will be useful for the matching to the EFT, which we will present below.

\subsection{Matching in the intermediate zone}
\label{subsec:scalar_matching_intermediate}

In this section we will discuss one of the most important aspect of this computation, namely matching of the \nz and the \fz solutions in the intermediate zone. The matching is possible only if both the solutions share an identical behaviour in the intermediate zone. In our case, it turns out that the \fz solution in the small $\omega r$ limit, see \ref{radial_far_ints}, and the near zone solution in the large $(r/M)$ limit, see \ref{radial_near_far}, have identical behaviour, i.e., they both include a growing piece $\sim r^{\ell}$, as well as a decaying piece $\sim r^{-\ell-1}$, modulo the factor involving $(\hl-\ell)\ln r$. Notice that we are computing the ratio between arbitrary constants and hence they cannot depend on $\ln r$, which suggests to express $\ln r\sim \ln (1/\omega)$, as $r\sim \omega^{-1}$ is the region where the intermediate zone ends. Therefore, the coefficient of $r^{\ell}$ as well as the coefficient of $r^{-\ell-1}$ coming from far and \nz must be matched. This yields both $A^{\infty}_{\ell(\rm reg)}$ and $A^{\infty}_{\ell(\rm irreg)}$ in terms of $A_{\rm H}^{\rm in}$. Thus we obtain the ratio $(A^{\infty}_{\ell(\rm irreg)}/A^{\infty}_{\ell(\rm reg)})$ to be, 
\begin{widetext}
\begin{align}\label{matching_scalar}
\frac{A^{\infty}_{\ell(\rm irreg)}}{A^{\infty}_{\ell(\rm reg)}}&=-\pi\left(\frac{\omega(r_{+}-r_{-})}{2}\right)^{2\ell+1}\frac{\Gamma(-1-2\hat{\ell})\Gamma(1+\hat{\ell})\Gamma(1+\hat{\ell}-2iP_{+})}{\Gamma(-\hat{\ell})\Gamma(1+2\hat{\ell})\Gamma(\ell+\frac{1}{2})\Gamma(\ell+\frac{3}{2})\Gamma(-\hat{\ell}-2iP_{+})}
\nonumber
\\
&\times \left\{1+2(\hl-\ell)\ln \left[\omega(r_{+}-r_{-})\right]+\mathcal{O}(M^{2}\omega^{2})\right\}~.
\end{align}
\end{widetext}
Due to $\hl$ being non-integer, c.f.~\ref{eq:lhat}, it follows that $\Gamma(-\hl)$ is finite and hence the above ratio is non-zero. The same argument holds for $\Gamma(-1-2\hl)$. As we will show later, the ratio of these two quantities will turn out to be finite and important for determining the dynamical response function.

At this stage, a few comments on the logarithmic term in the above expression are in order. Such a logarithmic term often arises in the analysis involving dynamical Love numbers of BHs, in particular, to describe the running of the tidal couplings with energy scale \cite{saketh2023dynamical-b30}. These terms are common in the scattering approaches, as well as in BH perturbation theory \cite{saketh2023dynamical-b30}. However, in the present context the logarithmic term scales as $\omega \ln \omega$, since $(\hl-\ell)\sim M\omega$, and (a) has the form of a tail term, (b) is higher order in $M\omega$ than considered here. The tail effect, on the other hand, comes from scattering of GW from the Newtonian potential and hence has no contribution to tides. Thus we ignore this logarithmic term in this analysis, while we expect that a better understanding of this term may arise from a proper matching procedure, e.g., the MST method~\cite{mino1997black-3ab, sasaki2003analytic-827}.

\subsection{Matching with EFT}
\label{subsec:EFTscalar}

In this section we match the results obtained from BH perturbation theory to EFT, as detailed in \ref{sec:EFT}. We note that the tidal response of BH perturbation theory is denoted by $\lambda_{\ell m}$, see \ref{eq:pert_response} and is related to the ratio $(A^{\infty}_{\ell(\rm irreg)}/A^{\infty}_{\ell(\rm reg)})$. To determine the exact relation we compare \ref{eq:asymptotic_EFT_sph} with \ref{asympradial}.
The BHPT scalar field for a given mode $\omega$ in \ref{asympradial} can be rewritten as 
\begin{equation}
\begin{split}
&\sqrt{\frac{\pi}{2\omega}}\Phi_\omega  = \frac{e^{-i\omega t}}{\omega r} \sum_{\ell,m} S_{\ell m}(\theta,\phi) R_{\ell m}
\end{split}
\end{equation}
Comparison of this expression with the EFT one in \ref{eq:asymptotic_EFT_sph}, along with the expression for $R_{\ell,m}$ from \ref{asympradial} then yields the relations
\begin{equation}
D_{\rm reg}^{\ell m} = (-1)^{\ell}A^{\infty}_{\ell(\rm reg)},~D_{\rm irreg}^{\ell m} = (-1)^{\ell+1}A^{\infty}_{\ell( \rm irreg)}\,,
\end{equation}
Thus, the ratio $D_{\rm irreg}/D_{\rm reg}$ is identical to that of $A$'s upto a minus sign. In particular, we have, $(D_{\rm irreg}^{\ell m}/D_{\rm reg}^{\ell m})=-(A^{\infty}_{\ell(\rm irreg)}/A^{\infty}_{\ell(\rm reg)})$, such that the BH perturbation theory response function reads,  
\begin{align}\label{matching_scalar}
\lambda_{\ell m}&=\pi\left(\frac{r_{+}-r_{-}}{2M}\right)^{2\ell+1}\frac{\Gamma(1+\hat{\ell}-2iP_{+})}{\Gamma(-\hat{\ell}-2iP_{+})}
\nonumber
\\
&\times \frac{\Gamma(-1-2\hat{\ell})\Gamma(1+\hat{\ell})}{\Gamma(-\hat{\ell})\Gamma(1+2\hat{\ell})\Gamma(\ell+\frac{1}{2})\Gamma(\ell+\frac{3}{2})}+\mathcal{O}(M^{2}\omega^{2})~.
\end{align}
Note that this result is in the spheroidal basis, which need to be converted to spherical basis in order to obtain the EFT response function. Even though the spheroidal to spherical transformation for the scalar perturbation appears at $\mathcal{O}(M^{2}\omega^{2})$, it follows that there are off-diagonal terms in the response function which actually contributes at those orders. As a consequence, since for scalar perturbation, $\ell$th mode couples to $(\ell\pm 2)$th modes, it follows that the EFT response function is equal to BH response function for $\ell=0,1$ modes. While for $\ell\geq 2$, the EFT response function is given by certain linear combination of $\ell=2$ and $\ell=0$ BH perturbation theory response function. The details of the computation follows from \ref{sec:EFT}. In what follows we will massage the BH perturbation theory response function, which, in turn can be related to EFT response function through appropriate transformations.

\subsection{The dynamical scalar tidal response of black holes}
\label{subsec:response_scalar}
Through matching with the point-particle EFT, we have been able to determine the EFT response function in terms of $\lambda_{\ell m}$. In this section, we will simplify the corresponding expression for $\lambda_{\ell m}$ and hence will determine the scalar Love number and dissipation number, which are defined as the real part and the imaginary part of the response function, respectively. For that purpose, we may first rewrite the dynamical response function, keeping only terms which are linear in $M\omega$, as follows (for a derivation see \ref{App:ScalarResponse}): 
\begin{align}
\label{eq:decompFtildescalar}
&\widetilde{\lambda}_{\ell m}^{\rm Kerr}(\omega)=\widetilde{\lambda}_{\ell m}^{\rm Kerr}|_{\rm stat}+\widetilde{\lambda}_{\ell m}^{\rm Kerr}|_{\rm dyn}\times M\omega\,,
\end{align}
where the rescaled response function $\widetilde{\lambda}$ is defined by 
\begin{equation}\label{scaling}
\widetilde{\lambda}_{\ell m}=\frac{1}{\pi}\left(\frac{2M}{r_{+}-r_{-}}\right)^{2\ell+1}\Gamma[\ell+(1/2)]\Gamma[\ell+(3/2)]\lambda_{\ell m}\,.
\end{equation}
Note that the above decomposition~\ref{eq:decompFtildescalar} neatly separates out the stationary and the dynamical rescaled response functions, each of which is given by, 
\begin{widetext}
\begin{align}
&\widetilde{\lambda}_{\ell m}^{\rm Kerr}|_{\rm stat}=iP^{a}_{+}\frac{\Gamma(1+\ell)^{2}}{\Gamma(2+2\ell)\Gamma(1+2\ell)}\prod_{k=1}^{\ell}\left(k^{2}+4\{P_{+}^{a}\}^{2}\right)\,,
\label{Kerr_static}
\\
&\widetilde{\lambda}_{\ell m}^{\rm Kerr}|_{\rm dyn}=\widetilde{\lambda}_{\ell m}^{\rm Kerr}|_{\rm stat}\left\{-i\coth(2\pi P_{+}^{a})
\left(-\frac{2m\pi a}{M(2\ell+1)}+\frac{4i\pi r_{+}}{r_{+}-r_{-}}\right)\right.
\nonumber
\\
&\qquad \quad +\left(-\frac{4ma}{M(2\ell+1)}\right)\Big[\Psi(1+\ell)-\Psi(2+2\ell)-\Psi(1+2\ell)+\frac{1}{2}\Psi(1+\ell+2iP^{a}_{+})+\frac{1}{2}\Psi(1+\ell-2iP^{a}_{+})\Big]
\nonumber
\\
&\left.\qquad \quad +\left(\frac{4ir_{+}}{r_{+}-r_{-}}\right)\Big[\Psi(1+\ell+2iP^{a}_{+})-\Psi(1+\ell-2iP^{a}_{+})\Big]\right\}\,,
\label{Kerr_dynamic}
\end{align}
\end{widetext}
where $P_{+}^{a}=-\{am/(r_{+}-r_{-})\}$ and $P_{+}^{\omega}=\{2Mr_{+}\omega/(r_{+}-r_{-})\}$. It is evident that the above expression has both real and imaginary parts for generic values of the rotation parameter $a$, the angular number $\ell$ and the frequency $\omega$, and hence will provide non-trivial scalar tidal responses. Before jumping onto the most general case, let us consider a few specific cases, which we summarize below.

\begin{itemize}

\item \emph{Static scalar response of Schwarzschild BHs} obtained by setting $a\to 0$, $P_{+}=2M\omega\to 0$, $r_{-}=0$, $\hat{\ell}=\ell\in \mathbb{Z}$. This leads to $\lambda_{\ell}^{\rm Sch}(0)=0$. For analytically continued $\ell$, and using \ref{gammatwiceid} from \ref{App:ScalarResponse} we can also express the rescaled response function as,
\begin{align}
\widetilde{\lambda}_{\ell}^{\rm Sch}|_{\rm stat}=\frac{1}{2^{4\ell+2}}\frac{\tan(\pi\ell)\left\{\Gamma(1+\ell)\right\}^{2}}{\Gamma\left(\ell+\frac{1}{2}\right)\Gamma\left(\ell+\frac{3}{2}\right)}~.
\end{align}
If we now consider $\ell$ to be an integer, the above response function identically vanishes, which is consistent with previous findings in the literature \cite{Chia:2020yla, LeTiec:2020bos, Bhatt:2024rpx}. Further, the above result matches with Eq.(3.122) of Ref.~\cite{creci2021tidal-42e} modulo our rescaling of the response function. Thus static TLNs as well as static dissipation numbers of a Schwarzschild BH vanish identically. 

\item \emph{Dynamical scalar response of Schwarzschild BHs} obtained by setting, $a\to0$, $P_{+}=2M\omega$, $r_{-}=0$, $\hat{\ell}=\ell\in \mathbb{Z}$. In that case, there is a non-zero dynamical piece given by
\begin{align}
\label{eq:scaledscalarresponse}
\widetilde{\lambda}_{\ell}^{\rm Sch}|_{\rm dyn}=\frac{2iM\omega\,\Gamma(1+\ell)^{2}}{\Gamma(2+2\ell)\Gamma(1+2\ell)}\prod_{k=1}^{\ell}\left(k^{2}\right)\,.
\end{align}
This result, using \ref{gammatwiceid} from \ref{App:ScalarResponse}, can also be expressed as,
\begin{align}
\widetilde{\lambda}_{\ell}^{\rm Sch}|_{\rm dyn}=\frac{2\pi iM\omega}{2^{4\ell+1}}\frac{\Gamma(\ell+1)^{2}}{\Gamma\left(\ell+\frac{1}{2}\right)\Gamma\left(\ell+\frac{3}{2}\right)}\,,
\label{eq:scaledscalarresponseAlt}
\end{align}
This result matches with Eq.~(3.125) of \cite{creci2021tidal-42e} provided we rescale the result of \ref{eq:scaledscalarresponseAlt} by using the factor introduced in \ref{scaling}. Further, as the response function is purely imaginary, it corresponds entirely to absorption effects and implies that Schwarzschild BHs have vanishing induced non-dissipative tidal deformations at linear order in the frequency. This is consistent with other results as well \cite{chakrabarti2013new-fa4, saketh2023dynamical-b30}. In the non-spinning limit, the dissipation function reduces to
\begin{align}
\widetilde{\nu}_{\ell}=\frac{2M\omega\,\Gamma(1+\ell)^{2}}{\Gamma(2+2\ell)\Gamma(1+2\ell)}\prod_{k=1}^{\ell}\left(k^{2}\right)\,.
\end{align}
For $\ell=2$ this becomes: $\nu_{2}=M\omega/90$, which is consistent with other results in the literature, e.g., those from the scattering-amplitude~\cite{Saketh:2023bul} and from the perturbative~\cite{Chakraborty:2025wvs} approaches.

\item \emph{Stationary scalar responses of Kerr BHs} are obtained by keeping $a\neq 0$, while setting $\omega=0$. Thus we have $P_{+}=P_{+}^{a}=-\{am/(r_{+}-r_{-})\}$, and $\hl=\ell\in \mathbb{Z}$. This yields a non-zero result, given by \ref{Kerr_static}. Using \ref{gammatwiceid} from \ref{App:ScalarResponse} we re-express the rescaled response function as,
\begin{align}
&\widetilde{\lambda}_{\ell m}^{\rm Kerr}|_{\rm stat}=-\frac{iam\pi}{2^{4\ell+1}\left(r_{+}-r_{-}\right)}
\nonumber
\\
&\times \frac{\prod_{k=1}^{\ell}\left[k^{2}+4\left(P_{+}^{a}\right)^{2}\right]}{\Gamma\left(\ell+\frac{1}{2}\right)\Gamma\left(\ell+\frac{3}{2}\right)}\,.
\end{align}
As the response function is imaginary, the stationary TLNs of Kerr BHs vanish identically, while the dissipation function becomes non-zero. From the imaginary part of the response function, we obtain, 
\begin{align}
\widetilde{\nu}_{\ell m}&=-\frac{am\pi}{2^{4\ell+1}\left(r_{+}-r_{-}\right)}
\nonumber
\\
&\quad \times\frac{\prod_{k=1}^{\ell}\left[k^{2}+4\left(P_{+}^{a}\right)^{2}\right]}{\Gamma\left(\ell+\frac{1}{2}\right)\Gamma\left(\ell+\frac{3}{2}\right)}\,.
\end{align}
Therefore, Kerr BHs have a non-zero dissipation number at zeroth order in frequency. This arises from the dragging of space due to rotation near the horizon, where for Kerr BHs one replaces $\omega$ by the co-rotating frequency $\bar{\omega}=\omega-m\Omega_{\rm H}$.

\item \emph{Dynamical scalar response of Kerr BHs} at linear order in the frequency. The coefficients of the linear-in-frequency piece of the conservative tidal response are given by half of the real part of the response function. Given the decomposition of the response function from \ref{eq:decompFtildescalar}, a similar decomposition applies for the conservative piece, with the leading order static piece vanishing. Thus from the dynamical response, we obtain the following non-dissipative tidal response for the scalar case, $\widetilde{k}_{\ell m}=M\omega\widetilde{k}^{\rm dyn}_{\ell m}$, while the dissipative part reads, $\widetilde{\nu}_{\ell m}=\widetilde{\nu}^{\rm stat}_{\ell m}+M\omega \widetilde{\nu}^{\rm dyn}_{\ell m}$ where, each of these coefficients are given by
\begin{widetext}
\begin{align}\label{dynamictlnscalar}
\widetilde{\nu}^{\rm stat}_{\ell m}&=P^{a}_{+}\frac{\Gamma(1+\ell)^{2}}{\Gamma(2+2\ell)\Gamma(1+2\ell)}\prod_{k=1}^{\ell}\left(k^{2}+4\{P_{+}^{a}\}^{2}\right)\,,
\\
\widetilde{k}^{\rm dyn}_{\ell m}&=-m\chi \frac{\pi }{(2\ell+1)}\widetilde{\nu}^{\rm stat}_{\ell m}
\\
\widetilde{\nu}^{\rm dyn}_{\ell m}&=
\widetilde{\nu}^{\rm stat}_{\ell m}\left\{\frac{4 r_{+}}{(r_{+}-r_{-})}\left[\pi\coth(2\pi P_{+}^{a})-\textrm{Im}\Big\{\Psi(1+\ell+2iP^{a}_{+})-\Psi(1+\ell-2iP^{a}_{+})\Big\}\right]\right.
\nonumber
\\
&\qquad -
\left.\frac{4ma}{M(2\ell+1)}\Big[\{\Psi(1+\ell)-\Psi(2+2\ell)-\Psi(1+2\ell)+\frac{1}{2}\textrm{Re}\left\{\Psi(1+\ell+2iP^{a}_{+})+\Psi(1+\ell-2iP^{a}_{+})\right\}\Big]\right\}\,,
\end{align}
\end{widetext}
Here, $\chi\equiv(a/M)$ is the dimensionless spin parameter of the Kerr BH. It is evident that the above expression for the scalar tidal response depends on the frequency linearly through the term $ma\omega$. Thus in the case of axi-symmetric perturbations ($m=0$), the scalar response vanishes identically. This dependence on $a$ is further boasted by the result that $P_{+}^{a}\coth(2\pi P_{+}^{a})\sim 1$, for small rotation.  

\end{itemize}

It is evident that the above expression is not well behaved in the extremal limit. This is because of the explicit $(r_{+}-r_{-})^{-1}$ terms in the response functions. The inverse powers of $(r_{+}-r_{-})$ terms in $P_{+}^{a}$, on the other hand are cancelled by the scaling factor in~\ref{scaling}, which involves $(r_{+}-r_{-})^{2\ell+1}$. This ill-behaved nature of the above dissipation and tidal Love numbers in the extremal limit arises due to the assumption that $P_{+}^{\omega}=(2Mr_{+}\omega/r_{+}-r_{-})\ll 1$, which does not hold in the extremal limit, even if $M\omega \ll 1$. Thus the above results are valid only in the non-extremal case. To rectify this issue we present another alternative form of the response function:
\begin{widetext}
\begin{align}
\lambda_{\ell m}^{\rm extremal}&=i\widetilde{P}_{+}\left(\frac{\pi}{\Gamma(\ell+\frac{1}{2})\Gamma(\ell+\frac{3}{2})}\right)\prod_{j=1}^{\ell}\left[j^{2}\left(\frac{r_{+}-r_{-}}{2M}\right)^{2}+4\widetilde{P}_{+}^{2}\right]\frac{(\ell!)^{2}}{(2\ell+1)!(2\ell)!}\left\{1+i\frac{2\pi am\omega}{2\ell+1}\coth(2\pi P_{+})\right\}
\nonumber
\\
&\qquad \times \left[1-\frac{2am\omega}{2\ell+1}\left\{2\Psi(1+\ell)-2\Psi(1+2\ell)-2\Psi(2+2\ell)
\right\}\right]\,,
\end{align}
\end{widetext}
where $\widetilde{P}_{+}=r_{+}\omega-(am/2M)$. In arriving at the above expression, we have expressed $\Gamma(1+\hat{\ell}\pm 2iP_{+})$ as $\Gamma(1+\ell\pm2iP_{+})$, since in the extremal limit $P_{+}\gg 1$, hence the $\mathcal{O}(M\omega)$ terms inside the Gamma functions can be ignored. Now in the extremal limit, $r_{+}\to r_{-}$, $\widetilde{P}_{+}$ is finite and appears in the expression for the response function as $\widetilde{P}_{+}^{2\ell+1}$, which can be expressed as, $-(am/2M)^{2\ell+1}\{1-(2\ell+1)(2M^{2}\omega/am)\}$. Similarly, in the extremal limit $P_{+}\to \infty$, leading to $\coth(2\pi P_{+})\to 1$. 
Thus the response function in the extremal limit becomes, 
\begin{align}\label{response_ext_scalar}
&\lambda_{\ell m}^{\rm extremal}=-im^{2\ell+1}\left(\frac{\pi}{2\Gamma(\ell+\frac{1}{2})\Gamma(\ell+\frac{3}{2})}\right)
\nonumber
\\
&\times \left(1-(2\ell+1)\frac{2M\omega}{m}\right)\left(1+i\frac{2\pi am\omega}{2\ell+1}\right)
\nonumber
\\
&\Big[1-\frac{2am\omega}{2\ell+1}\big\{2\Psi(1+\ell)-2\Psi(1+2\ell)-2\Psi(2+2\ell)\big\}\Big]\,.
\end{align}

In what follows we will show that this expression is identical to the one obtained from a separate calculation for extremal spins.

To summarize, the scalar Love number vanishes for both Schwarzschild and Kerr BH geometries in the stationary configuration. In the dynamical case, the conservative scalar Love number vanishes for Schwarzschild BHs, but is non-zero for Kerr BHs and scales as $am\omega$. On the other hand, the dissipation numbers are non-zero for Kerr BHs under stationary perturbations and for both Schwarzschild and Kerr BHs under dynamical perturbations. These results are consistent with previous findings in the literature~\cite{tiec2020tidal-db0, chia2020tidal-b97, charalambous2021vanishing-deb}. Note that the same holds for the EFT response functions as well.

\section{Generic spin perturbation of the non-extremal Kerr background}\label{genspin}

In this section, we discuss the spin-$s$ perturbations of a Kerr background due to external tidal fields and the associated dynamical response functions. Owing to the gauge invariance of scalar functions, the gravitational perturbations are best described by the Weyl scalars $\psi_0$ and $\psi_4$, in terms of which the perturbations are separable in the radial and the angular parts. The same approach will work for scalar and electromagnetic perturbations as well and all of them can be expressed in the following manner \cite{Teukolsky:1974yv, 1973ApJ...185..635T}: 
\begin{align}
\label{eq:Psidecomp}
\rho^{s-|s|}\Psi_{-s+|s|}=\sum_{\ell m}\int d\omega \underbrace{e^{-i\omega t}e^{im\phi}\,_{s}S_{\ell m}(\theta)\,_{s}R_{\ell m}(r)}_{\psi_{-s+|s|}}\,,
\end{align}
where $s=\pm 2$ denotes gravitational perturbation, $s=\pm 1$ denotes electromagnetic perturbation and $s=0$ denotes scalar perturbations. We have introduced the quantity $\psi_{-s+|s|}$, which is the frequency-space Weyl scalar. The radial part ${}_{s}R_{\ell m}(r)$ satisfies the following differential equation \cite{Teukolsky:1974yv, 1973ApJ...185..635T}: 
\begin{align}\label{radialgen}
\Delta \dfrac{d^{2}_{s}R_{\ell m}}{dr^{2}}&+(s+1)\Delta'\dfrac{d_{s}R_{\ell m}}{dr}
\nonumber
\\
&+\left[\frac{K^{2}-is\Delta'K}{\Delta}+4irs\omega-\lambda\right]\,_{s}R_{\ell m}=0~.
\end{align}
Here, $\lambda=E_{\ell m}-2am\omega-s(s+1)+\mathcal{O}(M^{2}\omega^{2})$ is the separation constant between radial and angular equations, and $K\equiv (r^{2}+a^{2})\omega-am$. The angular equation, on the other hand, can be expressed as \cite{Teukolsky:1974yv, 1973ApJ...185..635T, Press:1973zz}, 
\begin{align}
&\frac{1}{\sin \theta}\dfrac{d}{d\theta}\left(\sin \theta \dfrac{d_{s}S_{\ell m}}{d\theta}\right)
+\Big(a^{2}\omega^{2}\cos^{2}\theta-\frac{m^{2}}{\sin^{2}\theta}
\nonumber
\\
&-2a\omega s\cos \theta-\frac{2ms\cos \theta}{\sin^{2}\theta}-\frac{s^{2}}{\sin^{2}\theta}+E_{\ell m}\Big)\,_{s}S_{\ell m}=0~,
\end{align}
where $E_{\ell m}=\ell(\ell+1)-2ma\omega\{s^{2}/\ell(\ell+1)\}+\mathcal{O}(M^{2}\omega^{2})$ and the solutions of the above differential equation for the angular function are the spin-weighted spheroidal harmonics. These spheroidal harmonics can be expressed in terms of spherical harmonics, which we have already discussed in \ref{sec:EFT}.
\subsection{The near-zone solution}
\label{subsec:NZgeneric}

In this section, we derive the solution of the radial equation in the \nz and apply the appropriate boundary condition at the horizon. Introducing the radial coordinate $z$ defined in~\ref{eq:zdef}
it follows that, $\Delta=(r_{+}-r_{-})^{2}z(1+z)$, and $\Delta'=(d\Delta/dr)=(r_{+}-r_{-})(1+2z)$. On substituting the above results in~\ref{radialgen}, and ignoring terms $\mathcal{O}(M\omega z)$, we obtain the following radial differential equation, 
\begin{align}
&z(1+z)\dfrac{d^{2}_{s}R_{\ell m}}{dz^{2}}+(s+1)(1+2z)\dfrac{d_{s}R_{\ell m}}{dz}
\nonumber
\\
&+\left[\frac{P_{+}^{2}}{z(1+z)}-\frac{isP_{+}(1+2z)}{z(1+z)}-\lambda+4i\omega sr_{+}\right]\,_{s}R_{\ell m}=0\,.
\end{align}
Use of the near-zone approximation $M\omega z\ll 1$, allows us to write $K\approx (r_{+}^{2}+a^{2})\omega-am=2M\omega r_{+}-am$. Further, the quantity $P_{+}$ appearing in the radial equation is defined as $P_{+}\equiv (2Mr_{+}\omega-am)/(r_{+}-r_{-})$. We further introduce a quantity $\hat{\ell}_{s}$, which is defined as $\lambda-4i\omega sr_{+}\equiv \hat{\ell}_{s}(\hat{\ell}_{s}+1)-s(s+1)$, such that $E_{\ell m}=\lambda+2am\omega+s(s+1)+\mathcal{O}(a^{2}\omega^{2})=\hl_{s}(\hl_{s}+1)+2am\omega+4i\omega sr_{+}+\mathcal{O}(a^{2}\omega^{2})$. From the above expression, note that $\hl_{s=0}=\hl$, the quantity that we have used to describe the near-zone physics in the scalar case. Therefore, it follows that, 
\begin{align}
&\hat{\ell}_{s}(\hat{\ell}_{s}+1)=E_{\ell m}-2am\omega-4i\omega sr_{+}
\nonumber
\\
&=\ell(\ell+1)+M\omega\left(-\frac{4isr_{+}}{M}-\frac{2am}{M}-\frac{2ams^{2}}{M\ell(\ell+1)}\right)~,
\end{align}
implying, $\hat{\ell}_{s}=\ell+\mathcal{O}(M\omega)$. With this parametrization, the above equation can be solved exactly, yielding, 
\begin{widetext}
\begin{align}\label{radial_near}
\,_{s}R^{\rm (near)}_{\ell m}&=C_{1}\left(\frac{z}{1+z}\right)^{iP_{+}}\,_{2}F_{1}(-\hl_{s}+s,1+\hl_{s}+s,1+s+2iP_{+},-z)
\nonumber
\\
&+C_{2}\,z^{-iP_{+}-s}(1+z)^{-iP_{+}}\,_{2}F_{1}(-\hl_{s}-2iP_{+},1+\hl_{s}-2iP_{+};1-s-2iP_{+};-z)~.
\end{align}
\end{widetext}
The above solution involves two arbitrary constants $C_{1}$ and $C_{2}$, which need to be fixed by an appropriate boundary condition at the horizon. This is discussed in the next section. 

\subsubsection{Boundary condition at the horizon}

As we are studying perturbations of a BH spacetime, whose distinguishing feature is the presence of an event horizon, the appropriate boundary conditions are that there are only purely ingoing modes at the BH horizon. Considering the $z\to 0$ limit of \ref{radial_near}, we obtain, 
\begin{align}
\,_{s}R_{\ell m}^{\rm (near)}&=C_{1}z^{iP_{+}}+C_{2}z^{-iP_{+}-s}
\nonumber
\\
&\simeq C_{1}e^{i\bar{\omega}r_{*}}+\frac{C_{2}}{\Delta^{s}}e^{-i\bar{\omega}r_{*}}\,.
\end{align}
When combining this near-horizon radial behavior with the time-dependence in the decomposition of the Weyl scalars~\ref{eq:Psidecomp}, we see that the term $C_{1}$ describes purely outgoing waves near the horizon, while $C_{2}$ corresponds to purely ingoing waves \cite{Teukolsky:1974yv}. Thus, it follows that $C_{1}=0$ and the \nz solution becomes,
\begin{align}\label{radial_near_final}
&\,_{s}R_{\ell m}^{\rm (near)}=C_{\rm H}^{\rm in}z^{-iP_{+}-s}(1+z)^{-iP_{+}}
\nonumber
\\
&\times \,_{2}F_{1}(-\hl_{s}-2iP_{+},1+\hl_{s}-2iP_{+};1-s-2iP_{+};-z)~,
\end{align}
where $C_{\rm H}^{\rm in}$ is an overall arbitrary constant. This provides the \nz radial perturbation, consistent with the ingoing boundary condition at the horizon. 

\subsubsection{Solution in the intermediate zone}

Having derived the \nz solution, with appropriate boundary condition at the horizon, let us consider its asymptotic limit, in order to determine the solution to the radial equation in the intermediate zone. In the $z\to \infty$ limit, i.e., in the intermediate zone, the above solution, presented in \ref{radial_near_final}, yields,
\begin{align}\label{spinskerrinter}
&\,_{s}R_{\ell m\,\textrm{(int)}}^{\rm (near)}=\frac{C_{\rm H}^{\rm in}\,\Gamma(1-s-2iP_{+})\Gamma(1+2\hl_{s})}{\Gamma(1+\hl_{s}-2iP_{+})\Gamma(1+\hl_{s}-s)}\frac{r^{\hl_{s}-s}}{(r_{+}-r_{-})^{\hl_{s}-s}}
\nonumber
\\
&+\frac{C_{\rm H}^{\rm in}\,\Gamma(1-s-2iP_{+})\Gamma(-1-2\hl_{s})}{\Gamma(-\hl_{s}-2iP_{+})\Gamma(-s-\hl_{s})}\frac{r^{-1-\hl_{s}-s}}{(r_{+}-r_{-})^{-1-\hl_{s}-s}}\,.
\end{align}
Note that the above solution in the limit of $s\to 0$, exactly matches with \ref{radial_near_far}, derived in the context of scalar perturbations. Thus, the \nz solution has a growing mode $r^{\hat{\ell}_{s}-s}$ as well as a decaying mode $r^{-\hl_{s}-1-s}$, and the coefficients of these modes must be matched with the corresponding \fz solution. Thus, we next consider the \fz radial equation. 

\subsection{The solution in the far-zone}
\label{subsec:FZgeneric}

Consider now the radial Teukolsky equation for generic spin in the \fz 
\begin{align}
&r^{2}\dfrac{d^{2}_{s}R_{\ell m}}{dr^{2}}+2(s+1)r\dfrac{d_{s}R_{\ell m}}{dr}
\nonumber
\\
&+\left[\omega^{2}r^{2}+2is\omega r-2am\omega-2isM\omega-\lambda\right]\,_{s}R_{\ell m}=0~,
\end{align}
where $\lambda$ is defined below \ref{radialgen}. At this stage, we introduce $\tilde{\ell}$, which is defined as, $\tl(\tl+1)=\lambda+2am\omega+2isM\omega+s(s+1)$, such that we obtain,
\begin{align}\label{gens_ang}
\tl&(\tl+1)=E_{\ell m}+2isM\omega+\mathcal{O}(M^{2}\omega^{2})
\nonumber
\\
&=\ell(\ell+1)+M\omega\left(2is-\frac{2ams^{2}}{M\ell(\ell+1)}\right)+\mathcal{O}(M^{2}\omega^{2})~,
\end{align}
such that $\tl=\ell+\mathcal{O}(M\omega)$, as well as for $s=0$, $\tl=\ell+\mathcal{O}(M^{2}\omega^{2})$, which is consistent with our results derived in \ref{scalar_scattering}. Given the above definition of $\tl$, the solution of the radial Teukolsky equation in the \fz becomes, 
\begin{align}
\,_{s}R_{\ell m}^{\rm (far)}&=e^{-i\omega r}r^{\tl-s}\Big[B_{1}U(1+\tl-s,2+2\tl;2i\omega r)
\nonumber
\\
&+\widetilde{B}_{2}L(-1+s-\tl,1+2\tl;2i\omega r)\Big]~.
\end{align}
Here $U(a,b;x)$ is the confluent hypergeometric function, $L(a,b;x)$ is the Laguerre polynomial, and $B_1,\, \widetilde{B}_2$ are integration constants. It is possible to express the Laguerre polynomial in terms of the other confluent hypergeometric function $M(a,b;x)$, yielding the following expression for the radial function in the far-zone, 
\begin{align}\label{radial_far_old}
\,_{s}R_{\ell m}^{\rm (far)}&=e^{-i\omega r}r^{\tl-s}\Big[B_{1}U(1+\tl-s,2+2\tl;2i\omega r)
\nonumber
\\
&+B_{2}M(1-s+\tl,2+2\tl;2i\omega r)\Big]~.
\end{align}
The confluent hypergeometric function $U(a,b;x)$ can be expressed in terms of $M(a,b;x)$ using \ref{UasM} in \ref{AppA}, and hence we obtain the above far-zone radial function solely in terms of $M(a,b;z)$ as,
\begin{align}
&\,_{s}R_{\ell m}^{\rm (far)}=e^{-i\omega r}r^{\tl-s}
\nonumber
\\
&\times\Bigg[\Big\{B_{2}+B_{1}\frac{\Gamma(-1-2\tl)}{\Gamma(-\tl-s)}\Big\}M(1+\tl-s,2+2\tl;2i\omega r)
\nonumber
\\
&+B_{1}\frac{\Gamma(1+2\tl)}{\Gamma(1+\tl-s)}(2i\omega r)^{-1-2\tl}M(-s-\tl,-2\tl;2i\omega r)\Bigg]\,.
\end{align}
Anticipating that we will ultimately need to identify the regular and irregular pieces of the radial solution, we change the integration constants to two new constants $B^{\infty}_{\ell m,\,\rm reg}$ and $B^{\infty}_{\ell m,\,\rm irreg}$ defined as linear combinations of $B_{1}$ and $B_{2}$: 
\begin{align}
B_{2}&+B_{1}\frac{\Gamma(-1-2\tl)}{\Gamma(-\tl-s)}\equiv\frac{B^{\infty}_{\ell m, \,\rm reg}}{\Gamma(\bar{\nu}+1)}\left(\frac{\omega}{2}\right)^{\bar{\nu}}~,
\label{gens_regn}
\\
B_{1}&(2i\omega)^{-1-2\tl}\frac{\Gamma(1+2\tl)}{\Gamma(1+\tl-s)}\equiv-\left(\frac{B^{\infty}_{\ell m,\,\rm irreg}}{\pi}\right)\Gamma\left(\bar{\mu}\right)\left(\frac{\omega}{2}\right)^{-\bar{\mu}}~,
\label{gens_irregn}
\end{align}
where we have introduced the two constants $\bar{\nu}=\tl-|s|+(1/2)$ and $\bar{\mu}=\tl+|s|+(1/2)$, such that $\bar{\nu}+\bar{\mu}=2\tl+1$.

In terms of these newly defined constants, $B^{\infty}_{\ell m,\,\rm reg}$ and $B^{\infty}_{\ell m,\,\rm irreg}$, the radial solution for generic spin perturbation in the \fz reads,
\begin{align}\label{radial_far}
&\,_{s}R_{\ell m}^{\rm (far)}=e^{-i\omega r}r^{\tl-s}
\nonumber
\\
&\times \Bigg[\frac{B^{\infty}_{\ell m,\,\rm reg}}{\Gamma(\bar{\nu}+1)}\left(\frac{\omega}{2}\right)^{\bar{\nu}}M(1+\tl-s,2+2\tl;2i\omega r)
\nonumber
\\
&\quad -\frac{B^{\infty}_{\ell m,\,\rm irreg}}{\pi}\frac{\Gamma(\bar{\mu})}{r^{1+2\tl}}\left(\frac{2}{\omega}\right)^{\bar{\mu}}M(-s-\tl,-2\tl;2i\omega r)\Bigg]\,,
\end{align}
To determine the integration constants explicitly in terms of properties of the perturbed BH, we  first take the limit of the \fz solution to the intermediate zone, where we match it with the \nz solution. We also need the limit to the asymptotic region to match  with the EFT. 

Note that in the $s\to 0$ limit, $\bar{\nu}=\ell+(1/2)=\bar{\mu}$. Further, we can use \ref{zeroM} in \ref{AppA} in order to express the far-zone solution in terms of Bessel functions of order $\ell+(1/2)$. One can show that such a computation directly produces \ref{scalar_far}, the \fz solution for scalar perturbations. Thus our results are consistent with the scalar limit.
\subsubsection{Solution in the intermediate zone}

Having derived the general solution to the radial equation in the far-zone, let us consider the $\omega r\ll 1$ limit and determine the solution in the intermediate zone. For this purpose, we can use the results presented in \ref{limitzeroMU} of \ref{AppA}, such that \ref{radial_far} reduces to, 
\begin{align}
\,_{s}R_{\ell m\,\textrm{(int)}}^{\rm (far)}&=\frac{B^{\infty}_{\ell m,\,\rm reg}}{\Gamma(\bar{\nu}+1)}\left(\frac{\omega}{2}\right)^{\bar{\nu}}r^{\tl-s}
\nonumber
\\
&-\frac{B^{\infty}_{\ell m,\,\rm irreg}}{\pi}\Gamma(\bar{\mu})\left(\frac{\omega}{2}\right)^{-\bar{\mu}}r^{-1-\tl-s}\,.
\end{align}
Note that, alike the \nz solution, the above \fz solution also has a growing mode $(\sim r^{\tl-s})$ and a decaying mode $(\sim r^{-1-\tl-s})$ in the intermediate zone. Expressing, $\tl=\hl_{s}+\epsilon_{\rm s}M\omega$, with 
\begin{equation}
\epsilon_s=\left(\frac{1}{2\ell+1}\right)\left[2is\left(1-\frac{2r_{+}}{M}\right)+\frac{2am}{M}\right]\,,
\end{equation}
the above intermediate zone solution can be expressed as,
\begin{align}\label{radial_far_intn}
R_{\ell m\,\textrm{(int)}}^{\rm (far)}&=\frac{B^{\infty}_{\ell m,\,\rm reg}}{\Gamma(\bar{\nu}+1)}\left(\frac{\omega}{2}\right)^{\bar{\nu}}r^{\hl-s}(r_{+}-r_{-})^{\tl-\hl}
\nonumber
\\
&\qquad \times \left[1+\epsilon_{\rm s}M\omega \ln \left(\frac{r}{r_{+}-r_{-}}\right)\right]
\nonumber
\\
&-\frac{B^{\infty}_{\ell m,\,\rm irreg}}{\pi}\Gamma(\bar{\mu})\left(\frac{2}{\omega}\right)^{\bar{\mu}}r^{-1-\hl-s}(r_{+}-r_{-})^{-\tl+\hl}
\nonumber
\\
&\qquad \times \left[1-\epsilon_{\rm s}M\omega \ln \left(\frac{r}{r_{+}-r_{-}}\right)\right]~.
\end{align}
Note that the power of the growing mode and the decaying mode in the intermediate zone is identical for both \fz and \nz solutions and hence the respective coefficients can be properly matched to order $O(M\omega)$. The appearance of the $\ln r$ term is not new, as it had appeared in the case of scalar perturbations as well, which can be seen by noticing that $\epsilon_{s=0}\neq 0$. As usual, after matching between far and \nz the arbitrary constants cannot be a function of radius, and hence it seems appropriate to set $r\sim \omega^{-1}$. Thus, in the general spin case also there is a correction of $\mathcal{O}(\omega \ln \omega)$. As we have already emphasised earlier this term can be a tail effect, with no connection to the tidal field, and this is a term of higher order than what we have considered here. Hence, we will ignore the logarithmic piece and, hopefully, this can be better understood in the future using the MST method \cite{sasaki2003analytic-827}.

\subsubsection{Solution in the asymptotic region}

In order to find the matching of the perturbation theory approach with the EFT techniques we need to consider the asymptotic limit of the \fz solution. For this purpose, we will first re-express the \fz solution in terms of Bessel functions and then we will take the asymptotic limit. Therefore, to obtain the \fz solution in terms of Bessel functions one can use the identities within \ref{AppA}, presented in \ref{Mtwice1} and \ref{Mtwice2}. Using these relations, we obtain, for generic spin, 
\begin{align}\label{MIrelation1}
M(1+\tl-s,2+2\tl,2i\omega r)&=\Gamma(\bar{\nu}+1)e^{i\omega r}\left(\frac{i\omega r}{2}\right)^{-\bar{\nu}}
\nonumber
\\
&\times\sum_{k=0}^{2|s|}g_{k}\,I_{\bar{\nu}+k}(i\omega r)\,,
\end{align}
where, $I_{\alpha}(x)$ is the modified Bessel function, $\bar{\nu}=\tl-|s|+(1/2)$ has already been introduced earlier and the coefficient $g_{k}$ has the following expression:
\begin{align}
g_{k}&\equiv \left(\frac{s}{|s|}\right)^{k}\frac{\left(\bar{\nu}+k\right)}{\bar{\nu}}\frac{\Gamma(2\bar{\nu}+k)}{\Gamma(2\bar{\nu})}
\nonumber
\\
&\times\frac{\Gamma(2\bar{\nu}+1+2|s|)}{\Gamma(2\bar{\nu}+1+2|s|+k)}\frac{(-2|s|)_{k}}{k!}~,
\label{eq:gk1}
\end{align}
where $(a)_{k}$ is the Pochhammer symbol. Along identical lines, the other confluent hypergeometric function appearing in \ref{radial_far} can also be written as,
\begin{align}\label{MIrelation2}
M(-s-\tl,-2\tl;2i\omega r)&=\Gamma(1-\bar{\mu})e^{i\omega r}\left(\frac{i\omega r}{2}\right)^{\bar{\mu}}
\nonumber
\\
&\times \sum_{k=0}^{2|s|}\widetilde{g}_{k}I_{-\bar{\mu}+k}(i\omega r)\,,
\end{align}
where $\bar{\mu}=\tl+|s|+(1/2)$ has also been defined earlier, and the coefficient $\widetilde{g}_{k}$ reads, 
\begin{align}
\widetilde{g}_{k}&=\left(\frac{s}{|s|}\right)^{k}\frac{\left(-\bar{\mu}+k\right)}{-\bar{\mu}}\frac{\Gamma(-2\bar{\mu}+k)}{\Gamma(-2\bar{\mu})}
\nonumber
\\
&\times\frac{\Gamma(-2\bar{\mu}+1+2|s|)}{\Gamma(-2\bar{\mu}+1+2|s|+k)}\frac{(-2|s|)_{k}}{k!}
\nonumber
\\
&=\left(\frac{s}{|s|}\right)^{k}\frac{\left(\bar{\mu}-k\right)}{\bar{\mu}}\frac{\Gamma(1+2\bar{\mu})}{\Gamma(1+2\bar{\mu}-k)}\frac{\Gamma(2\bar{\mu}-2|s|-k)}{\Gamma(2\bar{\mu}-2|s|)}~.
\label{eq:gk2}
\end{align}
Therefore, using \ref{MIrelation1} and \ref{MIrelation2} in \ref{radial_far}, the far zone solution in the asymptotic regime takes the form, 
\begin{align}\label{radial_far_asymp1}
&\,_{s}R_{\ell m\,(\infty)}^{\rm (far)}=i^{-\bar{\nu}}B^{\infty}_{\ell m,\,\rm reg}r^{|s|-s-\frac{1}{2}}\,\sum_{k=0}^{2|s|}g_{k}\,I_{\bar{\nu}+k}(i\omega r)
\nonumber
\\
&-i^{\bar{\mu}}\frac{B^{\infty}_{\ell m,\,\rm irreg}}{\pi}\Gamma(\bar{\mu})\Gamma(1-\bar{\mu})\,r^{|s|-s-\frac{1}{2}}\,\sum_{k=0}^{2|s|}\widetilde{g}_{k}I_{-\bar{\mu}+k}(i\omega r)\,.
\end{align}
We next use the following relations between the modified Bessel functions and the Bessel functions of the first and second kinds given by $I_{\bar{\nu}+k}(i\omega r)=i^{k+\bar{\nu}}J_{\bar{\nu}+k}(\omega r)$ and $I_{-\bar{\mu}+k}(i\omega r)=i^{-\bar{\mu}+k}J_{-\bar{\mu}+k}(\omega r)$, and the identity $\Gamma(\bar{\mu})\Gamma(1-\bar{\mu})=\{\pi/\sin(\pi \bar{\mu})\}$. Upon using these identities, the \fz solution can be expressed in terms of Bessel functions as, 
\begin{align}\label{radial_far_asymp2}
\,_{s}R_{\ell m}^{\rm (far)}&=\frac{B^{\infty}_{\ell m,\,\rm reg}}{r^{-|s|+s+\frac{1}{2}}}\,\sum_{k=0}^{2|s|}i^{k}g_{k}\,J_{\bar{\nu}+k}(\omega r)
\nonumber
\\
&-\frac{B^{\infty}_{\ell m,\,\rm irreg}}{r^{-|s|+s+\frac{1}{2}}}\,\sum_{k=0}^{2|s|}i^{k}\widetilde{g}_{k}\frac{J_{-\bar{\mu}+k}(\omega r)}{\sin(\pi \bar{\mu})}\,.
\end{align}
This explains the labelling of the constants $B^{\infty}_{\ell m,\,\rm reg}$ and $B^{\infty}_{\ell m,\,\rm irreg}$, as these are the coefficients of $J_{\bar{\nu}+k}(\omega r)$ (regular in the $\omega r \to 0$ limit), and $J_{-\bar{\mu}+k}(\omega r)$ (irregular in the $\omega r\to 0$ limit), respectively. In the $s\to 0$ limit, only $k=0$ contributes to the sum, and $g_{0}=1=\widetilde{g}_{0}$. Moreover, in the case of $s\to 0$, $J_{\bar{\nu}+k}(\omega r)\to J_{\ell+1/2}(\omega r)$, as well as $J_{-\bar{\mu}+k}(\omega r)/\sin (\pi \bar{\mu})\to Y_{\ell+1/2}(\omega r)$. Thus it follows that we get back the results of \ref{scalar_far} and our results are consistent with the scalar perturbation derived earlier. 

Having expressed the \fz solution in terms of Bessel function, we will now consider the large $r$ limit. Expressing $\tl=\ell+\epsilon_{\ell}M\omega$, where, 
\begin{align}
\label{eq:epsilonldef}
\epsilon_{\ell}=\left(\frac{1}{2\ell+1}\right)\left(2is-\frac{2ams^{2}}{M\ell(\ell+1)}\right)\,,
\end{align}
we obtain the following asymptotic expansion of the Bessel functions:
\begin{align}
\label{eq:BesselJtosin}
&\medmath{J_{\bar{\nu}+k}(\omega r)=\sqrt{\frac{2}{\pi \omega r}}\Big\{\sin\left[\omega r-\frac{\pi}{2}(\ell-|s|)\right]\cos\left[\frac{\pi}{2}\left(k+\epsilon_{\ell}M\omega\right)\right]}
\nonumber
\\
&-\cos\left[\omega r-\frac{\pi}{2}(\ell-|s|)\right]\sin\left[\frac{\pi}{2}\left(k+\epsilon_{\ell}M\omega\right)\right] \Big\}\,,
\\
&\label{eq:BesselYtosin}\medmath{J_{-\bar{\mu}+k}(\omega r)=\sqrt{\frac{2}{\pi \omega r}}\Big\{\cos\left[\omega r-\frac{\pi}{2}(\ell-|s|)\right]\cos\left[\frac{\pi}{2}\left(k+\epsilon_{\ell}M\omega\right)\right]}
\nonumber
\\
&-\sin\left[\omega r-\frac{\pi}{2}(\ell-|s|)\right]\sin\left[\frac{\pi}{2}\left(k+\epsilon_{\ell}M\omega\right)\right]\Big\}\times (-1)^{\ell+k}\,.
\end{align}
Therefore, the asymptotic behaviour of the far-zone radial function can also be expressed in terms of sinusoidal functions. This is useful in the context of EFT in order to deal with the issue of mode mixing, as discussed below. Substituting the asymptotic expansions~\ref{eq:BesselJtosin} and~\ref{eq:BesselYtosin} into~\ref{radial_far_asymp2} leads to the following asymptotic behaviour of the radial function, 
\begin{widetext}
\begin{align}\label{radial_far_asymp}
\medmath{\,_{s}R_{\ell m\,(\infty)}^{\rm (far)}}&=\medmath{\sqrt{\frac{2}{\pi \omega}}\frac{\sin\left[\omega r-\frac{\pi}{2}(\ell-|s|)\right]}{r^{-|s|+s+1}}\,\Bigg\{B^{\infty}_{\ell m,\,\rm reg}\sum_{k=0}^{2|s|}i^{k}g_{k}\cos\left[\frac{\pi}{2}\left(k+\epsilon_{\ell}M\omega\right)\right]
+(-1)^{|s|}B^{\infty}_{\ell m,\,\rm irreg}\,\sum_{k=0}^{2|s|}(-i)^{k}\widetilde{g}_{k}\sin\left[\frac{\pi}{2}\left(k-\epsilon_{\ell}M\omega\right)\right]\Bigg\}}
\nonumber
\\
&\medmath{-\sqrt{\frac{2}{\pi \omega}}\frac{\cos\left[\omega r-\frac{\pi}{2}(\ell-|s|)\right]}{r^{-|s|+s+1}}\,\Bigg\{B^{\infty}_{\ell m,\,\rm reg}\sum_{k=0}^{2|s|}i^{k}g_{k}\sin\left[\frac{\pi}{2}\left(k+\epsilon_{\ell}M\omega\right)\right]+(-1)^{|s|}B^{\infty}_{\ell m,\,\rm irreg}\,\sum_{k=0}^{2|s|}(-i)^{k} \widetilde{g}_{k}\cos\left[\frac{\pi}{2}\left(k-\epsilon_{\ell}M\omega\right)\right]\Bigg\}\,.}
\end{align}
In the limit $s\to 0$, the summations in the above expression no longer exists and $g_{k}(s=0)=g_{0}=1=\widetilde{g}_{k}(s=0)=\widetilde{g}_{0}$. Further, as $\epsilon_{\ell}\to 0$ for scalar perturbation, as well as $\cos[\pi(\tl+|s|)]\to(-1)^{\ell}$, it follows that the asymptotic radial function reduces to the one presented in \ref{asympradial}, thus verifying that the $s\to 0$ limit correctly recovers the results of the previous section for scalar perturbation of Kerr BHs. 

\subsection{Matching in the intermediate zone}
\label{subsec:BHPTmatching}

The limit of the \fz solution in the intermediate zone, given by \ref{radial_far_intn}, can be matched with the limit of the near-zone solution in the intermediate zone, presented in \ref{spinskerrinter}. This yields the following relations between the coefficients of the growing and decaying modes in the near and the far-zone solutions: 
\begin{align}
&\frac{B^{\infty}_{\ell m,\,\rm reg}}{\Gamma(\bar{\nu}+1)}\left(\frac{\omega}{2}\right)^{\bar{\nu}}=\frac{C_{\rm H}^{\rm in}}{(r_{+}-r_{-})^{\tl-s}}\left(\frac{\Gamma(1-s-2iP_{+})\Gamma(1+2\hl_{s})}{\Gamma(1+\hl_{s}-2iP_{+})\Gamma(1+\hl_{s}-s)}\right)\,,
\\
&\frac{B^{\infty}_{\ell m,\,\rm irreg}}{\pi}\Gamma(\bar{\mu})\left(\frac{2}{\omega}\right)^{\bar{\mu}}=-\frac{C_{\rm H}^{\rm in}}{(r_{+}-r_{-})^{-1-\tl-s}}\left(\frac{\Gamma(1-s-2iP_{+})\Gamma(-1-2\hl_{s})}{\Gamma(-\hl_{s}-2iP_{+})\Gamma(-s-\hl_{s})}\right)~.
\end{align}
From the \fz solution~\ref{radial_far_intn}, we note that $D^{\infty}_{\ell m,\,\rm irreg}$ is the coefficient of the decaying mode in $r$ and $D^{\infty}_{\ell m,\,\rm reg}$ is the coefficient of the growing mode in $r$, and hence the ratio of the decaying to the growing mode must be related to the response function. Thus we explicitly state the corresponding ratio between the growing and the decaying mode:
\begin{align}\label{gens_ration}
\frac{B^{\infty}_{\ell m,\,\rm irreg}}{B^{\infty}_{\ell m,\,\rm reg}}&=(-1)^{s+1}\left[\frac{\omega(r_{+}-r_{-})}{2}\right]^{2\tl+1}\frac{\sin[\pi\hat{\ell}_{s}+2i\pi P_{+}]}{2\cos(\pi \hl_{s})}
\nonumber
\\
&\times \left(\frac{\Gamma(1+\hl_{s}-2iP_{+})\Gamma(1+\hl_{s}-s)\Gamma(1+\hl_{s}+s)\Gamma(1+\hl_{s}+2iP_{+})}{\Gamma(2+2\hl_{s})\Gamma(1+2\hl_{s})\Gamma(\bar{\mu})\Gamma(1+\bar{\nu})}\right)\,.
\end{align}
This provides the desired relation between the regular and the irregular piece for generic spin perturbations arising from BH perturbation theory alone. In the $s\to 0$ limit, we obtain $\tl\to \ell$, $\hat{\ell}_{s}\to \hat{\ell}$ and $\bar{\nu}=\bar{\mu}=\ell+(1/2)$, such that the right hand side of \ref{gens_ration} reproduces \ref{matching_scalar}. We now proceed to discuss the connection of the asymptotic limit of the \fz solution to the EFT.
\end{widetext}
\subsection{Matching with EFT}
\label{subsec:EFTgenerics}

We now want to match the EFT radial function in \ref{eq:psi4} with the asymptotic expression of the BH perturbation theory radial function in the far-zone in~\ref{radial_far_asymp}, for the gravitational case ($s=-2$). The electromagnetic case follows in a similar fashion and is thus not discussed in detail here. In order to match BH perturbation theory and EFT we note that \ref{radial_far_asymp} can also be expressed as,
\begin{eqnarray}\label{Match_Grav_01}
{}_{-2}R^{\rm (far)}_{\ell m(\infty)} &=& i^{-\ell}\omega^2\frac{e^{-i\omega (t-r)}}{\omega r} \left(B_{\ell m,\rm reg}^{\infty}\frac{32i\ell(\ell-1)}{(3+8\ell+4\ell^2)\omega}\right.\nonumber\\
&&\left.+B_{\ell m,\rm irreg}^{\infty}\frac{32 \ell(\ell+1)}{(3-8\ell+4\ell^2)\omega}\right)\,.
\end{eqnarray}
The regular and irregular pieces  match directly with those in \ref{eq:psi4} after expanding spin weighted ($s=-2$) spheroidal harmonics into spin weighted spherical harmonics.

To proceed further, it is convenient to redefine the integration constants in \ref{Match_Grav_01} as,
\begin{equation}
\begin{split}    
\bar{B}^{\infty}_{\ell m, \rm reg} &= B_{\ell m,\rm reg}^{\infty}\frac{64\ell(\ell-1)}{(3+8\ell+4\ell^2)\omega}\,, 
\\
\bar{B}^{\infty}_{\ell m, \rm irreg} &= - B_{\ell m,\rm irreg}^{\infty}\frac{64\ell(\ell+1)}{(3-8\ell+4\ell^2)\omega}\,.
\end{split}
\end{equation}
Then, we can write 
\begin{equation}
\begin{split}
\psi_4 =(-1)^{\ell} \omega^2 \frac{e^{-i\omega(t-r)}}{\omega r}\sum_{\ell=2}^{\infty} i^{\ell-2}\bar{B}^{\ell m}_{\rm out}{}_{-2}S_{\ell m}(\theta,\phi)\,,
\end{split}
\end{equation}
where $\bar{B}^{\ell m}_{\rm out}=(1/2)(\bar{B}^{\ell m}_{\rm irreg}-i\bar{B}^{\ell m}_{\rm reg})$. The coefficients can now be directly identified by comparing with \ref{eq:psi4}. This yields
\begin{equation}
\begin{split}
D_{\ell m}&=(-1)^{\ell}\bar{B}_{\ell m}\,,
\end{split}
\label{eq:gravity_reverse}
\end{equation}
which is valid for both regular and irregular coefficients. One can then follow the analysis in \ref{sec:eft_grav_disc} to compute the tidal-response coefficients in EFT which are a priori defined in spherical harmonic basis. Here, we will restrict to computing the tidal response coefficients in the spheroidal harmonic basis.

\subsection{Tidal response function}\label{subsec:tidalresponsegeneric}

The determination of the dynamical tidal response function of a Kerr BH under generic spin perturbation follows an identical route to that adopted in the scalar case. The regular solution encodes the tidal field, while the irregular solution, which is sourced by a delta function at the origin, corresponds to the quadrupole moment. Using the results obtained by matching with the EFT, provided in the previous section, BH perturbation theory tidal response function reads, 
\begin{widetext}

\begin{align}\label{dyn_resp_fn_01}
{}_{-2}\mathcal{\lambda}_{\ell m}&=(M\omega)^{-2\ell-1}\frac{D^{\ell m}_{\rm irreg}}{D^{\ell m}_{\rm reg}}=(M\omega)^{-2\ell-1}\frac{\bar{B}^{\ell m}_{\rm irreg}}{\bar{B}^{\ell m}_{\rm reg}}=-(M\omega)^{-2\ell-1}\frac{(\ell+1)}{(\ell-1)}\left(\frac{3+8\ell+4\ell^{2}}{3-8\ell+4\ell^{2}}\right)\frac{B^{\ell m}_{\rm irreg}}{B^{\ell m}_{\rm reg}}
\nonumber
\\
&=\left(\frac{r_{+}-r_{-}}{2M}\right)^{2\ell+1}\left[\frac{\omega(r_{+}-r_{-})}{2}\right]^{2(\tl-\ell)}\frac{(\ell+1)}{(\ell-1)}\left(\frac{3+8\ell+4\ell^{2}}{3-8\ell+4\ell^{2}}\right)\frac{\sin[\pi\hat{\ell}_{s}+2i\pi P_{+}]}{2\cos(\pi \hl_{s})}
\nonumber
\\
&\times \left(\frac{\Gamma(1+\hl_{s}-2iP_{+})\Gamma(1+\hl_{s}-s)\Gamma(1+\hl_{s}+s)\Gamma(1+\hl_{s}+2iP_{+})}{\Gamma(2+2\hl_{s})\Gamma(1+2\hl_{s})\Gamma(\bar{\mu})\Gamma(1+\bar{\nu})}\right)\,.
\end{align}
\end{widetext}
Note that $\{\omega(r_{+}-r_{-})\}^{2(\tl-\ell)}=1+2\epsilon_{\ell}M\omega\ln \{\omega(r_{+}-r_{-})\}\simeq 1$, as we are ignoring terms of $\mathcal{O}(\omega \ln \omega)$. Given the above, we first consider a rescaled response function ${}_{s}\widetilde{\mathcal{F}}_{\ell m}$, related to the original response function as:
\begin{eqnarray}
{}_{-2}\widetilde{\lambda}_{\ell m}&=&\frac{1}{\pi}\left(\frac{2M}{r_{+}-r_{-}}\right)^{2\ell+1}\frac{(\ell-1)}{(\ell+1)}\left(\frac{3-8\ell+4\ell^{2}}{3+8\ell+4\ell^{2}}\right)\nonumber\\
&&\times \Gamma(\ell+5/2)\Gamma(\ell-1/2){}_{-2}\mathcal{\lambda}_{\ell m}\,.
\end{eqnarray}
Using the results from \ref{App:GensResponse}, and ignoring terms of $\mathcal{O}(\omega \ln \omega)$ and $\mathcal{O}(M^{2}\omega^{2})$ we obtain upto linear order in $M\omega$,
\begin{align}\label{responsedecomp}
{}_{-2}\widetilde{\lambda}_{\ell m}={}_{-2}\widetilde{\lambda}^{\rm stat}_{\ell m}+{}_{-2}\widetilde{\lambda}^{\rm dyn}_{\ell m}M\omega\,,
\end{align}
where the stationary response function yields (see \ref{App:GensResponse}), 
\begin{align}\label{gen_response_static}
{}_{-2}\widetilde{\lambda}^{\rm stat}_{\ell m}=iP_{+}^{a}\frac{(\ell-s)!(\ell+s)!}{(2\ell+1)!(2\ell)!}\prod_{n=1}^{\ell}\{n^{2}+4(P_{+}^{a})^{2}\}\,, 
\end{align}
with $P_+^a$ defined in~\ref{eq:Pplusadef}. This result matches with earlier results in the literature \cite{LeTiec:2020bos, chia2020tidal-b97, Bhatt:2024rpx}. As evident, the above expression is purely imaginary and hence the tidal Love numbers of a Kerr BH under general perturbations identically vanish. The dynamical response function, as in \ref{responsedecomp}, reads (see \ref{App:GensResponse}): 
\begin{widetext}

\begin{align}\label{dyn_resp_gens}
&{}_{-2}\widetilde{\lambda}^{\rm dyn}_{\ell m}=P_{+}^{a}\frac{(\ell-s)!(\ell+s)!}{(2\ell+1)!(2\ell)!}\prod_{n=1}^{\ell}\{n^{2}+4(P_{+}^{a})^{2}\}\Big[\left(\frac{\pi \Delta \ell+2i\pi P_{+}^{\omega}}{M\omega}\right)\coth(2\pi P_{+}^{a})
\nonumber
\\
&-\left(\frac{2P_{+}^{\omega}}{M\omega}\right)\left\{\Psi(1+\ell+2iP^{a}_{+})-\Psi(1+\ell-2iP_{+}^{a})\right\}-i\epsilon_{\ell}\left\{\Psi(\bar{\mu}_{e})+\Psi(1+\bar{\nu}_{e})\right\}
\nonumber
\\
&+i\left(\frac{\Delta \ell}{M\omega}\right)\left\{\Psi(1+\ell-2iP_{+}^{a})+\Psi(1+\ell-s)+\Psi(1+\ell+s)+\Psi(1+\ell+2iP^{a}_{+})-2\Psi(2+2\ell)-2\Psi(1+2\ell)\right\}\Big]\,.
\end{align}
\end{widetext}
Given the above response functions, the conservative and dissipative tidal responses are the real and imaginary parts of ~\ref{responsedecomp}. We substitute the expressions for the quantities $\epsilon_{\ell}$, $\Delta\ell$, $P_{+}^{\omega}$ from~\ref{eq:epsilonldef}, ~\ref{eq:Deltaelldef}, and~\ref{eq:Pplusadef} respectively, and define the dissipative coefficients to linear order in frequency by
\begin{equation}
{}_{-2}\widetilde{\nu}_{\ell m}^{\rm stat}+M\omega {}_{-2}\widetilde{\nu}_{\ell m}^{\rm dyn}=\textrm{Im}[{}_{-2}\widetilde{\lambda}_{\ell m}].
\end{equation}
We find that the stationary part of the response function is purely imaginary and
\begin{align}
{}_{-2}\widetilde{\nu}_{\ell m}^{\rm stat}=P_{+}^{a}\frac{(\ell-s)!(\ell+s)!}{(2\ell+1)!(2\ell)!}\prod_{n=1}^{\ell}\{n^{2}+4(P_{+}^{a})^{2}\}\,.
\end{align}
The dynamical part of the dissipation function from \ref{dyn_resp_gens} is,
\begin{widetext}
\begin{align}
&{}_{-2}\widetilde{\nu}_{\ell m}^{\rm dyn}={}_{-2}\widetilde{\nu}_{\ell m}^{\rm stat}\Bigg[\left(\frac{4\pi sr_{+}}{M(2\ell+1)}+\frac{4\pi r_{+}}{r_{+}-r_{-}}\right)\coth(2\pi P_{+}^{a})
-\left(\frac{4r_{+}}{r_{+}-r_{-}}\right)\textrm{Im}\left\{\Psi(1+\ell+2iP^{a}_{+})-\Psi(1+\ell-2iP_{+}^{a})\right\}\nonumber\\
&\qquad+\frac{2ams^{2}}{M\ell(\ell+1)(2\ell+1)}\left\{\Psi(\bar{\mu}_{e})+\Psi(1+\bar{\nu}_{e})\right\}-\frac{2am}{M(2\ell+1)}\left(1+\frac{s^{2}}{\ell(\ell+1)}\right)\left\{\Psi(1+\ell-2iP_{+}^{a})+\Psi(1+\ell-s)\right.\nonumber\\
& \qquad \left.+\Psi(1+\ell+s)+\Psi(1+\ell+2iP^{a}_{+})-2\Psi(2+2\ell)-2\Psi(1+2\ell)\right\}\Bigg]\,.
\end{align}
Similarly, the conservative part of the response has the low-frequency expansion: 
\begin{align}
{}_{-2}\widetilde{k}_{\ell m}=\frac{1}{2} \, {\rm Re}[_{-2}{\widetilde \lambda}_{\ell m}]={}_{-2}\widetilde{k}_{\ell m}^{\rm stat}+M\omega {}_{-2}\widetilde{k}_{\ell m}^{\rm dyn}\,.
\end{align}
As discussed above, ${}_{-2}\widetilde{k}_{\ell m}^{\rm stat}=0$ for a Kerr background, while the dynamical coefficient reads:
\begin{align}\label{dyn_cons_tide}
{}_{-2}\widetilde{k}^{\rm dyn}_{\ell m}&=
_{-2}\nu_{\ell m}^{\rm stat}\Big[
-\frac{m\chi}{(2\ell+1)}\left(1+\frac{s^{2}}{\ell(\ell+1)}\right)\pi\coth(2\pi P_{+}^{a})
\nonumber
\\
&
+\frac{s}{2\ell+1}\left\{\Psi(\bar{\mu}_{e})+\Psi(1+\bar{\nu}_{e})\right\}
+\frac{2sr_{+}}{M(2\ell+1)}\Big\{\Psi(1+\ell-s)+\Psi(1+\ell+s)
\nonumber
\\
&+\textrm{Re}[\Psi(1+\ell-2iP_{+}^{a})+\Psi(1+\ell+2iP^{a}_{+})]-2\Psi(2+2\ell)-2\Psi(1+2\ell)\Big\}\Big] \,.
\end{align}
\end{widetext}
Here $\Psi$ denotes the digamma function. One can straightforwardly check that the above expression reduces to \ref{dynamictlnscalar} in the limit $s\to 0$. Moreover, note that in the limit of $a\to 0$, the dynamical coefficient at linear order in $M\omega$ vanishes, which is also consistent with the previous results in the literature \cite{Chakraborty:2025wvs, Katagiri:2024wbg, saketh2023dynamical-b30}. In addition, for axisymmetric perturbations, i.e., for $m=0$, we obtain $P_{+}^{a}=0$, and hence the dynamic tidal response function vanishes identically up to linear order in $M\omega$. 
Note that unlike the case of the dynamical tidal Love number, the dynamical dissipation number depends explicitly on $(r_{+}-r_{-})^{-1}$ and hence is ill-behaved in the extremal limit. We will discuss these issues in a subsequent paragraph. All in all, the above provides the complete expressions for the stationary as well as dynamic Love numbers and dissipation numbers for a Kerr BH under arbitrary perturbations.

We would like to point out one interesting feature of the above dynamical response function in this regard. First of all, note that the term involving sinusoidal functions in \ref{dyn_resp_fn_01} can be expressed as $(1/2)\sinh(2\pi P_{+})[1+\mathcal{O}(M\omega)]$. Similarly, the following combination of Gamma functions becomes $\Gamma(1+\hl_{s}+2iP_{+})\Gamma(1+\hl_{s}-2iP_{+})=\{2\pi P_{+}/\sinh(2\pi P_{+})\}\prod_{k=1}^{\ell}(k^{2}+4P_{+}^{2})\left[1+\mathcal{O}(M\omega)\right]$. Hence the overall scaling of the dynamical response function and of the dynamical TLNs becomes $iP_{+}\prod_{k=1}^{\ell}(k^{2}+4P_{+}^{2})$. Recalling the definition of $P_+$ from~\ref{eq:Pplusdef}, this shows that the response function changes sign as the frequency crosses the superradiant bound $m\Omega_{\rm H}$.

Finally note that the expression for the dynamical response function as in \ref{dyn_resp_gens} is ill-behaved in the extremal limit. This is a consequence of the fact that in the extremal limit, $\{P_{+}^{a},P_{+}^{\omega}\}\to \infty$, and $r_{+}\to r_{-}$. The divergent behaviour of the response function in \ref{dyn_resp_gens} in the extremal limit arises from the fact that in deriving \ref{dyn_resp_gens} we had assumed $P_{+}^{\omega}\ll 1$, which is not true in this limit. However, the physical response is finite for extremal spins, as can be seen from an alternative expression  for the dynamical response function we calculated:
\begin{widetext}
\begin{align}\label{response_ext_gens}
&{}_{-2}\lambda_{\ell m}=i\widetilde{P}_{+}\left(\frac{\pi}{\Gamma(\ell+5/2)\Gamma(\ell-1/2)}\right)\frac{(\ell+1)}{(\ell-1)}\left(\frac{3+8\ell+4\ell^{2}}{3-8\ell+4\ell^{2}}\right)\frac{(\ell-s)!}{(2\ell+1)!(2\ell)!}\prod_{j=1}^{\ell}\left[j^{2}\left(\frac{r_{+}-r_{-}}{2M}\right)^{2}+4\widetilde{P}_{+}^{2}\right]
\nonumber
\\
&\qquad \times \left[1-i\pi \Delta \ell_{s}\coth(2\pi P_{+})\right] \Big[1+\Delta \ell_{s}\left\{2\Psi(1+\ell-s)-2\Psi(2+2\ell)-2\Psi(1+2\ell)
\right\} 
\nonumber
\\
&\qquad -\epsilon_{\ell}\left\{\Psi(\ell+5/2)+\Psi(\ell-1/2)\right\}\Big]\,.
\end{align}
\end{widetext}
The well-behaved-ness can be first apprehended from the definition of the quantity $\widetilde{P}_{+}=r_{+}\omega-(am/2M)$, which approaches finite values in the extremal limit. In fact, in the extremal limit, $r_{+}\to r_{-}\to M$, along with $a=M$, the above response function for gravitational perturbation becomes, 
\begin{widetext}
\begin{align}\label{response_ext_gens}
&{}_{-2}\lambda_{\ell m}=-im^{2\ell+1}\left\{1-(2\ell+1)\frac{2M\omega}{m}\right\}\left(\frac{\pi}{2\Gamma(\ell+5/2)\Gamma(\ell-1/2)}\right)\frac{(\ell+1)}{(\ell-1)}\left(\frac{3+8\ell+4\ell^{2}}{3-8\ell+4\ell^{2}}\right)\frac{(\ell-s)!}{(2\ell+1)!(2\ell)!}
\nonumber
\\
&\times \left[1-i\pi \Delta \ell_{s}\right] \Big[1+\Delta \ell_{s}\left\{2\Psi(1+\ell-s)-2\Psi(2+2\ell)-2\Psi(1+2\ell)
\right\} 
-\epsilon_{\ell}\left\{\Psi(\ell+5/2)+\Psi(\ell-1/2)\right\}\Big]\,,
\end{align}
\end{widetext}
where we have used the result that in the $P_{+}\to \infty$ limit, $\coth(2\pi P_{+})\to 1$, and at the same time we have replaced $\Gamma(1+\hat{\ell}_{s}\pm2iP_{+})$ by $\Gamma(1+\ell\pm2iP_{+})$, where we have ignored the $\mathcal{O}(M\omega)$ term as $P_{+}\gg 1$. We will compute now the extremal case in an explicit manner and shall match the respective results with the expression of the dynamical response function as presented above. 
\section{Generic spin perturbation of an extremal Kerr black hole}
\label{sec:extremal}

In this section we would like to briefly touch upon the dynamical tides for an extremal Kerr BH. In this case, the spin length $a$ is the same as the mass $a=M$ and the two horizons coincide. Thus, we have $\Delta=(r-r_{+})^{2}$, with $r_{+}=M=a$. Note that, since for an extremal Kerr BH $r_{+}=r_{-}$, the coordinate $z$ introduced in~\ref{eq:zdef} and used in all previous calculations vanishes and thus cannot be used for studying the extremal case. Instead, we introduce a new radial coordinate $y\equiv (r/r_{+})-1$ and express the Teukolsky equation in the Boyer-Lindquist coordinate as,
\begin{widetext}
\begin{align}
\label{eq:radialTeukBL}
y^{2}\dfrac{d^{2}{}_{s}R^{(\rm BL)}_{\ell m}}{dy^{2}}+2(s+1)y\dfrac{d{}_{s}R^{(\rm BL)}_{\ell m}}{dy}+\left[\frac{K^{2}-2isKr_{+}y}{r_{+}^{2}y^{2}}+4is\omega r_{+}(1+y)-\lambda\right]{}_{s}R^{(\rm BL)}_{\ell m}=0~.
\end{align}
Here, the superscript to the radial coordinate denotes that the radial function is expressed in the Boyer-Lindquist coordinate system.
It turns out that in the near-zone region, the above equation does not admit an analytic solution. However, when switching to  advanced null coordinates $(v,r,\theta,\widetilde{\phi})$, where $dv=dt+\{(r^{2}+a^{2})/\Delta\}dr$ and $d\widetilde{\phi}=d\phi+(a/\Delta)dr$, the Teukolsky equation can be solved in the near-zone region. On the other hand, the solution of the Teukolsky equation in the far-zone is best solved in Boyer-Lindquist coordinates, rather than in advanced null coordinates. Thus we will solve the near-zone Teukolsky equation in null coordinates, while the far-zone Teukolsky equation is considered in Boyer-Lindquist coordinates.  

\subsection{Near-zone Solution}

Under the near-zone approximation ($M\omega\ll1$ and $M\omega y\ll 1$), we have, $K\approx (r_{+}^{2}+a^{2})\omega-am$, and $\omega r\approx \omega r_{+}$, such that the radial Teukolsky equation~\ref{eq:radialTeukBL} in the ingoing null coordinates becomes,
\begin{align}
y^{2}\dfrac{d^{2}{}_{s}R^{(\rm EF)}_{\ell m}}{dy^{2}}+\left[2(s+1)y-2i\widetilde{P}_{+}\right]\dfrac{d{}_{s}R^{(\rm EF)}_{\ell m}}{dy}+\left[-\frac{4is\widetilde{P}_{+}}{y}
-\lambda\right]{}_{s}R^{(\rm EF)}_{\ell m}=0~.
\end{align}
Here, the superscript implies the use of Eddington-Finkelstein like coordinates. In addition, we have introduced $\widetilde{P}_{+}=2M\bar{\omega}$, where $\bar{\omega}=\omega-(ma/2M^{2})$ is the frequency in the co-rotating frame of reference. Moreover, the radial function ${}_{s}R^{(\rm EF)}_{\ell m}$ in ingoing null coordinates is related to the radial function ${}_{s}R^{(\rm BL)}_{\ell m}$ in the Boyer-Lindquist coordinate system through the following transformation: ${}_{s}R^{(\rm BL)}_{\ell m}=e^{-i\omega r_{*}}e^{im\widetilde{r}}{}_{s}R^{(\rm EF)}_{\ell m}$, where $r_{*}$ is the Tortoise coordinate and $d\widetilde{r}\equiv (a/\Delta)dr$. Further note that, the separation constant $\lambda$ is given by $\lambda=\hat{\ell}_{e}(\hat{\ell}_{e}+1)-s(s+1)$.
Therefore, the radial Teukolsky equation in the near zone, in ingoing null coordinates, takes the following final form:
\begin{align}
y^{2}\dfrac{d^{2}{}_{s}R^{(\rm EF)}_{\ell m}}{dy^{2}}+\left[2(s+1)y-2i\widetilde{P}_{+}\right]\dfrac{d{}_{s}R^{(\rm EF)}_{\ell m}}{dy}+\left[-\frac{4is\widetilde{P}_{+}}{y}-\hat{\ell}_{e}(\hat{\ell}_{e}+1)+s(s+1)\right]{}_{s}R^{(\rm EF)}_{\ell m}=0~,
\end{align}
which has the following solution in terms of confluent hypergeometric functions: 
\begin{align}\label{extrem_near_zone}
{}_{s}R^{(\rm EF)}_{\ell m}&=Ay^{\hat{\ell}_{e}-s}M\left(-\hl_{e}-s,-2\hl_{e};-\frac{2i\widetilde{P}_{+}}{y}\right)+By^{-1-s-\hl_{e}}M\left(1+\hl_{e}-s,2+2\hl_{e};-\frac{2i\widetilde{P}_{+}}{y}\right)~.
\end{align} 
\end{widetext}
Here, $M(a,b;z)$ is the confluent hypergeometric function and $\hl_{e}$ is related to the angular momentum such that,
\begin{align}
\hl_{e}=\ell-\frac{2M\omega}{2\ell+1}\left[
m\chi\left(1+\frac{s^{2}}{\ell(\ell+1)}\right)\right]+\mathcal{O}(M^{2}\omega^{2})\,.
\end{align}
Note that, $\hat{\ell}_{e}-\ell=\hat{\ell}_{s}-\ell$, which for $s=0$ reduces to $\hat{\ell}-\ell$. Thus the renormalized angular momentum in the extremal case behaves identically to that in the non-extremal case.
Given the near-zone solution we will now take the near-horizon limit and hence impose appropriate boundary conditions at the horizon, and then we take the asymptotic limit for matching with the far-zone solution at the intermediate zone.
\subsubsection{The limit to the horizon}

We first consider the near-horizon limit, which corresponds to $y\to 0$. In this case, using \ref{Minfty} in \ref{AppA}, the radial function in the limit towards the horizon becomes, 
\begin{align}
&{}_{s}R^{(\rm EF)}_{\ell m}|_{\rm hor}=y^{-2s}\left[A\frac{\Gamma(-2\hl_{e})}{\Gamma(s-\hl_{e})}e^{\pm i\pi(\hl_{e}+s)}(-2i\widetilde{P}_{+})^{\hl_{e}+s}\right.\nonumber\\
&\quad\left.+B\frac{\Gamma(2+2\hl_{e})}{\Gamma(1+\hl_{e}+s)}e^{\mp i\pi(1+\hl_{e}-s)}(-2i\widetilde{P}_{+})^{-1-\hl_{e}+s}\right]
\nonumber
\\
&\quad+e^{-\frac{2i\widetilde{P}_{+}}{y}}\left[A\frac{\Gamma(-2\hl_{e})}{\Gamma(-s-\hl_{e})}(-2i\widetilde{P}_{+})^{\hl_{e}-s}\right.\nonumber\\
&\quad\left.+B\frac{\Gamma(2+2\hl_{e})}{\Gamma(1+\hl_{e}-s)}(-2i\widetilde{P}_{+})^{-1-\hl_{e}-s} \right]\,.
\end{align}
For an extremal BH, the ingoing solution of the radial Teukolsky function behaves as $\Delta^{-s}\approx y^{-2s}$, and the outgoing solution takes the form $e^{-2i\widetilde{P}_{+}/y}$. Since for BHs there are no outgoing solutions at the horizon, it follows that the coefficient of $\exp(-2i\widetilde{P}_{+}/y)$ must vanish. Hence the ratio $(B/A)$ takes the following form, 
\begin{align}
\frac{B}{A}=-\frac{\Gamma(-2\hl_{e})\Gamma(1+\hl_{e}-s)}{\Gamma(-\hl_{e}-s)\Gamma(2+2\hl_{e})}\left(-2i\widetilde{P}_{+}\right)^{1+2\hl_{e}}~.
\end{align}
Therefore, by substituting the ratio $(B/A)$ in \ref{extrem_near_zone}, the radial function ${}_{s}R^{(\rm EF)}_{\ell m}$, purely ingoing at the BH horizon, becomes, 
\begin{align}\label{extnearfinal}
&{}_{s}R^{(\rm EF)}_{\ell m}=Ay^{\hl_{e}-s}\Bigg[M\left(-\hl_{e}-s,-2\hl_{e};-\frac{2i\widetilde{P}_{+}}{y}\right)
\nonumber
\\
&\qquad  -\frac{\Gamma(-2\hl_{e})\Gamma(1+\hl_{e}-s)}{\Gamma(-\hl_{e}-s)\Gamma(2+2\hl_{e})}\left(\frac{-2i\widetilde{P}_{+}}{y}\right)^{1+2\hl_{e}}\nonumber\\
&\qquad \quad\times M\left(1+\hl_{e}-s,2+2\hl_{e};-\frac{2i\widetilde{P}_{+}}{y}\right)\Bigg]~.
\end{align}
This yields the solution of the radial Teukolsky equation in the near-zone, which is purely ingoing at the BH horizon. The next step is to take the asymptotic limit and hence determine the behaviour of the radial solution in the intermediate-zone, to be matched with the far-zone solution. 
\subsubsection{Extension to the intermediate zone}

In order to arrive at the intermediate zone, we take the limit $y\to \infty$ of \ref{extnearfinal}, which implies that the argument of the confluent hypergeometric functions must vanish, and since $\lim_{z\to 0}M(a,b;z)=1$, we obtain, 
\begin{align}\label{final_near_int}
{}_{s}R^{\textrm{(EF)}\,\textrm{(int)}}_{\ell m}&=A\left(\frac{r}{r_{+}}\right)^{\hl_{e}-s}\Bigg[1+\frac{\Gamma(-2\hl_{e})\Gamma(1+\hl_{e}-s)}{\Gamma(-\hl_{e}-s)\Gamma(2+2\hl_{e})}\nonumber\\
&\quad \times\left(\frac{r_{+}}{r}\right)^{1+2\hl_{e}}\left(2i\widetilde{P}_{+}\right)^{1+2\hl_{e}}\Bigg]~.
\end{align}
Here, we have expressed $y\approx (r/r_{+})$, leading to the standard behaviour of the radial function, with one part growing as $r^{\hl_{e}-s}$, and another part is decaying as $r^{-1-\hl_{e}-s}$. Since $\hl_{e}$ is a complex quantity, the analytical continuation is automatically satisfied, and hence the above neatly separates the tidal part from the response. 

\subsection{Far-zone Solution and matching at intermediate zone}

The far-zone solution, on the other hand, is identical for non-extremal and extremal BHs, as the asymptotic behaviour is identical with the rotation parameter $a$ set to be equal to the mass $M$ everywhere. Therefore, the far-zone solution is still given by \ref{radial_far}, but in Boyer-Lindquist coordinates. The small $r$ limit of the far-zone solution is then explicitly given by 
\begin{align}
\,_{s}R_{\ell m\,\textrm{(int)}}^{\textrm{(BL)}}&=\frac{C^{\infty}_{\rm reg}}{\Gamma(\bar{\nu}+1)}\left(\frac{\omega}{2}\right)^{\bar{\nu}}r^{\tl-s}\nonumber\\
&
-\frac{C^{\infty}_{\rm irreg}}{\pi}\Gamma(\bar{\mu})\left(\frac{\omega}{2}\right)^{-\bar{\mu}}r^{-1-\tl-s}\,,
\end{align}
where $\bar{\mu}=\widetilde{\ell}+|s|+(1/2)$ and $\bar{\nu}=\widetilde{\ell}-|s|+(1/2)$, with $\widetilde{\ell}$ being defined by \ref{gens_ang}. The above small $r$ expansion of the far-zone solution can now be matched with \ref{final_near_int}, yielding, 
\begin{align}
&\frac{C^{\infty}_{\rm reg}}{\Gamma(\bar{\nu}+1)}\left(\frac{\omega}{2}\right)^{\bar{\nu}}=AM^{s-\tl}\,,
\\
&\frac{C^{\infty}_{\rm irreg}}{\pi}\Gamma(\bar{\mu})\left(\frac{\omega}{2}\right)^{-\bar{\mu}}M^{-1-\tl-s}=A\frac{\Gamma(1-s+\hat{\ell}_{e})}{\Gamma(-s-\hat{\ell}_{e})}\nonumber\\
&\qquad \qquad \quad\times\frac{\Gamma(-2\hat{\ell}_{e})}{\Gamma(2+2\hat{\ell}_{e})}\left(-2i\widetilde{P}_{+}\right)^{2\hl_{e}+1}\,.
\end{align}
We note that there are extra factors of $\exp(-i\omega r_{*})$ and $\exp(im\tilde{r})$ in the connection formulae, due to converting the radial function in Boyer-Lindquist to advanced null coordinates, where $(dr_{*}/dr)=\{(r^{2}+a^{2})/\Delta\}$ and $(d\tilde{r}/dr)=(a/\Delta)$. However, in the far-zone, neither of these contributes. From the above connection formulae, we finally obtain, 
\begin{widetext}
\begin{align}
\frac{C^{\infty}_{\rm irreg}}{C^{\infty}_{\rm reg}}=\frac{\Gamma(1-s+\hat{\ell}_{e})}{\Gamma(-s-\hat{\ell}_{e})}\frac{\Gamma(-2\hat{\ell}_{e})}{\Gamma(2+2\hat{\ell}_{e})}\left(-4iM\bar{\omega}\right)^{2\hat{\ell}_{e}+1}
\frac{\pi}{\Gamma(\bar{\mu})\Gamma(\bar{\nu}+1)}
\left(\frac{M\omega}{2}\right)^{2\tl+1}\,.
\end{align}
Using the identities involving Gamma functions, we can re-express the above expression as,
\begin{align}\label{irregregextrem}
\frac{C^{\infty}_{\rm irreg}}{C^{\infty}_{\rm reg}}=\frac{\Gamma(1-s+\hat{\ell}_{e})}{\Gamma(2+2\hat{\ell}_{e})}\frac{\Gamma(1+s+\hat{\ell}_{e})}{\Gamma(1+2\hat{\ell}_{e})}\left(-4iM\bar{\omega}\right)^{2\hat{\ell}_{e}+1}
\frac{\pi(-1)^{s+\ell}}{2\Gamma(\bar{\mu})\Gamma(\bar{\nu}+1)}
\left(\frac{M\omega}{2}\right)^{2\tl+1}\,.
\end{align}

This provides the matching between the two zones in BH perturbation theory. We now consider matching of the far-zone in BH perturbation theory with the EFT solution and hence obtain the gauge invariant dynamical response function.

\subsection{EFT and asymptotic matching: The response functions}

The EFT action and the subsequent computations performed in~\ref{sec:EFT} can be directly applied here, or, in other words, the EFT of an extremal BH is identical to that in the context of non-extremal BHs. Hence all the results from the EFT calculations also hold in the extremal case. Below we discuss the case for scalar, as well as for gravitational perturbation. 

\subsubsection{Scalar Perturbation}

For scalar perturbations, using the EFT results from \ref{sec:EFT}, the BH perturbation theory tidal response function turns out to be, 
\begin{align}\label{dyn_resp_ext_01_0s}
\lambda^{\rm ext}_{\ell m}&=\left(\frac{1}{M\omega}\right)^{2\ell+1}\frac{\Gamma(1+\hat{\ell}_{e})}{\Gamma(2+2\hat{\ell}_{e})}\frac{\Gamma(1+\hat{\ell}_{e})}{\Gamma(1+2\hat{\ell}_{e})}\left(-4iM\bar{\omega}\right)^{2\hat{\ell}_{e}+1}
\frac{\pi(-1)^{\ell+1}}{2\Gamma(\ell+1/2)\Gamma(\ell+3/2)}
\left(\frac{M\omega}{2}\right)^{2\ell+1}
\nonumber
\\
&=\frac{(-1)^{\ell+1}\pi}{2^{2\ell+2}}\times \left(\frac{\Gamma(1+\hat{\ell}_{e})}{\Gamma(2+2\hat{\ell}_{e})}\frac{\Gamma(1+\hat{\ell}_{e})}{\Gamma(1+2\hat{\ell}_{e})} \right)
\times \frac{\left(2im-4iM\omega\right)^{1+2\hat{\ell}_{e}}}{\Gamma(\ell+1/2)\Gamma(\ell+3/2)}\,.
\end{align}
Expanding $\hat{\ell}_{e}$ in powers of $M\omega$, we can rewrite the above response function upto linear orders in $M\omega$ as,
\begin{align}\label{dyn_resp_ext_02_0s}
\lambda^{\rm ext}_{\ell m}&=-im^{2\ell+1}\times\frac{\pi}{2}\times \left(\frac{(\ell!)^{2}}{(2\ell)!(2\ell+1)!}\right)\times \frac{\left\{1-(2\ell+1)\frac{2M\omega}{m}\right\}}{\Gamma(\ell+1/2)\Gamma(\ell+3/2)}
\nonumber
\\
&\times\left[1+\Delta \ell_{e}\left\{2\Psi(1+\ell)-2\Psi(1+2\ell)-2\Psi(2+2\ell)\right\} \right]\left(1-i\pi \Delta \ell_{e}\right)\,.
\end{align}
Further, since $\Delta \ell_{e}=\Delta \ell$, a comparison between \ref{dyn_resp_ext_02_0s} and \ref{response_ext_scalar} shows that the scalar response for extremal BH matches with the extremal limit of non-extremal BH. 

\subsubsection{Gravitational Perturbation}

The response function associated with the gravitational perturbation of an extremal Kerr BH, using the EFT matching used in \ref{subsec:EFTgenerics}, takes the following form,
\begin{align}\label{dyn_resp_ext_01}
{}_{-2}\lambda^{\rm ext}_{\ell m}&=-(M\omega)^{-2\ell-1}\frac{(\ell+1)}{(\ell-1)}\left(\frac{3+8\ell+4\ell^{2}}{3-8\ell+4\ell^{2}}\right)\frac{C^{\infty}_{\rm irreg}}{C^{\infty}_{\rm reg}}
\nonumber
\\
&=-(M\omega)^{-2\ell-1}\frac{(\ell+1)}{(\ell-1)}\left(\frac{3+8\ell+4\ell^{2}}{3-8\ell+4\ell^{2}}\right)\left(\frac{\Gamma(3+\hl_{e})\Gamma(\hl_{e}-1)}{\Gamma(2+2\hl_{e})\Gamma(1+2\hl_{e})}\right)
\nonumber
\\
&\times \left(-4iM\bar{\omega}\right)^{2\hat{\ell}_{e}+1}
\frac{\pi(-1)^{\ell}}{2\Gamma(\bar{\mu})\Gamma(\bar{\nu}+1)}
\left(\frac{M\omega}{2}\right)^{2\tl+1}\,.
\end{align}
This can be further simplified to yield, 
\begin{align}\label{dyn_resp_ext_02}
{}_{-2}\lambda^{\rm ext}_{\ell m}&=\frac{\pi(-1)^{\ell+1}}{2^{2\ell+2}}\frac{\left(2im-4iM\omega\right)^{1+2\hat{\ell}_{e}}}{\Gamma(\ell+5/2)\Gamma(\ell-1/2)}\frac{(\ell+1)}{(\ell-1)}\left(\frac{3+8\ell+4\ell^{2}}{3-8\ell+4\ell^{2}}\right)\left(\frac{(\ell+2)!(\ell-2)!}{(1+2\ell)!(2\ell)!}\right)
\nonumber
\\
&\times \left[1+\Delta \ell_{e}\left\{\Psi(3+\ell)+\Psi(\ell-1)-2\Psi(1+2\ell)-2\Psi(2+2\ell)\right\} \right]\left[1-\epsilon_{\ell}M\omega\left\{\Psi(\Psi(\ell+5/2)+\Psi(\ell-1/2)\right\} \right]
\nonumber
\\
&=-i\frac{\pi m^{2\ell+1}}{2}\frac{(\ell+1)}{(\ell-1)}\left(\frac{3+8\ell+4\ell^{2}}{3-8\ell+4\ell^{2}}\right)\left(\frac{(\ell+2)!(\ell-2)!}{(1+2\ell)!(2\ell)!}\right)
\times \frac{1}{\Gamma(\ell+5/2)\Gamma(\ell-1/2)}
\nonumber
\\
&\times \left[1+\Delta \ell_{e}\left\{\Psi(3+\ell)+\Psi(\ell-1)-2\Psi(1+2\ell)-2\Psi(2+2\ell)\right\} \right]\left[1-\epsilon_{\ell}M\omega\left\{\Psi(\ell+5/2)+\Psi(\ell-1/2)\right\} \right]
\\
&\times \left[1-i\pi\Delta \ell_{e}\right]\left[1-\frac{2M\omega(1+2\ell)}{m}\right]\,.
\end{align}
Note that the above expression exactly matches with the result obtained in \ref{response_ext_gens}, and hence for gravitational perturbation as well the response function for extremal BH coincides with the response function of non-extremal BH in the extremal limit. 

The above response function can also be expressed as ${}_{-2}\lambda^{\rm ext}_{\ell m}={}_{-2}\lambda^{\rm ext\, (stat)}_{\ell m}+M\omega{}_{-2}\lambda^{\rm ext\,(dyn)}_{\ell m}$. Substituting $\omega=0$ in the above expression for the dynamical response function, we obtain the stationary case, for which the response function becomes,
\begin{align}
{}_{-2}\lambda^{\rm ext\, (stat)}_{\ell m}=-i\frac{\pi m^{2\ell+1}}{2}\frac{(\ell+1)}{(\ell-1)}\left(\frac{3+8\ell+4\ell^{2}}{3-8\ell+4\ell^{2}}\right)\left(\frac{(\ell+2)!(\ell-2)!}{(1+2\ell)!(2\ell)!}\right)\times \frac{1}{\Gamma(\ell+5/2)\Gamma(\ell-1/2)}\,.
\end{align}
As evident, this response function is purely imaginary, thereby leading to vanishing stationary TLNs for extremal Kerr BH.
The dynamical response function, on the other hand, is given by, 
\begin{align}
\label{eq:responsegenericextBHspin}
{}_{-2}\lambda^{\rm ext\, (dyn)}_{\ell m}&={}_{-2}\lambda^{\rm ext\, (stat)}_{\ell m}\Big[-\frac{2(2\ell+1)}{m}-\left(\frac{1}{2\ell+1}\right)\left(-4i-\frac{8m}{\ell(\ell+1)}\right)\left\{\Psi(\ell+5/2)+\Psi(\ell-1/2) \right\}
\nonumber
\\
&\qquad -\frac{2m}{(2\ell+1)}\left(1+\frac{4}{\ell(\ell+1)}\right)\left\{\Psi(3+\ell)+\Psi(\ell-1)-2\Psi(1+2\ell)-2\Psi(2+2\ell)-i\pi\right\}\Big]\,.
\end{align}
Since the static response function is purely imagniary, it follows that the dynamical response function has both real and imaginary parts, i.e., at $\mathcal{O}(M\omega)$, extremal BH has non-zero TLNs as well as non-zero dissipation numbers. Starting with the TLNs, we obtain, 
\begin{align}
{}_{-2}k_{\ell m}^{\rm ext}={}_{-2}k_{\ell m}^{\rm ext\,(stat)}+M\omega {}_{-2}k_{\ell m}^{\rm ext\,(dyn)}\,,
\end{align}
where, ${}_{-2}k^{\rm ext}_{\ell m}=(1/2)\textrm{Re}\,_{-2}\lambda^{\rm ext}_{\ell m}$. As already discussed, the static part of the response function is purely imaginary, and hence the static TLNs vanish identically. The dynamical part is given by, 
\begin{align}
{}_{-2}k_{\ell m}^{\rm ext\,(dyn)}&=\frac{\pi m^{2\ell+1}}{2}\frac{(\ell+1)}{(\ell-1)}\left(\frac{3+8\ell+4\ell^{2}}{3-8\ell+4\ell^{2}}\right)\left(\frac{(\ell+2)!(\ell-2)!}{(1+2\ell)!(2\ell)!}\right)\times \frac{1}{\Gamma(\ell+5/2)\Gamma(\ell-1/2)}
\nonumber
\\
&\times \left[\left(\frac{8}{2\ell+1}\right)\left\{\Psi(\ell+5/2)+\Psi(\ell-1/2)\right\}+\frac{2m\pi}{(2\ell+1)}\right]\,.
\end{align}
The dissipation numbers, on the other hand, are defined as, ${}_{-2}\nu_{\ell m}^{\rm ext}=\textrm{Im}\,_{-2}\lambda_{\ell m}^{\rm ext}$. This has a non-zero static and dynamical parts, these are given by,
\begin{align}
&\,_{-2}\nu_{\ell m}^{\rm ext\,(stat)}=-\frac{\pi m^{2\ell+1}}{2}\frac{(\ell+1)}{(\ell-1)}\left(\frac{3+8\ell+4\ell^{2}}{3-8\ell+4\ell^{2}}\right)\left(\frac{(\ell+2)!(\ell-2)!}{(1+2\ell)!(2\ell)!}\right)\times \frac{1}{\Gamma(\ell+5/2)\Gamma(\ell-1/2)}\,,
\\
&\,_{-2}\nu_{\ell m}^{\rm ext\,(dyn)}={}_{-2}\nu^{\rm ext\, (stat)}_{\ell m}\Big[-\frac{2(2\ell+1)}{m}+\left(\frac{1}{2\ell+1}\right)\frac{8m}{\ell(\ell+1)}\left\{\Psi(\ell+5/2)+\Psi(\ell-1/2) \right\}
\nonumber
\\
&\qquad -\frac{2ma}{M(2\ell+1)}\left(1+\frac{4}{\ell(\ell+1)}\right)\left\{\Psi(3+\ell)+\Psi(\ell-1)-2\Psi(1+2\ell)-2\Psi(2+2\ell)\right\}\Big]\,.
\end{align}
Note that the vanishing of static TLNs, and existence of non-zero dynamical TLNs hold for Kerr BHs in general, irrespective of the spin of the BH. The above analysis is for gravitational perturbation, but also holds for $s=0$. Furthermore, the dynamical TLNs vanish for either zero rotation ($a=0$) or for axisymmetric perturbations ($m=0$). Otherwise, the dynamical TLNs of an extremal/non-extremal Kerr BH under generic tidal perturbations are non-zero.
   
\end{widetext}
 
\section{Summary of key results and discussion}
\label{sec:discussion}

In summary, we have calculated the dynamical tidal response function characterizing the ratio of a BH's induced multipoles to the strength of the tidal perturbation to linear order in frequency. Our method is based on defining the response via a worldline EFT and then linking it to BH perturbation theory for fields with generic spin $s$. The worldline EFT describes a (perturbed) compact object on large scales and is the essential foundation for further calculations of GWs from binary systems. The matching between EFT and strong-field BH perturbation is an important step towards computing GW signatures on a binary system due to tidal effects. In particular, this determines the tidal response function in terms of the ratio of wave amplitudes that are regular and irregular at the location of the BH, taken to be the origin, where the regular parts correspond to modes of the external tidal field, and the irregular parts are related to the induced multipole moments. While similar calculations had already been performed in previous literature, our work filled in gaps in the methodology of linking the worldline EFT with the BH perturbation theory for generic spins, both in term of angular momentum of the BH and the spin weights of the tidal perturbation fields, and providing explicit, ready-to-use results for the relevant response coefficients.  

A complication in relating BH perturbation theory results to those of the EFT arose from the fact that the EFT formulation is based on symmetric-trace-free tensors, corresponding to a spherical-harmonic basis, while BH perturbation theory calculations are naturally decomposed into spheroidal harmonics. Connecting the different bases generally leads to mode mixing. In addition, the off-diagonal components of the EFT tidal response couples with these mode mixing coefficients in such a manner, that even after keeping only linear order terms in $M\omega$, the BH perturbation theory response function gets related to the diagonal and off-diagonal EFT response function through mixing coefficients. 

As far as BH perturbation theory is considered, one can relate the regular and irregular coefficients of BH perturbation theory in the far-zone ($\omega r\gg 1$) to the regular and irregular coefficients in the EFT. Keeping in mind the contribution of both diagonal and off diagonal response coefficients of EFT to the BH perturbation theory response coefficient, one can invert the relation and hence each EFT response coefficient can be determined in terms of diagonal and off-diagonal BH response coefficients. Following this which we computed the BH response coefficients for scalar and gravitational perturbations, both for non-extremal and extremal Kerr BH.

Throughout the text, we have discussed several checks of our results in the limiting cases, such as against the zero-frequency computations by Refs.~\cite{Chia:2020yla, LeTiec:2020bos, Ivanov:2022qqt}, the non-spinning results of Refs.~\cite{creci2021tidal-42e, Combaluzier--Szteinsznaider:2025eoc, Chakraborty:2025wvs, bhatt2023addressing-56b}, and the dissipative coefficients in Refs.~\cite{Saketh:2022xjb, Saketh:2023bul, bhatt2023addressing-56b}. We also note here that our results differ from those of Refs.~\cite{Perry:2023wmm} and in the case of extremal BHs from~\cite{Perry:2024vwz}, though some structures in the final expressions are identical. We attribute this discrepancy to different definitions of the response. Specifically, the work of~\cite{Perry:2023wmm, Perry:2024vwz} extracted the response based on BH perturbations, while here, we defined it at the level of the worldline EFT. As mentioned, the matching to the worldline EFT involves non-linearities such as mode mixing, leading to non-trivial mappings between results from BH perturbation theory and relevant EFT quantities. 
\begin{figure}
\includegraphics[width=.45\textwidth]{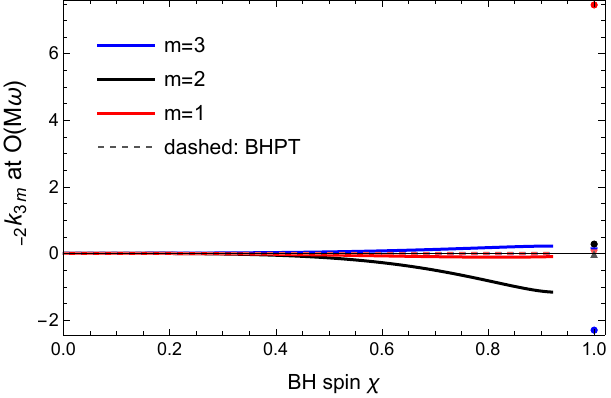}
\caption{The figure depicts {\it Octupolar conservative tidal responses}($\ell=3$) for gravitational perturbation ($s=-2$) for different choices of $m$. Dashed curves indicate the BHPT results (identical to the EFT for $|m|=l$), see \ref{dyn_cons_tide}. The other conservative responses are obtained from EFT, and follow from substitution of \ref{dyn_cons_tide} in \ref{grav_eft_bh} and then in \ref{grav_eft_scale}, respectively. Dots indicate the values for extremal spins, obtained from the above procedure.}
\label{fig:resultscomparison}
\end{figure}

\subsection{Features of the dynamical tidal response function}

Our main results are that to linear order in frequency, the tidal response function for a non-extremal BH computed from perturbation theory is given by~\ref{dyn_resp_fn_01}, while for an extremal BH the response function becomes~\ref{dyn_resp_ext_01} for gravitational perturbations. The corresponding cases for scalar perturbations can be obtained from \ref{matching_scalar} and \ref{dyn_resp_ext_01_0s}, respectively. The EFT response, which is closely linked to physical effects in the binary system, is then obtained by using the results of~\ref{grav_eft_bh} and~\ref{grav_eft_scale} for the gravitational case, and those described in~\ref{subsec:EFTscalar} for the scalar case.

We also find that an interesting pattern that was previously noticed for the $\mathcal{O}(\omega^2)$ response of a Schwarzschild BH, namely that all the response coefficients are directly proportional to the leading-order dissipation coefficient, also holds for the $\mathcal{O}(\omega)$ response of spinning BHs. To elucidate the parameter dependencies of the response coefficients, we exhibit results in a few limiting cases, recalling that our results are also truncated at $\mathcal{O}(M\omega)$. The dissipative piece of the response for the tensor quadrupole case with $s=-2, \ell=2$ has the following limiting behavior
\begin{subequations}
\label{eq:quadspinexpdiss}
\begin{eqnarray}
&&\lim_{\chi\to 1}{}_{-2}\nu_{2m}(\omega)=-\frac{2m^{5}}{45}-\frac{2m^{5}}{45}(M\omega)\nonumber\\
&&\quad \times\left[-\frac{10}{m}+\frac{4m}{15}\left\{\Psi(9/2)+\Psi(3/2)\right\}\right.\nonumber\\
&&\quad \left.-\frac{2m}{3}\left\{\Psi(1)-\Psi(5)-2\Psi(6)\right\} \right]\,,
\\
&&\lim_{\chi\to 0}{}_{-2}\nu_{2m}(\omega)=\frac{2^{5} M\omega}{225}\,.
\end{eqnarray}
\end{subequations}
The EFT response, which underpins physical effects in the dynamics and gravitational waves, is then obtained by using the identification~\ref{eq:lambdatofquad} and the normalization factor~\ref{grav_eft_scale}. Note that in the $M\omega\to 0$ limit, the dissipation number for Schwarzschild vanishes, while the dissipation number for extremal Kerr BH is non-zero. For the conservative response associated with the tensor quadrupole moments are obtained by substituting $s=-2$ and $\ell=2$ in respective equations, they yield,
\begin{subequations}
\label{eq:quadspinexp}
\begin{eqnarray}
\lim_{\chi\to 1}{}_{-2}k_{2m}(\omega)&=&M\omega\left(\frac{2m^{5}}{45}\right)\left[\frac{8}{5}\left\{\Psi(9/2)+\Psi(3/2)\right\}\right.\nonumber\\
&&\left.\qquad +\frac{2m\pi}{5} \right]\,,
\label{eq:quadhighspin}
\\
\lim_{\chi\to 0}{}_{-2}k_{2m}(\omega)&=&0\,.
\end{eqnarray}
\end{subequations}
This shows that for spinning BHs, a non-vanishing conservative response first appears at linear order in $M\omega$. For non-spinning BHs, this contribution vanishes and a non-trivial conservative response starts only at $\mathcal{O}(M^{2}\omega^2)$~\cite{Chakraborty:2025wvs, Combaluzier--Szteinsznaider:2025eoc}. We present below a few for interesting results arising out of the above expressions.

\begin{figure}
\includegraphics[width=.45\textwidth]{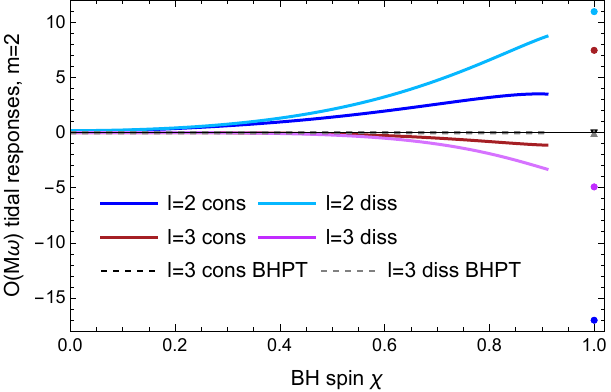}
\caption{\emph{Dependence on $\ell$ of the $O(M\omega)$ coefficients} of the tensor responses for $m=2$. Blue tones indicate the quadrupole, red-pink the octupole, with darker/lighter hues corresponding to the conservative/dissipative coefficients. Dashed lines indicate the coefficients from black hole perturbation theory (BHPT). Circles indicate the values for extremal spins. Note that the conservative coefficient (darker blue and red curves) has a non-trivial behaviour near extremality, including a change in sign.}
\label{fig:tensor_comparisons}
\end{figure}

\emph{Effect of the azimuthal number $m$:} 
From~\ref{eq:quadspinexp}, we see that the sign of $_{-2}k_{2m}$ is primarily the negative of the sign of $m$ and that $|_{s}k_{\ell m}|$ is larger for larger $|m|$, i.e. for a fixed multipole, the largest conservative response for a given $\chi$ occurs for $|m|=\ell$. For high BH spins, the factor of $m^5$ in the conservative response, see~\ref{eq:quadhighspin}, causes lower $m$ contributions to be very strongly suppressed. The dependence of the dissipation numbers on $m$ is more complicated, but the $\mathcal{O}(M\omega)$ term remains the same under $m\to -m$. Another interesting feature is depicted by \ref{fig:resultscomparison}\footnote{We would like to emphasize that the limiting expressions in \ref{eq:quadspinexpdiss}, \ref{eq:quadspinexp} are for the BH Perturbation Theory response, while the plots show the EFT response}, as it clearly shows for $\ell=3$, $m=2$ mode, the conservative response dominates all the other $m$ modes, including $\ell=3$, $m=3$ mode. This shows that the off-diagonal entries are more important than diagonal entries in tidal response for $\ell\geq 3$. 

\smallskip

\emph{Dependence on BH spin}: The largest conservative response, i.e., the largest dynamical TLNs arise for extremal BHs. In the tensor quadrupole case, it is given by \ref{eq:quadhighspin}. The spin dependence of the conservative and dissipative numbers are complicated, while the leading order spin effect in the dissipation number is due to superradiance, 
as already shown in previous work and also in our discussion, see \ref{subsec:tidalresponsegeneric}. 

\smallskip

\emph{Effect of the multipolar number $\ell$}: The dependence of conservative and dissipative parts on the multipole $\ell$ is shown in~\ref{fig:tensor_comparisons}. We see that in the tensor case, the responses of higher multipoles are suppressed compared to the cases of $|m|=\ell=2$, which maximize the response in each cases. Also from~\ref{fig:tensor_comparisons} it is clear that the magnitude of the dissipative part of the response is always larger than the conservative part for moderately high spins but interestingly not for the extremal case. Moreover, in the conservative as well as in the dissipative case, the response coefficient changes sign between different multipoles, with the even-$\ell$ quadrupole/odd-$\ell$ octupole leading to a positive/negative sign for $m=2$. Finally the response of the conservative sector always crosses zero to reach the extremal limit.

\subsection{Outlook}

Our work made several simplifying assumptions that we plan to relax in future work. On the side of BH perturbation calculations, we took the approach to solve the radial Teukolsky equation both in the near-horizon regime and the far-zone to determine the relation between integration constants in the far-zone to  imprints of near-horizon properties on asymptotic fields. We informally verified that this agrees with the leading-order results using the Mano-Suzuki-Takasugi (MST) method, which provides much more refined solutions with wider validity. More detailed full calculations using the MST solutions are an important next step left to future work, and will also elucidate logarithmic frequency dependencies and tail effects, all of which could be neglected at the leading-order to which we were working. 

On the side of the matching with the worldine EFT, we have fully established the leading-order link to the BH perturbation results based on wave amplitudes for scalar as well as gravitational perturbations. Our matching involves diagonal as well as non-diagonal and mixing coefficients from spherical to spheroidal transformation. This leaves next-to-leading order matching in $\mathcal{O}(M^{4}\omega^{4})$ as an important aspect of future work. 

Our aim in this work was not only to obtain explicit results for the tidal response of spinning black holes but also to flesh out the methodology and address issues such as the mode mixing. The methodology has broad applicability and several immediate extensions, for instance, to higher order in frequency and more generally, to describe full tidal resonance effects, and to go beyond the linear response. Applications to other kinds of compact objects, effects of dark matter, and scenarios beyond four-dimensional General Relativity are also left for future work.   

Finally, a main objective of our work is to provide inputs for use in GW measurements of BH binary inspirals. The results developed here can already be used in existing parameterized waveform models, and will serve as a basis for future refinements and inclusion of more realistic tidal effects in full inspiral-merger-ringdown waveform models. 

\section*{Acknowledgements}

The research of SC is supported by MATRICS (MTR/2023/000049) and Core Research Grants (CRG/2023/000934) from SERB, ANRF, Government of India and also in part by the International Centre for Theoretical Sciences (ICTS) for participating in the program --- The Future of Gravitational-Wave Astronomy 2025 (code: ICTS/FGWA2025/10). SC also acknowledges the hospitality of the Albert Einstein Institute, as a part of this work was done during a visit to the Albert Einstein Institute. The research of M.V.S.S. is supported by the National Post-Doctoral Fellowship (PDF/2025/004764), ANRF, Government of India. TH acknowledges funding from the Dutch Research Council (NWO).

\appendix
\labelformat{section}{Appendix #1} 
\labelformat{subsection}{Appendix #1}
\begin{widetext}

\section{Computation of scalar tidal fields using worldline EFT}
\label{app:scalartidescomp}

The part of scalar perturbations which are regular at origin are given by
\begin{align}
    \phi_{\rm reg}=e^{-i\omega t} \sum_{\ell=0}^\infty \frac{1}{\omega^\ell}C^{I_\ell}_{\rm reg}\partial_{I_\ell}\psi_{\rm reg},\quad \psi_{\rm reg}=\frac{\sin(\omega r)}{\omega r},
\end{align}
where $C^{I_\ell}_{\rm reg}$ are symmetric trace-free (STF) tensors. Now, we use the formula given in \cite{creci2021tidal-42e}, which is
\begin{align}
    \partial_{\langle I_\ell \rangle} f(r) = x_{\langle I_\ell \rangle}\left(\frac{1}{r}\frac{\partial}{\partial r}\right)^\ell f(r).
\end{align}
This yields, 
\begin{align}
    \phi = e^{-i\omega t}\sum_{\ell=0}^\infty \frac{1}{\omega^\ell}C^{I_\ell}_{\rm reg} x_{I_\ell} \left(\frac{1}{r}\frac{\partial}{\partial r}\right)^\ell \psi_{\rm reg}.
\end{align}
Now, we use the taylor expansion $\sin(\rho)/\rho=\sum_{k=0}^\infty (-1)^k \rho^{2k}/(2k+1)!$, which leads to 
\begin{align}
    \phi = e^{-i\omega t}\sum_{k=0}^\infty \sum_{\ell=0}^k \omega^\ell C^{I_\ell}_{\rm reg} x_{I_\ell} \frac{(-1)^k 2^\ell k!}{(k-l)!(2k+1)!}(\omega r)^{2k-2\ell}
\end{align}
Now, the tidal fields are given by 
\begin{align}
    E_{J_{\ell'}} = \partial_{\langle J_{\ell'}
    \rangle}\phi_{\rm reg}|_{\vec{x}=0} =   e^{-i\omega t}\sum_{k=0}^\infty \sum_{\ell=0}^k \omega^\ell C^{I_\ell}_{\rm reg} \frac{(-1)^k2^\ell k!}{(k-l)!(2k+1)!}\lim_{\vec{x}\rightarrow 0} \partial_{\langle J_{\ell'}\rangle } x_{\langle I_\ell \rangle} (\omega r)^{2k-2\ell}.
    \label{eq:temp}
\end{align}
It is clear that only those terms which satisfy $\ell'+\ell=2k$ can contribute by counting powers of $x$ and derivatives. Moreover, in the limit where we take it to origin, the result can only depend on invariant tensorial combinations or that of $\hat{n}^i=x^i/r$, but the latter would be ill-defined at origin, and thus eliminated (in favour of invariant tensorial combinations) upon a suitable averaging procedure on a sphere. Thus, only those terms with $\ell'=\ell$ can contribute, as otherwise, the quantity $\lim_{\vec{x}\rightarrow 0} \partial_{\langle J_{\ell'}\rangle } x_{\langle I_\ell \rangle} (\omega r)^{2k-2\ell}$, would need to depend on terms involving the unit vector $\hat{n}_i$. Now, if $\ell'=\ell=k$, then we simply have 
\begin{align}
    E_{J_{\ell}} = \partial_{\langle J_{\ell}
    \rangle}\phi_{\rm reg}|_{\vec{x}=0} =   e^{-i\omega t}\omega^\ell C^{I_\ell}_{\rm reg} \frac{(-1)^\ell 2^\ell \ell!}{(2\ell+1)!}\lim_{\vec{x}\rightarrow 0} \partial_{\langle J_{\ell}\rangle } x_{\langle I_\ell \rangle}.
    \label{eq:temp}
\end{align}

We can proceed as follows: In \ref{eq:temp}, we use the formula 
\begin{align}
    \partial_{\langle I_\ell  \rangle} = \partial_{I_\ell} - \partial^2 \partial_{(I_{\ell-2}} \delta_{i_{\ell-1}i_\ell)}+\dots,
\end{align}
where $(...)$ represents symmetrization. Now, note that these derivatives act upon $x_{I_\ell}$ (where we drop the $\langle...\rangle$ because it is contracted with an STF tensor $C^{I_\ell}_{\rm reg}$). Thus, when any term with fewer than $\ell$ derivatives acts on $x_{I_\ell}$, it will leave behind some powers of $x$ and thus such terms vanish at the origin. Thus, we can simply write
\begin{align}
\begin{split}
        E_{J_{\ell}} &= \partial_{\langle J_{\ell}
    \rangle}\phi_{\rm reg}|_{\vec{x}=0} =   e^{-i\omega t}\omega^\ell C^{I_\ell}_{\rm reg} \frac{2^\ell (-1)^\ell \ell!}{(2\ell+1)!}\lim_{\vec{x}\rightarrow 0} \partial_{\langle J_{\ell}\rangle } x_{ I_\ell } =  e^{-i\omega t}\omega^\ell C^{I_\ell}_{\rm reg} \frac{2^\ell (-1)^\ell \ell!}{(2\ell+1)!}\lim_{\vec{x}\rightarrow 0} \partial_{J_{\ell} } x_{ I_\ell } \\&= e^{-i\omega t}\omega^\ell C^{I_\ell}_{\rm reg} \frac{2^\ell (-1)^\ell \ell!\ell!}{(2\ell+1)!}\delta_{J_\ell,(I_\ell)} = e^{-i\omega t}\omega^\ell C^{J_\ell}_{\rm reg} \frac{2^\ell (-1)^\ell \ell!\ell!}{(2\ell+1)!}.
\end{split}
    \label{eq:temp2}
\end{align}
This concludes the computation. We now show the analogous computation for gravitational case.

\section{Computation of gravitational tidal fields for worldline EFT}
\label{app:gravity}

Given metric perturbations on flat spacetime 
\begin{align}
\begin{split}
            h_{ij} &= \sum_{\ell=2}^\infty \Bigg(C^{K_\ell}_{\mce,\rm reg}\Pi^{k_{\ell-1}k_{\ell}}_{ij}\hat{\partial}_{K_{\ell-2}}\\&+\omega^{-1}C_{\mcb}^{K_{\ell}}\Pi^{k_\ell m}_{ij}\epsilon_{k_{\ell-1}mn}\hat{\partial}_{K_{\ell-2}n} \Bigg)\psi_{\rm reg}+(\rm reg\leftrightarrow \rm irreg),
\end{split}
\end{align}
where $\hat{\partial}_i=\partial_i/\omega$. Note that since $\psi_{\rm reg}$ satisfies the homogeneous wave equation, we have $\hat{\partial}_i\hat{\partial}_i=-1$.
In this appendix, we will show how the electric tidal field may be computed from the expression for the linearized metric perturbation. An analogous procedure can be followed for magnetic tidal fields. We begin by rewriting the regular and electric part  (terms containing $C_{\mce,\rm reg}$) of the above expression in a suitable form. 

For convenience, we drop the `reg' label, and keep all indices as subscripts going forward. Now, using $\Pi^{kl}_{ij}\equiv\Pi_{kl,ij}=(1/2)(P_{ki}P_{lj}+P_{kj}P_{li}-P_{kl}P_{ij})$, where $P_{ij}=\delta_{ij}+\hat{\partial}_i\hat{\partial}_{j}$, and the fact that $C_{\mce,\rm reg}^{K_\ell}\equiv C_{K_{\ell}}$ is an STF tensor, we can write (for a given multipolar mode $\ell$)
\begin{align}
    \begin{split}
    C_{K_\ell}\Pi_{k_{\ell-1}k_{\ell},ij}\hat{\partial}_{K_{\ell-2}}&=C_{K_\ell}(P_{k_{\ell-1}i}P_{k_\ell j}-\frac{1}{2}\hat{\partial}_{k_{\ell-1}}\hat{\partial}_{k_{\ell}}P_{ij})\hat{\partial}_{K_{\ell-2}}, \\
    &= C_{K_{\ell-2}ij}\hat{\partial}_{K_{\ell-2}}+2 C_{K_{\ell-1}(i}\hat{\partial}_{j)}\hat{\partial}_{K_{\ell-1}} -  C_{K_\ell}\frac{1}{2}\delta_{ij}\hat{\partial}_{K_{\ell}} +C_{K_\ell}\frac{1}{2}\hat{\partial}_{i}\hat{\partial}_{j}\hat{\partial}_{K_{\ell}},
    \end{split}
    \label{eq:master}
\end{align}
where $(...)$ implies symmetrization over contained indices. We need to simplify the second and fourth terms in the above expression. Let us simplify the second term first as follows:
\begin{align}
    \begin{split}
        C_{K_{\ell-1}(i}\hp_{j)}\hp_{\langle K_{\ell-1} \rangle} \psi_{\rm reg} &= C_{K_{\ell-1}(i}\hp_{j)} \hat{x}_{\langle K_{\ell-1} \rangle}\left(\frac{\hp_r}{\hat{r}}\right)^{\ell-1} \psi_{\rm reg} \\ &= (l-1)C_{K_{\ell-2}ij}\hat{x}_{K_{\ell-2}}\left(\frac{\hp_r}{\hat{r}}\right)^{\ell-1}\psi_{\rm reg}  + C_{K_{\ell-1}(i}\hat{x}_{j)}\hat{x}_{K_{\ell-1}}\left(\frac{\hp_r}{\hat{r}}\right)^{\ell} \psi_{\rm reg}.
    \end{split}
\end{align}
Similarly, the last (fourth) term can be simplified as 
\begin{align}
    \begin{split}
        C_{K_{\ell}}\hp_i\hp_j\hp_{\langle K_{\ell}\rangle } \psi_{\rm reg} &= C_{K_{\ell}}\hp_i\hp_j \hat{x}_{ K_\ell } \left(\frac{\hp_r}{\hat{r}}\right)^\ell \psi_{\rm reg}
        \\& =  2\ell C_{K_{\ell-1}(i}\hat{x}_{j)}\hat{x}_{K_{\ell-1}}\left(\frac{\hp_r}{\hat{r}}\right)^{\ell+1} \psi_{\rm reg} + (\ell)(\ell-1) C_{K_{\ell-2} ij}\hat{x}_{K_{\ell-2}}\left(\frac{\hp_r}{\hat{r}}\right)^{\ell} \psi_{\rm reg} \\& + \delta_{ij}C_{K_{\ell}}\hat{x}_{K_{\ell}}\left(\frac{\hp_r}{\hat{r}}\right)^{\ell+1} \psi_{\rm reg}+ C_{K_{\ell}}\hat{x}_i \hat{x}_j \hat{x}_{K_{\ell}}\left(\frac{\hp_r}{\hat{r}}\right)^{\ell+2} \psi_{\rm reg}.
    \end{split}
\end{align}
We can now rewrite the last line of \ref{eq:master} as 
\begin{align}
    \begin{split}
        C_{K_\ell}\Pi_{k_{\ell-1}k_{\ell},ij}\hat{\partial}_{K_{\ell-2}} &=  C_{K_{\ell-2}ij}\hat{x}_{K_{\ell-2}}\left(\frac{\hp_r}{\hat{r}}\right)^{\ell-2}\left[1+2(l-1)\left(\frac{\hp_r}{\hat{r}}\right)+\frac{1}{2}\ell(\ell-1)\left(\frac{\hp_r}{\hat{r}}\right)\right] 
        \\
        & + 2 C_{K_{\ell-1}(i}\hat{x}_{j)}\hat{x}_{K_{\ell-1}}\left(\frac{\hp_r}{\hat{r}}\right)^{\ell} \left[1+\frac{\ell}{2} \left(\frac{\hp_r}{\hat{r}}\right) \right] - \frac{1}{2}\delta_{ij}C_{K_\ell}\hat{x}_{K_\ell}\left(\frac{\hp_r}{\hat{r}}\right)^\ell\left[1-\left(\frac{\hp_r}{\hat{r}}\right)\right] 
        \\
        & +\frac{1}{2}C_{K_\ell}\hat{x}_i\hat{x}_j\hat{x}_{K_\ell} \left(\frac{\hp_r}{\hat{r}}\right)^{\ell+2}\,.
    \end{split}
    \label{eq:master2}
\end{align}

Now, we want to argue that the tidal fields only receive contributions from the first term. The definition of electric tidal field is given by
\begin{align}
    \begin{split}
        E_{I_\ell} = \lim_{\vec{x}\rightarrow 0}\partial_{\langle I_{\ell-2}}R_{i_{\ell-1}|0|i_{\ell}\rangle|0} = -\lim_{\vec{x}\rightarrow 0}\frac{\omega^2}{2}\partial_{\langle I_{\ell-2}} h_{i_{\ell-1}i_\ell\rangle}.
    \end{split}
  \label{eq:tidfield}
\end{align}
Now, consider how the derivatives will act on the expression in \ref{eq:master2}. Just as in the scalar case, non-zero contributions in the limit $\vec{x}\rightarrow 0$ can only arise if the number of derivatives ($\ell-2$) match the number of position vectors $\hat{x}_i$. However, when the position vectors contain $\hat{x}_i$ or $\hat{x}_j$, then the derivative acting on them will lead to delta functions of the derivative-index with $i$ or $j$, which will be eliminated when we symmetrize and remove traces. Similarly, the term with $\delta_{ij}$ can also be dropped as it is eliminated upon trace-removal. The magnetic part of the metric perturbations will not contribute due to parity considerations. Thus, only the first term in \ref{eq:master2} will contribute and we can write
\begin{align}
    E_{I_{\ell_0}} = - \frac{\omega^{\ell_0}}{2}\sum_{\ell=2}^\infty\lim_{\vec{x}\rightarrow 0} C_{K_{\ell-2}\langle i_{\ell_0-1} i_{\ell_0}}\hp_{I_{\ell_0-2}\rangle} \hat{x}_{K_{\ell-2}}\left(\frac{\hp_r}{\hat{r}}\right)^{\ell-2}\left[1+2(l-1)\left(\frac{\hp_r}{\hat{r}}\right)+\frac{1}{2}\ell(\ell-1)\left(\frac{\hp_r}{\hat{r}}\right)\right]\psi_{\rm reg}.
\end{align}
We can evaluate this similar to the scalar case. Once again, let us first consider evaluating the above expression without symmetrization and trace-removal, and take it at the final stage. We have 
\begin{align}
    \begin{split}
        - \frac{\omega^{\ell_0}}{2}\sum_{\ell=2}^\infty\lim_{\vec{x}\rightarrow 0} C_{K_{\ell_0-2} i_{\ell_0-1} i_{\ell_0}}\hp_{I_{\ell_0-2}} \hat{x}_{K_{\ell-2}}\left(\frac{\hp_r}{\hat{r}}\right)^{\ell-2}\left[1+2(l-1)\left(\frac{\hp_r}{\hat{r}}\right)+\frac{1}{2}\ell(\ell-1)\left(\frac{\hp_r}{\hat{r}}\right)\right]\psi_{\rm reg}.
    \end{split}
\end{align}
Now, as reasoned before in the scalar case, only terms where $\ell_0=\ell$ can contribute at the origin, as any other terms will produce terms that will be eliminated upon symmetrization and trace-removal later, and thus we can drop the sum. Further more, using the Taylor expansion of $\sin(x)/x=\sum_{k=0}^\infty (-1)^k \rho^{2k}/(2k+1)!$ as before, we have
\begin{align}
     - \frac{\omega^{\ell_0}}{2} \lim_{\vec{x}\rightarrow 0} C_{K_{\ell_0-2} i_{\ell_0-1} i_{\ell_0}}(\hp_{I_{\ell_0-2}} \hat{x}_{K_{\ell_0-2}})\sum_{k=\ell_0-2}^\infty \frac{(-1)^k k!}{(2k+1)!} \hat{r}^{2k-2\ell_0}\left(\frac{2^{\ell_0-2}}{(k-\ell_0+2)!}\hat{r}^{4}+\frac{(\ell_0-1)2^{\ell_0}}{(k-\ell_0+1)!}\hat{r}^{2}+\frac{\ell_0(\ell_0-1)2^{\ell_0-1}}{(k-\ell_0)!}\right).
\end{align}
Now, again, the relevant terms at origin are obtained when $k=\ell_0-2$ in first term, $k=\ell_0-1$ in second, and $k=\ell_0$ in the last term, yielding
\begin{align}
     - \frac{\omega^{\ell_0}}{2} C_{K_{\ell_0-2}i_{\ell_0-1}i_{\ell_0}}(\hp_{I_{\ell_0-2}} \hat{x}_{K_{{\ell_0}-2}})(-1)^{\ell_0} 2^{-\ell_0-2}\frac{(1+\ell_0)(2+{\ell_0})\sqrt{\pi}}{\Gamma(\frac{3}{2}+\ell_0)}.
\end{align}
Finally, evaluating the derivatives (and removing traces and symmetrization which is trivial) we get
\begin{align}
    E_{I_{\ell_0}}=(-1)^{\ell_0+1} \omega^{\ell_0 }C_{I_{\ell_0}}2^{-\ell_0-3}\frac{(1+\ell_0)(2+\ell_0)\sqrt{\pi}}{\Gamma(\frac{3}{2}+\ell_0)}.
\end{align}
The magnetic field can be evaluated similarly.

\section{Dynamical scalar response function}\label{App:ScalarResponse}

From the EFT matching with the BH perturbation theory computation of the dynamical response function for Kerr BH under scalar tidal perturbation, one obtains the expression in \ref{matching_scalar}. Also here, $\hl=\ell+\{2ma\omega/(2\ell+1)\}+\mathcal{O}(M^{2}\omega^{2})$. To simplify the above expression, we may use the following identities, 
\begin{align}
\Gamma(-1-2\hl)&=\frac{\pi}{\sin(2\pi\hl)\Gamma(2+2\hl)}~,
\\
\Gamma(-\hl)&=-\frac{\pi}{\sin(\pi\hl)\Gamma(1+\hl)}~;
\\
\Gamma(-\hl-2iP_{+})&=-\frac{\pi}{\sin(\pi\hl+2i\pi P_{+})\Gamma(1+\hl+2iP_{+})}~;
\end{align}
to obtain the following expression for rescaled response function,
\begin{align}
\widetilde{\lambda}_{\ell}^{\rm Kerr}(\omega)&
=\frac{\sin(\pi\hl+2i\pi P_{+})}{2\pi\cos(\pi\hl)}
\times \frac{\Gamma(1+\hat{\ell}+2iP_{+})\Gamma(1+\hat{\ell}-2iP_{+})\Gamma(1+\hat{\ell})^{2}}{\Gamma\left(2\hat{\ell}+1\right)\Gamma\left(2\hl+2\right)}~.
\end{align}
Expressing $(\hl-\ell)=-2am\omega/(2\ell+1)\equiv\delta\ell$, and also decomposing $P_{+}=P_{+}^{a}+P_{+}^{\omega}$, where $P_{+}^{a}=-\{am/(r_{+}-r_{-})\}$ and $P_{+}^{\omega}=\{2Mr_{+}\omega/(r_{+}-r_{-})\}$, such that $P_{+}^{\omega}\ll 1$, we obtain the ratio of the sinusoidal functions appearing above as, 
\begin{align}
\frac{\sin(\pi\hl+2i\pi P_{+})}{2\pi\cos(\pi\hl)}&=\frac{\sin(\pi\ell+\pi \delta \ell+2i\pi P_{+}^{a}+2i\pi P_{+}^{\omega})}{2\pi\cos(\pi\ell+\pi \delta \ell)}
=\frac{\sin(\pi\ell+2i\pi P_{+}^{a})+\cos(\pi\ell+2i\pi P_{+}^{a})(\pi \delta \ell+2i\pi P_{+}^{\omega})}{2\pi \cos(\pi \ell)}
\nonumber
\\
&=\frac{1}{2\pi}\left[\sin(2i\pi P_{+}^{a})+\cos(2i\pi P_{+}^{a})(\pi \delta \ell+2i\pi P_{+}^{\omega})\right]\,.
\end{align}
Similarly, the following combination of Gamma functions yield, 
\begin{align}
&\Gamma(1+\hat{\ell}+2iP_{+})\Gamma(1+\hat{\ell}-2iP_{+})
=\Gamma(1+\ell+\delta \ell+2iP_{+}^{a}+2iP_{+}^{\omega})\Gamma(1+\ell+\delta \ell-2iP^{a}_{+}-2iP^{\omega}_{+})
\nonumber
\\
&=\Gamma(1+\ell+2iP^{a}_{+})\Gamma(1+\ell-2iP^{a}_{+})\Big[1+(\delta \ell+2iP_{+}^{\omega})
\Psi(1+\ell+2iP^{a}_{+})+(\delta \ell-2iP_{+}^{\omega})\Psi(1+\ell-2iP^{a}_{+})\Big]
\nonumber
\\
&=\frac{2\pi P_{+}^{a}}{\sinh(2\pi P_{+}^{a})}\prod_{k=1}^{\ell}\left(k^{2}+4\{P_{+}^{a}\}^{2}\right)\Big[1+(\delta \ell+2iP_{+}^{\omega})
\Psi(1+\ell+2iP^{a}_{+})+(\delta \ell-2iP_{+}^{\omega})\Psi(1+\ell-2iP^{a}_{+})\Big]\,.
\end{align}
Other Gamma functions can also be expanded as $\Gamma(\ell+\Delta \ell)=\Gamma(\ell)\{1+\delta \ell\Psi(\ell)\}$. Using all of these relations, we can express the redefined/rescaled response function as, 
\begin{align}
\widetilde{\lambda}_{\ell}^{\rm Kerr}(\omega)&=\left[i\sinh(2\pi P_{+}^{a})+\cosh(2\pi P_{+}^{a})(\pi \delta \ell+2i\pi P_{+}^{\omega})\right]\frac{P_{+}^{a}}{\sinh(2\pi P_{+}^{a})}\prod_{k=1}^{\ell}\left(k^{2}+4\{P_{+}^{a}\}^{2}\right)
\nonumber
\\
&\times \Big[1+(\delta \ell+2iP_{+}^{\omega})
\Psi(1+\ell+2iP^{a}_{+})+(\delta \ell-2iP_{+}^{\omega})\Psi(1+\ell-2iP^{a}_{+})\Big]
\nonumber
\\
&\times \frac{\Gamma(1+\ell)^{2}}{\Gamma(2+2\ell)\Gamma(1+2\ell)}\times \Big[1+2\delta \ell \Psi(1+\ell)-2\delta \ell\Psi(2+2\ell)-2\delta \ell\Psi(1+2\ell) \Big]\,.
\end{align}
This is the result used in the main text. Further, using the identity: 
\begin{align}
\Gamma(2z)&=\frac{2^{2z-1}}{\sqrt{\pi}}\Gamma(z)\Gamma\left(z+\frac{1}{2}\right)~,
\end{align}
we can express the Gamma functions appearing above as, 
\begin{align}\label{gammatwiceid}
\frac{\Gamma(1+\ell)^{2}}{\Gamma(2+2\ell)\Gamma(1+2\ell)}&=\frac{\Gamma(1+\ell)^{2}}{\left[\frac{2^{2\ell+1}}{\sqrt{\pi}}\Gamma(\ell+1)\Gamma\left(\ell+\frac{3}{2}\right)\right]\left[\frac{2^{2\ell}}{\sqrt{\pi}}\Gamma\left(\ell+\frac{1}{2}\right)\Gamma(\ell+1)\right]}
\nonumber
\\
&=\frac{\pi}{2^{4\ell+1}\Gamma\left(\ell+\frac{3}{2}\right)\Gamma\left(\ell+\frac{1}{2}\right)}\,.
\end{align}

\section{Useful Mathematical Identities}\label{AppA}

\begin{enumerate}

\item The confluent hypergeometric function $U(a,b;z)$ can be expressed in terms of the other function $M(a,b;z)$, as follows [DLMF (13.2.42)],
\begin{align}\label{UasM}
U(a,b;z)=\frac{\Gamma(1-b)}{\Gamma(a-b+1)}M(a,b;z)+\frac{\Gamma(b-1)}{\Gamma(a)}z^{1-b}M(a-b+1,2-b;z)\,.
\end{align}

\item The limit of the confluent hypergeometric functions in the zero limit takes the following form, 
\begin{align}\label{limitzeroMU}
\lim_{z\to 0} M(a,b;z)\sim 1~;
\qquad
\lim_{z\to 0} U(a,b;z)\sim \frac{\pi}{\sin(\pi b)}\left[\frac{1}{\Gamma(b)\Gamma(1+a-b)}-\frac{z^{1-b}}{\Gamma(a)\Gamma(2-b)}\right]\,.
\end{align}

\item The asymptotic limit of the confluent hypergeometric function $M(a,b;z)$ becomes, 
\begin{align}\label{Minfty}
\lim_{z\to \infty}M(a,b;z)=\frac{\Gamma(b)}{\Gamma(b-a)}e^{\mp i\pi a}z^{-a}
+\frac{\Gamma(b)}{\Gamma(a)}e^{z}z^{a-b}~.
\end{align}

\item For a confluent hypergeometric function $M(a,b;z)$, if the coefficients $a$ and $b$ satisfies the relation $b=2a$, then we have
\begin{align}
M\left(\nu+\frac{1}{2},2\nu+1;2z\right)=\Gamma(1+\nu)e^{z}\left(\frac{z}{2}\right)^{-\nu}I_{\nu}(z)~,
\label{zeroM}
\\
U\left(\nu+\frac{1}{2},2\nu+1;2z\right)=\frac{1}{\sqrt{\pi}}e^{z}\left(2z\right)^{-\nu}K_{\nu}(z)~.
\label{zeroU}
\end{align}

\item For a confluent hypergeometric function $M(a,b;z)$, if the coefficients $a$ and $b$ satisfies the relation $b-2a$ is an integer, then we have two useful identities: 
\begin{align}
M\left(\nu+\frac{1}{2},2\nu+1+n;2z\right)&=\Gamma(\nu+1)e^{z}\left(\frac{z}{2}\right)^{-\nu}
\nonumber
\\
&\times \sum_{k=0}^{n}\frac{(-n)_{k}}{k!}\frac{\Gamma(2\nu+k)}{\Gamma(2\nu)}\frac{\Gamma(2\nu+1+n)}{\Gamma(2\nu+1+n+k)}\frac{\nu+k}{\nu}\,I_{\nu+k}(z)\,,
\label{Mtwice1}
\\
M\left(\nu+\frac{1}{2},2\nu+1-n;2z\right)&=\Gamma(\nu-n+1)e^{z}\left(\frac{z}{2}\right)^{n-\nu}
\nonumber
\\
&\times \sum_{k=0}^{n}(-1)^{k}\frac{(-n)_{k}}{k!}\frac{\Gamma(2\nu-2n+k)}{\Gamma(2\nu-2n)}\frac{\Gamma(2\nu+1-n)}{\Gamma(2\nu+1-n+k)}\frac{\nu-n+k}{\nu-n}\,I_{\nu+k}(z)\,.
\label{Mtwice2}
\end{align}

\item These identities relate the modified Bessel functions to Bessel functions of first and second kind.
\begin{align}\label{identity_Bessel}
I_{\nu}(z)=e^{\mp i\pi \nu/2}J_{\nu}(\pm iz)~;
\quad
K_{\nu}(z)=\frac{\pi}{2}\left(\frac{I_{-\nu}(z)-I_{\nu}(z)}{\sin(\pi\nu)}\right)~;
\quad 
J_{-(m+\frac{1}{2})}(x)=(-1)^{m+1}Y_{m+\frac{1}{2}}(x)~,
\end{align}

\end{enumerate}

\section{Dynamical response function: Generic Spin}\label{App:GensResponse}

In this appendix, we will present all the results for the simplification of the dynamical response function of a Kerr BH under generic spin perturbation. The starting point is \ref{dyn_resp_fn_01} in the main text, and here we will simplify parts of that result. First of all, the collection of the Gamma functions yield,
\begin{align}
&\frac{\Gamma(1+\hl_{s}-2iP_{+})\Gamma(1+\hl_{s}-s)\Gamma(1+\hl_{s}+s)\Gamma(1+\hl_{s}+2iP_{+})}{\Gamma(2+2\hl_{s})\Gamma(1+2\hl_{s})\Gamma(\bar{\mu})\Gamma(1+\bar{\nu})}=\frac{\Gamma(1+\ell-2iP^{a}_{+})\Gamma(1+\ell-s)\Gamma(1+\ell+s)\Gamma(1+\ell+2iP^{a}_{+})}{\Gamma(2+2\ell)\Gamma(1+2\ell)\Gamma(\bar{\mu}_{e})\Gamma(1+\bar{\nu}_{e})}
\nonumber
\\
&\quad \times\Big[1+\Delta \ell\left\{\Psi(1+\ell-2iP_{+})+\Psi(1+\ell-s)+\Psi(1+\ell+s)+\Psi(1+\ell+2iP^{a}_{+})-2\Psi(2+2\ell)-2\Psi(1+2\ell)\right\} 
\nonumber
\\
&\qquad +2iP_{+}^{\omega}\left\{\Psi(1+\ell+2iP^{a}_{+})-\Psi(1+\ell-2iP_{+})\right\}-\epsilon_{\ell}M\omega\left\{\Psi(\bar{\mu}_{e})+\Psi(1+\bar{\nu}_{e})\right\}\Big]\,,
\end{align}
where, 
\begin{equation}
\label{eq:Pplusadef}
P^{a}_{+}=-am/(r_{+}-r_{-}), \qquad  P_{+}^{\omega}=2Mr_{+}\omega/(r_{+}-r_{-}),
\end{equation}
and $\Delta \ell_{s}$ has the following expression:
\begin{align}
\label{eq:Deltaelldef}
\Delta \ell_{s}=\omega \left[\frac{4isr_{+}}{2\ell+1}-\frac{2am}{2\ell+1}-\frac{2ams^{2}}{\ell(\ell+1)(2\ell+1)}\right]\,.
\end{align}
The leading order contribution from the Gamma function turns out to be, 
\begin{align}
\frac{\Gamma(1+\ell-2iP^{a}_{+})\Gamma(1+\ell-s)\Gamma(1+\ell+s)\Gamma(1+\ell+2iP^{a}_{+})}{\Gamma(2+2\ell)\Gamma(1+2\ell)\Gamma(\bar{\mu}_{e})\Gamma(1+\bar{\nu}_{e})}=\frac{2\pi P_{+}^{a}}{\sinh(2\pi P_{+}^{a})}\frac{(\ell-s)!(\ell+s)!}{(2\ell+1)!(2\ell)!}\frac{\prod_{n=1}^{\ell}\{n^{2}+4(P_{+}^{a})^{2}\}}{\Gamma(\bar{\mu}_{e})\Gamma(1+\bar{\nu}_{e})}\,.
\end{align}
Similarly, the combination of the sinusoidal functions appearing in the response function, given by \ref{dyn_resp_fn_01}, can be re-expressed as,
\begin{align}
\frac{\sin[\pi\hat{\ell}_{s}+2i\pi P_{+}]}{2\cos(\pi \hl_{s})}=\frac{i}{2}\sinh(2\pi P_{+}^{a})+\left(\frac{\pi \Delta \ell+2i\pi P_{+}^{\omega}}{2}\right)\cosh(2\pi P_{+}^{a})\,.
\end{align}
Combining all of these, the dynamical response function of a Kerr BH under generic spin perturbation reduces to, 
\begin{align}\label{dyn_resp_fn_02}
{}_{-2}\widetilde{\lambda}_{\ell m}&=\frac{(\ell-s)!(\ell+s)!}{(2\ell+1)!(2\ell)!}\prod_{n=1}^{\ell}\{n^{2}+4(P_{+}^{a})^{2}\}\left[iP_{+}^{a}+\left(\pi \Delta \ell+2i\pi P_{+}^{\omega}\right)P_{+}^{a}\coth(2\pi P_{+}^{a})\right]
\nonumber
\\
&\times\Big[1+\Delta \ell\left\{\Psi(1+\ell-2iP_{+})+\Psi(1+\ell-s)+\Psi(1+\ell+s)+\Psi(1+\ell+2iP^{a}_{+})-2\Psi(2+2\ell)-2\Psi(1+2\ell)\right\} 
\nonumber
\\
&\qquad +2iP_{+}^{\omega}\left\{\Psi(1+\ell+2iP^{a}_{+})-\Psi(1+\ell-2iP_{+})\right\}-\epsilon_{\ell}M\omega\left\{\Psi(\bar{\mu}_{e})+\Psi(1+\bar{\nu}_{e})\right\}\Big]\,.
\end{align}
This is the result we have used in the main text. 



\section{
Normalization relevant for spheroidal harmonic decomposition
}\label{normalization}

Here, we show how to derive \ref{eq:legend}, restated below.
\begin{equation}
\begin{split}
       &\int d\Omega N_{\langle I_{\ell-2}}m_{i_{\ell-1}}m_{i_\ell\rangle}  N^{\langle J_{\ell'-2}}\bar{m}^{j_{\ell'-1}}\bar{m}^{j_\ell'\rangle } \\ & = \delta_{\ell \ell'} \delta^{J_{\ell}}_{I_\ell} \frac{8\pi(\ell-2)!}{\ell(\ell-1)(2\ell+1)(2\ell-3)!!},
\end{split}
\label{eq:legendapp}
\end{equation}
where $\delta^{J_\ell}_{I_\ell} = \delta^{\langle j_1}_{i_1} \delta^{j_2}_{i_2} \dots  \delta^{j_\ell\rangle}_{i_\ell}$.
We will use the following facts:
\begin{equation}
    \begin{split}
        N^iN_i = 1,~N^i m_i = 0,~ m^im_i = 0, ~m^i \bar{m}_i = 1.
    \end{split}
    \label{eq:identities}
\end{equation}
We start by establishing 
\begin{equation}
\begin{split}
       &\int d\Omega N_{\langle I_{\ell-2}}m_{i_{\ell-1}}m_{i_\ell\rangle}  N^{\langle J_{\ell'-2}}\bar{m}^{j_{\ell'-1}}\bar{m}^{j_\ell'\rangle } =\alpha \delta^{\ell'}_{\ell}\delta^{J_{\ell}}_{I_\ell}.
\end{split}
\end{equation}
This follows simply from symmetries. The answer must be STF in $I_\ell$ and $J_\ell'$ separately. It must comprise of invariant tensors $\delta_{ij}$ and $\epsilon^{ijk}$ with coefficients determined by the dot products between the various vectors. These two statements, along with the orthogonality of $N^i$ and $m^i$ are sufficient to estalish this result. Now, let us contract both sides for $\ell'=\ell$ as
\begin{equation}
    \begin{split}
        \int d\Omega N_{\langle I_{\ell-2}}m_{i_{\ell-1}}m_{i_\ell\rangle}  N^{\langle I_{\ell-2}}\bar{m}^{i_{\ell-1}}\bar{m}^{i_\ell\rangle } &=\alpha \delta^{I_{\ell}}_{I_\ell} = \alpha (2\ell+1), \\
        \int d\Omega N_{ I_{\ell-2}}m_{i_{\ell-1}}m_{i_\ell}  N^{\langle I_{\ell-2}}\bar{m}^{i_{\ell-1}}\bar{m}^{i_\ell\rangle } &=\alpha(2\ell+1), \\
        \int d\Omega N_{ I_{\ell-2}}m_{i_{\ell-1}}m_{i_\ell}  N^{(I_{\ell-2}}\bar{m}^{i_{\ell-1}}\bar{m}^{i_\ell)} &=\alpha(2\ell+1), \\
        \frac{(\ell-2)!2!}{\ell!}\int d\Omega N_{ I_{\ell-2}} N^{I_{\ell-2}} &=\alpha(2\ell+1),
    \end{split}
    \label{eq:inter}
\end{equation}
where in the third line we used the fact that $N_{I_{\ell-2}}$ is already STF, and second relation from the identities in \ref{eq:identities} to replace the STF brackets with symmetrization. In the last line, we once again used the orthogonality of $\vec{N}$ and $\vec{m}$, along with the last relation in \ref{eq:identities}.

Finally, we use the equation in Ref.~\cite{RevModPhys.52.299}
\begin{equation}
    \int d\Omega N_{I_\ell}N^{I_\ell} =  \frac{4\pi\ell!}{(2\ell+1)!!},
\end{equation}
to further evaluate \ref{eq:inter} as
\begin{equation}
    \begin{split}
         \frac{(\ell-2)!2!}{\ell!}\int d\Omega N_{ I_{\ell-2}} N^{I_{\ell-2}} &=\alpha(2\ell+1),\\
         \frac{(\ell-2)!2!}{\ell!} \frac{4\pi (\ell-2)!}{(2\ell-3)!!}&=\alpha(2\ell+1) \implies \alpha=\frac{8\pi(\ell-2)!}{\ell(\ell-1)(2\ell+1)(2\ell-3)!!}.
    \end{split}
\end{equation}

\end{widetext}


\bibliographystyle{./utphys1}
\end{document}